\documentclass[preprint]{aastex631}
\usepackage[utf8]{inputenc}
\usepackage{array}
\usepackage{enumitem} % For lists inside table cells

\usepackage{subfiles}
\usepackage[normalem]{ulem}
\usepackage{multirow}
\usepackage{graphicx,xspace}
\usepackage{ltablex} % Ensure compatibility for longtable*

\usepackage{amsmath} % Added for alignment and mathematical symbols
\def\ero{\textit{eROSITA}\xspace}
\def\gai{\textit{Gaia}\xspace}
\DeclareUnicodeCharacter{0301}{\'e}
\DeclareUnicodeCharacter{0300}{\`{}}
\begin{document}
%\linenumbers
\title{Sloan Digital Sky Survey-V: Pioneering Panoptic Spectroscopy}
\author[0000-0001-9852-1610]{Juna A. Kollmeier}
\affiliation{The Observatories of the Carnegie Institution for Science, 813 Santa Barbara Street, Pasadena, CA 91101, USA}
\affiliation{Canadian Institute for Theoretical Astrophysics, 
University of Toronto, Toronto, ON M5S-98H, Canada}
\affiliation{Canadian Institute for Advanced Research, 661 University Avenue, Suite 505, Toronto, ON M5G 1M1 Canada}
\email{jak@carnegiescience.edu}

\author[0000-0003-4996-9069]{Hans-Walter Rix}
\affiliation{Max-Planck-Institut f\"ur Astronomie, K\"onigstuhl 17, D-69117 Heidelberg, Germany}

\author[0000-0003-1822-7126]{Conny Aerts}
\affiliation{Institute of Astronomy, KU Leuven, Celestijnenlaan 200D, B-3001 Leuven, Belgium}
\affiliation{Max-Planck-Institut f\"ur Astronomie, K\"onigstuhl 17, D-69117 Heidelberg, Germany}

\author[0000-0003-1908-8463]{James Aird}
\affiliation{Institute for Astronomy, University of Edinburgh, Royal Observatory, Edinburgh EH9 3HJ, UK}

\author{Pablo Vera Alfaro}
\affiliation{Las Campanas Observatory, Ra\'ul Bitr\'an 1200, La Serena, Chile}

\author[0009-0000-0733-2479]{Andr\'es Almeida}
\affiliation{Department of Astronomy, University of Virginia, Charlottesville, VA 22904-4325, USA}

\author[0000-0002-6404-9562]{Scott F. Anderson}
\affiliation{Department of Astronomy, University of Washington, Box 351580, Seattle, WA 98195, USA}

\author[0000-0001-7434-5165]{\'Oscar Jim\'enez Arranz}
\affiliation{Institut de Ci\'encies del Cosmos, Universitat de Barcelona, Mart\'ii Franqu\`es 1, 08028 Barcelona, Spain}

\author[0000-0002-6270-8624]{Stefan M. Arseneau}
\affiliation{Center for Astrophysical Sciences, Department of Physics and Astronomy, Johns Hopkins University, 3400 North Charles Street, Baltimore, MD 21218, USA}
\affiliation{Department of Astronomy \& Institute for Astrophysical Research, Boston University, 725 Commonwealth Ave., Boston, MA 02215, USA}

\author{Roberto Assef}
\affiliation{Instituto de Estudios Astrof\'isicos, Universidad Diego Portales, Facultad de Ingenier\'ia y Ciencias, Av. Ej\'ercito, Libertador 441, Santiago, Chile}

% ...

%\collaboration{20}{(AAS Journals Data Editors)}
 
\begin{abstract} %250 words
The \textit{Sloan Digital Sky Survey V} (SDSS-V) is pioneering panoptic spectroscopy: it is the first all-sky, multi-epoch, optical-to-infrared spectroscopic survey. SDSS-V is mapping the sky with multi-object spectroscopy (MOS) at telescopes in both hemispheres (the 2.5-m Sloan Foundation Telescope at Apache Point Observatory and the 100-inch du Pont Telescope at Las Campanas Observatory), where 500 zonal robotic fiber positioners feed light from a wide-field focal plane to an optical (R$\sim 2000$, 500 fibers) and a near-infrared (R$\sim 22,000$, 300 fibers) spectrograph. In addition to these MOS capabilities, the survey is pioneering ultra wide-field ($\sim$ 4000~deg$^2$) integral field spectroscopy enabled by a new dedicated facility (LVM-I) at Las Campanas Observatory, where an integral field spectrograph (IFS) with 1801 lenslet-coupled fibers arranged in a 0.5 degree diameter hexagon feeds multiple R$\sim$4000 optical spectrographs that cover 3600-9800 \AA.  SDSS-V's hardware and multi-year survey strategy are designed to decode the chemo-dynamical history of the Milky Way Galaxy and tackle fundamental open issues in stellar physics in its \textit{Milky Way Mapper} program, trace the growth physics of supermassive black holes in its \textit{Black Hole Mapper} program, and understand the self-regulation mechanisms and the chemical enrichment of galactic ecosystems at the energy-injection scale in its \textit{Local Volume Mapper} program.  The survey is well-timed to multiply the scientific output from major all-sky space missions. The SDSS-V MOS programs began robotic operations in 2021; IFS observations began in 2023 with the completion of the LVM-I facility.  SDSS-V builds upon decades of heritage of SDSS's pioneering advances in data analysis, collaboration spirit, infrastructure, and product deliverables in astronomy. 
%Join us or die jealous.
\end{abstract}

\section{Introduction}
\label{sec:intro}
Humans have long been mapping the sky in ever greater fidelity and detail, with telescopy being a critical, but relatively recent \citep{Galileo1610}, invention.  The recognition by modern scientists that the Universe itself and its contents are not static, but inherently dynamic \citep[e.g.][]{Lemaitre1927}, along with the theoretical, methodological, and industrial techniques to probe this dynamical state, naturally culminates with all-sky time-domain surveys across different photon and gravitational wave energies as the ultimate empirical basis to understand our universe. 

Scientifically, and practically, there are at least three simple and generic arguments for {\it all-sky} astrophysical surveys. 

\begin{itemize} 
\item{Both the most nearby sources (e.g., exoplanets and their hosts, low-mass dwarf stars) and cosmologically distant sources are distributed uniformly across the whole sky. Therefore, exoplanet research and cosmology, as cornerstones of modern astrophysics, {\it natively} rely on all-sky surveys. Whenever one is seeking the best, brightest, or most interesting example of such astrophysical phenomena, it could exist anywhere in the sky.}
\item{ The Milky Way (MW) Galaxy is the single best ``local laboratory'' we have to study the inner workings of galaxy formation and evolution, as well as the physics of stars from their birth to their end. Because we observe it from within, the MW is the ultimate all-sky study subject, where neither the northern nor the southern hemisphere alone can possibly provide the full perspective.}
\item{The orbital requirements and scanning geometry of multiyear space survey missions naturally enable coverage of the whole sky. This has been realized from microwaves ({\it COBE}\;\citep{COBE1992}, {\it WMAP}\; \citep{WMAP2003}, {\it Planck}\;\citep{Planck2016}), to the infrared ({\it IRAS}\;\citep{IRAS1984}, {\it WISE}\;\citep{WISE2010}), optical ({\it Gaia}\;\citep{Gaia2016}, {\it TESS}\;\citep{TESS2015}), UV ({\it Galex}\;\citep{galex2005}), all the way to X-rays ({\it ROSAT}\;\citep{ROSAT1999}, and {\it eROSITA}\;\citep{eROSITA2021}) and gamma rays ({\it Fermi}\;\citep{atwood2009fermi}). The scientific value of these costly but valuable data sets is greatly enhanced by complementary ground-based observations, which naturally should then be all-sky.}
\end{itemize}

However, to date, there has never been an all-sky survey providing comprehensive optical and IR spectra\footnote{In the optical, ESA's {\it Gaia} mission has two spectroscopic instruments that have delivered spectroscopic all-sky optical data. Gaia's BP/RP (XP) instrument provides spectral information across the optical range, but at very low resolution ($R\lesssim 100$). Gaia's RVS instrument provides optical spectra for bright targets, but only across a very narrow spectral range.  The utility of these spectra is enormously enhanced by high-quality ``training" spectra provided, e.g. by previous phases of SDSS \citep{Andrae2023}.}, let alone an optical IFU program that attempts to ``spectroscopize" \textcolor{red}{}{enormous, contiguous sectors of the sky}.  {\bf The SDSS-V Project is now pioneering such ``panoptic\footnote{panoptic: presenting a comprehensive or encompassing view of the whole.} spectroscopy'' by providing the first optical plus infrared survey of multi-object spectroscopy (MOS) for millions of sources spread across the entire sky and the first ultra-wide-field optical IFU coverage of a significant fraction (about 1/10th) of the celestial sphere.}

All astronomical sources are changing with time if one looks closely enough: they are moving through space and/or changing energy amplitudes and power spectra for a vast range of physically interesting reasons. Therefore, it makes sense to probe these objects in the time domain, to understand the processes that drive their evolution. Examples of periodic changes include planets that occult their hosts, stars that oscillate, close binary components that deform each other, etc. Transient phenomena include the incessant ringing and death throes of stars, magnetic flaring, variations in black hole accretion rates, etc. The time domain has been recognized as one of the great astrophysical frontiers of modernity. This is reflected in the large investments in space missions such as {\it Kepler}, {\it Gaia}, {\it TESS}, {\it PLATO} and {\it Roman}, and ground-based surveys such as PTF, PS1, ZTF, and LSST. All of these projects are, at root, time-domain imaging surveys. Yet, there has been no comparably coherent effort to systematically explore and exploit the time domain with spectroscopy across the sky. {\it The second key aspect of SDSS-V involves expanding its panoptic spectroscopy to include multi-epoch survey spectroscopy.}

\begin{figure}[!ht]\centering\includegraphics[width=0.9\linewidth]{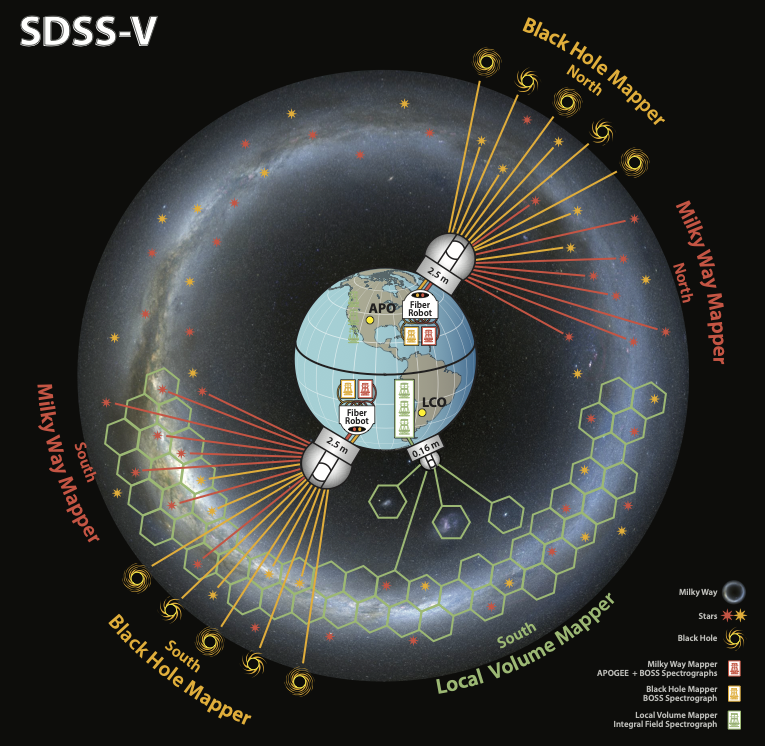}\caption{\footnotesize\linespread{1.2}\selectfont\textbf{A schematic 
  representation of SDSS-V:} an all-sky, multi-epoch spectroscopic facility and its science programs. Dual-hemisphere survey operations are undertaken at Apache Point Observatory (APO) and Las Campanas Observatory (LCO). Multi-object fiber spectroscopy is being carried out with two 2.5~m telescopes, each feeding an near-IR APOGEE spectrograph (300 fibers, $R \sim 22,000$) and an optical BOSS spectrograph (500 fibers, $R \sim 2,000$). This enables a sky-survey rate of $\gtrsim$ 25~deg$^2$/hour. Ultra-wide field integral field spectroscopy is being carried out by the LVM-I at LCO using a smaller 0.16~m telescope, with a $\sim$2,000-fiber IFU feeding three optical $R\sim4,000$ spectrographs. This schematic also outlines the three primary science programs: the Milky Way Mapper (red), drawing on both APOGEE (red) and BOSS (yellow) spectra; the Black Hole Mapper (orange), taking BOSS spectra of fainter targets; and the Local Volume Mapper (green), performing IFU mapping of the ionized ISM in the MW and nearby galaxies. (Image Credit: M. Seibert)}
  \label{fig:SDSSV_schematic}
\end{figure}

\begin{figure}[!ht]\centering\includegraphics[width=0.9\linewidth]{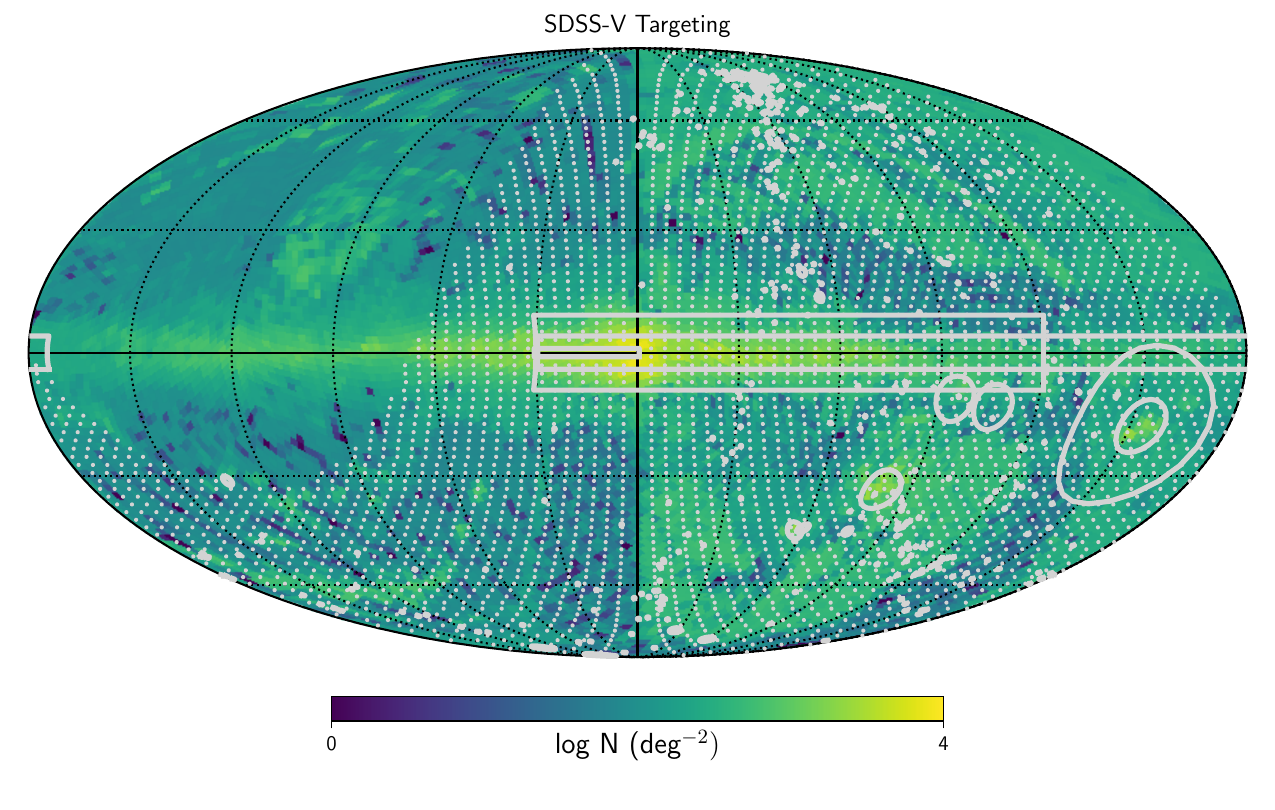}\caption{\footnotesize\linespread{1.2}\selectfont{}{\bf Target density of SDSS-V's two MOS mappers:} 
  The MOS target densities (MWM and BHM) across the sky are illustrated in a logarithmic color scale, ranging from 30 to 15,000 targets / deg$^2$.   While the target {\it density} is not uniform across the entire sky, the reache of the program is clear. The LVM survey footprint is overlaid in grey.}
  \label{fig:SDSSV_targets_1}
\end{figure}

However, many astrophysical phenomena do not lend themselves to being parsed into sets of discrete, 
compact sources to be observed by multiobject spectroscopy. These phenomena call for spatially contiguous spectral mapping, or integral field spectroscopy (IFS), over a wide range of angular and physical 
scales. Pushing IFS to the all-sky regime is even more daunting than all-sky MOS, 
as evidenced by the fact that the largest  contiguous (optical) IFS maps of the sky before SDSS-V covered only $0.001$\% of the celestial sphere.

\emph{SDSS-V's third central goal is to bring IFS to a regime of mapping an appreciable portion} ($\gtrsim$4,000~deg$^2$, $\sim$10\%)  \emph{of the sky with comprehensive optical spectroscopy.}

Most ground-based telescopes have tiny fields of view and cannot see the entire sky, which creates seeming tension with SDSS-V's three central goals: to produce high-quality, near-IR and optical MOS of 
$\sim 6\times 10^6$ objects across the entire sky; to produce homogeneous
multi-epoch spectroscopy for $10^6$ objects across the sky; and to carry out ultra-wide-field IFS mapping across more than 4,000~deg$^2$ of the sky. These advances can only be achieved with the right combination of hardware and survey strategy. They require wide-field telescopes and the possibility to re-acquire new spectroscopic targets rapidly (e.g., every 15 minutes). Additionally, 
they require a focus on brighter sources  ($H<11$ and $G\lesssim 17$) 
for MOS and on emission lines for IFS. SDSS-V has succeeded in meeting these requirements: it has drawn from the infrastructure, instrumentation, survey operational expertise, and collaboration heritage of the previous phases of Sloan Digital Sky Surveys (SDSS) I-IV.  It has combined this with an ambitious new suite of hardware, drawing where possible on proven technology for efficient construction at manageable costs (A schematic overview of the SDSS-V dual-hemisphere, program is shown in Figure~\ref{fig:SDSSV_schematic}).

In this paper, we provide a high-level overview of the SDSS-V science program and its enabling hardware, software, data distribution and archiving, and organizational infrastructure.  Each of these elements is described in more detail in other publications that are published or in preparation.  A tabular summary of the SDSS-V program can be seen in Table~\ref{tab:SDSSV_program_table}

\begin{table*}
\centering
\begin{tabular}{|l|l|l|l|l|}
\hline
\multicolumn{5}{|c|}{{\bf SDSS-V Program Overview}}\\
\hline
\hline
 {\bf Program}  & {\bf Main Science}  & {N$_{\rm Objects}$/Sky Area} & {\bf Spectrographs} & {\bf Primary Science}\\
 & {\bf Targets}  & & & {\bf Goals}\\
 \hline 
 \hline

 {\bf B}lack {\bf H}ole {\bf M}apper & Supermassive Black & $>$700,000 unique  & BOSS (optical) & Probing black hole \\

 ({\bf BHM}) & Holes & sources; all-sky &  at APO and LCO & growth and mapping \\
 & & &  LCO & The X-ray Sky \\
\hline
  {\bf M}ilky {\bf W}ay {\bf M}apper  & Milky Way  & $>$7M unique stars;  & IR; APOGEE   LCO & Understanding Milky \\
{\bf(MWM)} & Way and Magellanic  & all-sky & (near-IR) and &Way formation and \\
 & Cloud Stars & & BOSS (optical)  & stellar physics\\
& & & at APO and LCO  & \\
 \hline
 
 \hline
  {\bf L}ocal {\bf V}olume {\bf M}apper  & Ionized gas and & $>$55M contiguous  & LVM-I (optical)  & Physics of the ISM, \\ 
({\bf LVM}) & stellar populations & spectra over  & at LCO & stellar feedback, \\
  &in the MW, LMC,&  4300 deg$^2$ & & chemical enrichment, \\
 & and Local  & &  & and stellar\\
 &Volume galaxies & & & populations \\
 & & & &   \\
\hline
\hline
\end{tabular}
\caption{A summary of the SDSS-V Mapper programs: Milky Way Mapper, Black Hole Mapper, and Local Volume Mapper.  The numbers given for $N_{Objects}$ for the Milky Way and Black Hole Mappers show planned unique sources not total number of spectra.  Owing to the time-domain nature of SDSS-V, the total number of spectra is much larger than the total number of objects.  This table doesn't include our Target-of-Opportunity sources.  For the Local Volume Mapper, the anticipated total spectral count is provided rather than unique source counts as we are tiling the sky.}
\label{tab:SDSSV_program_table}
\end{table*}

\begin{figure}[!ht]
    \centering
  \includegraphics[width=0.9\linewidth]{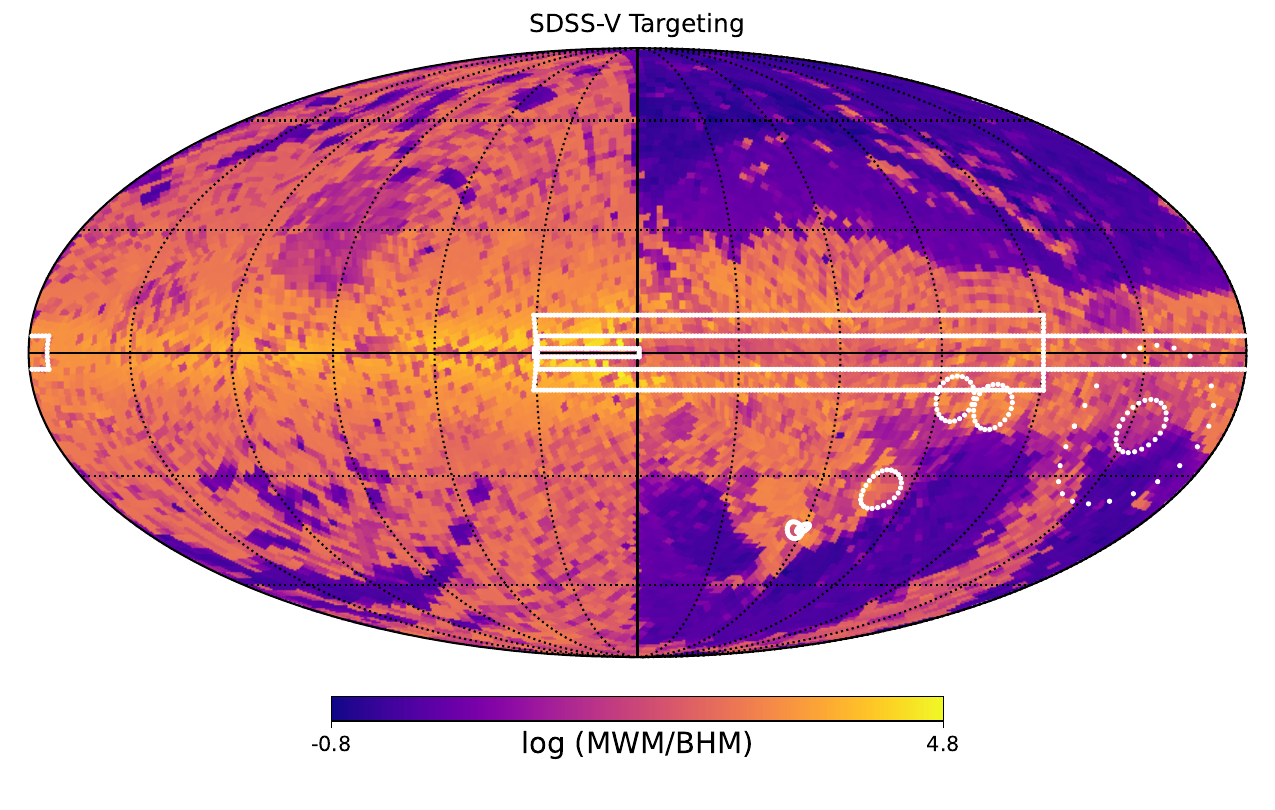}
\caption{\footnotesize\linespread{1.2}\selectfont{}{{\bf Ratio of MWM to BHM targets across the sky in Galactic coordinates.} This ratio highlights where each mapper is driving the survey implementation. The primary LVM footprint (excluding the high-latitude and nearby galaxy survey) is overlaid in white.}}
  \label{fig:SDSSV_targets_2}
\end{figure}

The COVID-19 pandemic arrived at a critical time for the SDSS-V project, resulting in a schedule modified from that envisioned in \citet{Kollmeier_2017_sdss5} due to the closure of vendor manufacturing facilities, the declaration of \emph{force majeur} and corporate bankruptcy in construction contracts, the lack of access to laboratory space, and restructuring of assembling procedures to obey social distancing. Undeterred, but building under higher viscosity conditions, SDSS-V found creative solutions to move the project forward during this unusual circumstance.  The APO Sloan Telescope corrector was delivered at APO in July 2021, with commissioning completed in November 2021. The first FPS unit was delivered to APO in November of 2021; first fiber light was achieved in December 2021, and the commissioning was completed in March 2022. The second FPS unit was delivered to LCO in July 2022, first fiber light was achieved in August 2022, and commissioning ran through December 2022. 
The site assembly, integration, and testing of the LVM instrument took place between February and April 2023. After a short commissioning phase, the science team took over the facility in July to carry out a science verification campaign with targeted programs to demonstrate the required sensitivity, efficiency, etc. Formal survey operations for the Local Volume Mapper started in November 2023 and fully robotic operations began in 2025.

This paper continues with an overview of our three primary survey programs in Section~\ref{sec:mappers}.  We then provide an overview of our hardware infrastructure in Sections~\ref{sec:instrument_mos} and \ref{sec:lvmi}, followed by our survey planning methodology and associated enabling software innovations in Sections~\ref{sec:SurveyOperations_MOS} and ~\ref{sec:SurveyOperations_LVM}. We describe the data reduction and analysis pipelines Section~\ref{sec:pipelines}, and the data processing, archiving and data release procedures that have been at the heart of all SDSS generations and central to SDSS success in Section~\ref{sec:data}.  We describe the organizational structure of the SDSS-V collaboration, and its core policies in Section~\ref{sec:Org}.  Readers interested in further details will be referred to more specialized SDSS-V publications throughout.

\section{SDSS-V's Tripartite Science: Three Mappers}
\label{sec:mappers}
%\textcolor{purple}{Now, I get to tell you about the amazing mappers.  Buckle up buttercup.  SDSS-V is a tale of three mappers.  Three pillars of science that represent everything that our cosmologist friends consider "nuissance parameters".  In other words, all the physics we actually have a prayer to understand in this decade.}

At the core of ``Sloan Surveys'' \citep{York_2000_sdss, Eisenstein_11_sdss3overview, Blanton_2017_sdss4}, has been the existence of clear and focused core science objectives that are both sufficiently ambitious to be singularly transformative, but at the same time enable broad astrophysical exploitations as behooves astronomical \emph{surveys}.  The experimental design flows from these scientific considerations.  SDSS-V began with the boundary condition that if no superlative science questions could be answered with affordable\footnote{Here, we mean mid-scale affordability which is of order \$50M United States Dollars circa 2016} modifications and enhancements to the existing infrastructure, then there would be no SDSS program. By ``infrastructure," we refer not only to the telescopes and instruments, but also to the control software, the data analysis, archive, and delivery systems, and the policies and procedures that have been honed through SDSS generations to yield a healthy collaborative machinery.  SDSS-V answered the question: how can elements of existing hardware and the exceptional project and collaboration culture be taken to new frontiers that promise transformational results in the 2020's with telescopes of modest size? The final science scope and vision of SDSS-V arose from a rigorous, multi-year community-wide process that reviewed all of these factors and weighed a multitude of competing ideas\footnote{This began first in an internal Futures committee in 2015 followed by an ``After-SDSS-IV Steering Committee'' in 2016.}.  With the natural move of cosmological redshift surveys, which had been core driving elements of SDSS I-IV, to larger-aperture telescopes, and with the rebirth of stellar astrophysics and black hole astrophysics owing to a combination of new experiments and new theoretical work, SDSS-V has also moved its central focus beyond the once-defining themes of the ``Sloan Surveys.'' Expanding upon its industrial mapping capacity, SDSS-V's three central science areas are now articulated as three science \emph{mappers}\footnote{In his dialogues, ca. 370 B.C.E.,  Plato divided the soul into three parts: the rational part, the spirited part, and the appetitive part.  A similar partitioning can be found in the Bahgavad Gita, written during a similar period, which refers to three Gunas). We leave it to the reader to make a formal mapping between our mappers and Platonic/Gitanic and other tripartite ideas.}  as follows:

The {\bf Black Hole Mapper} (BHM; Section~\ref{sec:bhm}) will focus on the physics of supermassive black hole growth by carrying out a) long-term, time-domain studies of AGN, including direct measurement of black hole masses, and b) optical follow-up of {\it eROSITA} X-ray sources.  The {\bf Milky Way Mapper} (MWM; Section~\ref{sec:mwm}) will provide a) a global spectroscopic orbit-abundance-age map of the stellar MW, primarily using near-IR MOS concentrated at low Galactic latitudes; b) a multi-epoch stellar astrophysics survey, focused on follow-up of {\it Gaia} and TESS; c) a multi-epoch survey of young, massive stars throughout the Galaxy; and d) a survey of white dwarfs, both single and in compact binary configurations, the former selected from {\it Gaia}, the latter from {\it eROSITA} and {\it GALEX} jointly with {\it Gaia}.  The {\bf Local Volume Mapper} (LVM; Section~\ref{sec:lvm}) will focus on understanding the physics of the ionized ISM and of star formation and stellar feedback, by providing the first integral field spectral map that spans a) the full optical window, b) the bulk of the MW disk, and c) the Magellanic Clouds and other nearby galaxies in the Local Volume.  
\subsection{The Black Hole Mapper}
\label{sec:bhm}

The Black Hole Mapper (BHM) provides data to better understand the physical processes surrounding the growth of supermassive black holes (SMBH) at the centers of galaxies. For the most part, this growth is affected by the accretion of material onto these SMBHs. During these often brief ($\Delta t\ll t_{Hubble}$) phases, galactic centers appear as active galactic nuclei (AGN) or even quasars, the sources of the highest sustained luminosity in the Universe.\footnote{Historically, ``quasars'' and ``AGN'' describe different classes of objects, but here we use them interchangeably in recognition of the fact that they both describe accreting supermassive black holes \citep[e.g.,][]{Merloni_2016}.}
Quasars are astrophysical beacons bright enough to mark and trace the growth of black holes across cosmic distance and time. The tight correlations between the masses of central SMBHs and the properties of their hosts \citep[e.g.,][]{Magorrian1998,HaeringRix04} demonstrate a clear connection between the formation of the stellar component of a galaxy and the growth of its central black hole. In particular, it is clear that black hole feedback plays a central role in regulating the star formation history of massive galaxies \citep[e.g.][]{Croton2006}.  This connection means that quasar studies are not only critical for understanding SMBHs and their accretion physics but are also closely linked to the questions of galaxy formation and evolution being explored by SDSS-V's MWM (Section~\ref{sec:mwm}) and LVM (Section~\ref{sec:lvm}). 

\begin{figure}[ht!]
%\vspace{0.2in}
\centering
\includegraphics[width=1.0\linewidth]{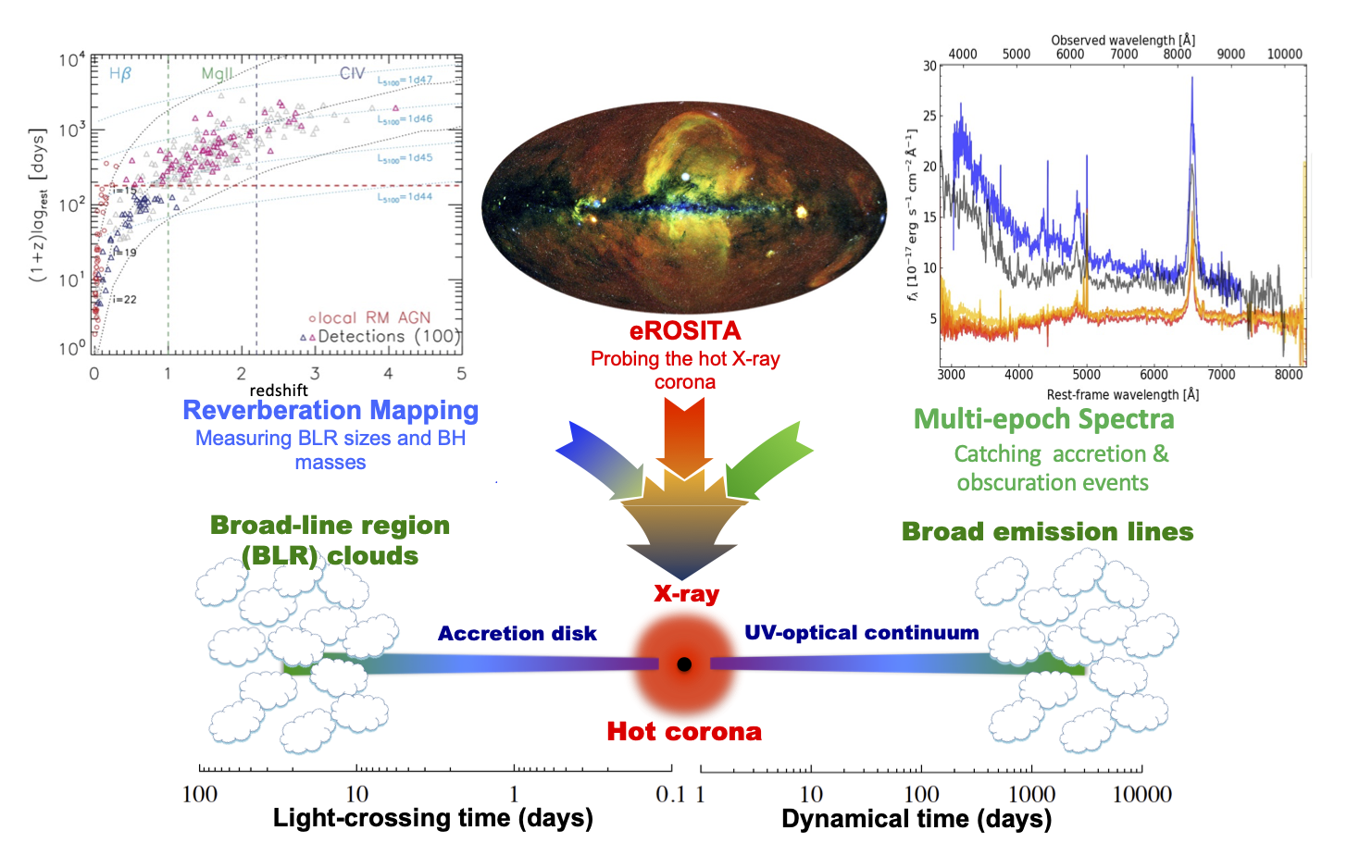}
  \caption{\footnotesize\linespread{1.2}\selectfont{} {\bf Schematic of the innermost regions around a quasar's
central supermassive black hole (BH):} the X-ray corona, accretion disk, and broad-line region (BLR). SDSS-V explores the physics of supermassive BH accretion and dynamics with three parallel approaches: reverberation mapping, {\it eROSITA} follow-up, and multi-epoch spectroscopy (represented by the top three panels). Top left: a simulation of the anticipated reverberation mapping (RM) yield 
of time delays  
(here in just one of four BHM/RM fields) between UV/optical continua from the accretion disk and emission line flux from the BLR, which yields the BH mass. Top center: an early  and representative {\it eROSITA} X-ray all-sky image [J. Sanders, H. Brunner and the eSASS team (MPE); E. Churazov, M. Gilfanov (on behalf of IKI)], depicting high-energy data that -- combined with SDSS-V spectra --  allow BHM to conduct a census of the X-ray/optical
properties for $>10^5$ quasars.  Top right: an example of multi-epoch spectra \citep{Zeltyn_etal_2022} that reveals a marked change in observed continuum and emission line spectral character in a BHM/SDSS-V ``changing look" quasar, even extending down to months-long timescales; similar spectral changes are seen already among more than a hundred BHM-studied quasars \citep{Zeltyn_etal_2024} and are probes of accretion state and obscuration events.}
\label{fig:agn_schematic}
\end{figure}

BHM's observational approach to understanding the physics of SMBH growth is broadly two-fold. First,  BHM is obtaining time-domain spectroscopy of AGN, covering time scales from days to decades, since the temporal variability of the accretion luminosity and of the AGN emission lines encode information about the structure, dynamics, and evolution of emitting regions, even in areas often too small to resolve spatially with any telescope for the foreseeable future \citep[e.g.][]{Gravity_BLR_2024}. Second, BHM is obtaining the first comprehensive follow-up of X-ray sources from the {\it eROSITA} survey \citep{Predehl_etal_2021,Merloni_etal_2024}. These are mostly AGNs whose X-ray emission probes regions much closer to the actual SMBH than the optical; X-ray emission is generally less affected by intervening obscuration, and thus provides diagnostics complementary to those at longer wavelengths. 

Observational tests of the BH growth physics therefore require three primary measurements: precise mass constraints, multi-wavelength SEDs, and a detailed characterization of variability (Figure~\ref{fig:agn_schematic}). The basic Black Hole Mapper (BHM) observing strategy is summarized in 
Table~\ref{tab:bhmtargetclasses},  designed to provide these measurements for a large sample of quasars by adding wide-area, multi-epoch optical spectroscopy to the current and upcoming time-domain optical imaging projects and to the current and next generation of X-ray surveys (e.g., ZTF, Rubin/LSST, and eROSITA). Some further details on BHM can be found in the SDSS-V Data Release 18 paper, by \cite{sdss_dr18}.

\begin{table*}
 \centering
    \begin{tabular}{|l|l|l|l|l|}
    \hline
    \multicolumn{5}{|c|}{{\bf SDSS-V Black Hole Mapper Core Targeting Summary}} \\
    \hline
    {\bf Science Goals}  & {\bf Primary Selection} & {\bf Density [deg$^{-2}$]} & {\bf N$_{\rm targets}$} & {\bf N$_{\rm epochs}$}\\
    \hline
    \hline
    Reverberation mapping (RM), & Optical QSOs, & 80 & 1,500 & 100-150\\
           BH Masses  &  $i<$20--21 &  &   &  \\
        \hline
        \hline
AQMES lower-cadence multi- & Optical QSOs, & 10 & 20,000 & 1-10 \\
epoch spectra: accretion events \& &  $i<19$ & & &\\
outflow astrophysics; line profile  & & & & \\
changes and dynamics; &  & & &  \\
\hline
\hline

SPIDERS: follow-up spectra of & $f_{\rm X-ray} \geq 2\times 10^{-14}$ & 30--40  & 300,000 & 1 (effectively)\\
{\it eROSITA} AGN, X-ray binaries,  &  erg~s$^{-1}$~cm$^{-2}$, $i<21.4$ &  &  & \\
galaxy clusters & & & & \\
\hline
\hline
\hline
    \end{tabular}
    \caption{Summary of planned scope of the BHM core programs being carried out, specifically with optical BOSS spectrographs in both hemispheres.} 
 
\label{tab:bhmtargetclasses}
       
\end{table*}

\subsubsection{BHM's Time-Domain Spectroscopy}

Over past few decades, the temporal flux variations of quasars have proven successful in constraining basic aspects of quasar models. A prime example is \emph{reverberation mapping} (RM) of the broad emission line region (BLR). RM measures the time delay between variations in the light curve of the (ionizing) continuum that arises from the accretion disk and their ``echo'' in the variations of the BLR fluxes \citep[e.g.,][]{Blandford_McKee_1982,Peterson_1993}. Such RM delays measure the typical sizes of the BLR via light travel time arguments and, when combined with the velocity width of the broad emission lines, allow a virial estimate of the BH mass, the most fundamental of all BH parameters. Large representative samples of quasars with robust spectroscopic variability studies are still being assembled \citep[e.g.,][]{Vestergaard_2011}, and the full power of {\it spectral variability} to constrain quasar models is still being explored.

The earlier incarnation of Sloan Sky Surveys, SDSS-III and -IV, have demonstrated the feasibility and potential of this approach \citep[e.g.,][]{Shen_etal_2015a,King_etal_2015,Shen_etal_2023}. SDSS-V's BHM is taking RM observations to an ``industrial scale.'' Spectroscopic RM sampling with of order one hundred epochs each to determine robust BH masses will be obtained for $\sim$1,000--1,500 quasars with a range of redshifts ($0.1<z<4.5$) and luminosities ($L_{\rm bol}\sim 10^{45}-10^{47}\,{\rm erg\,s^{-1}}$). This is a $\sim 10$-fold increase in number, compared to the historical sample of nearby low-luminosity AGNs with RM-determined masses. 

In its All-Quasar Multi-Epoch Spectroscopy (AQMES) program, BHM is also characterizing the optical spectral variability of approximately 20,000 quasars with a few to nearly a dozen epochs per target that span baselines from months to a decade. AQMES' goal is to understand why quasars vary, to illuminate the astrophysics of SMBH accretion disks and BLRs, to search for signatures of binary BHs, and to characterize quasar outflows. In addition, studies of \emph{changing-look} AGN, in which the broad lines around AGN either appear or disappear (see e.g. the upper right panel of Figure~\ref{fig:agn_schematic}),
comprise a burgeoning field that challenges standard accretion disk theory 
%\citep[e.g.,][]{LaMassa_2017_changinglookAGN}
\citep[e.g.,][]{LaMassa_etal_2015,Runnoe_etal_2016}; more than 100 new examples of these intriguing objects have already been discovered in the BHM's AQMES program \citep[e.g., as shown in the upper right panel of Figure~\ref{fig:agn_schematic}, with further and more comprehesive details in][]{Zeltyn_etal_2022,Zeltyn_etal_2024}. 

\begin{figure}[!ht]
%\vspace{+1.0in}
\centering
\includegraphics[width=0.49\linewidth]{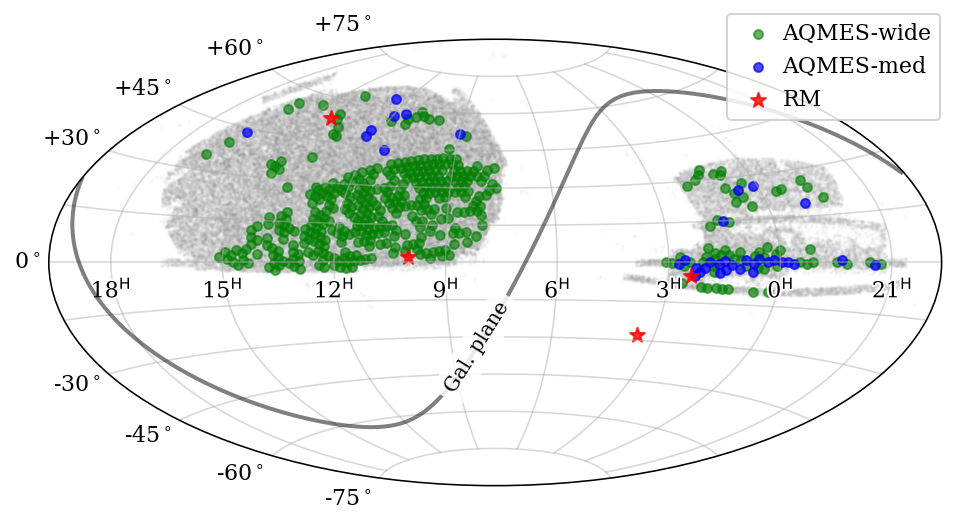}
\includegraphics[width=0.49\linewidth]{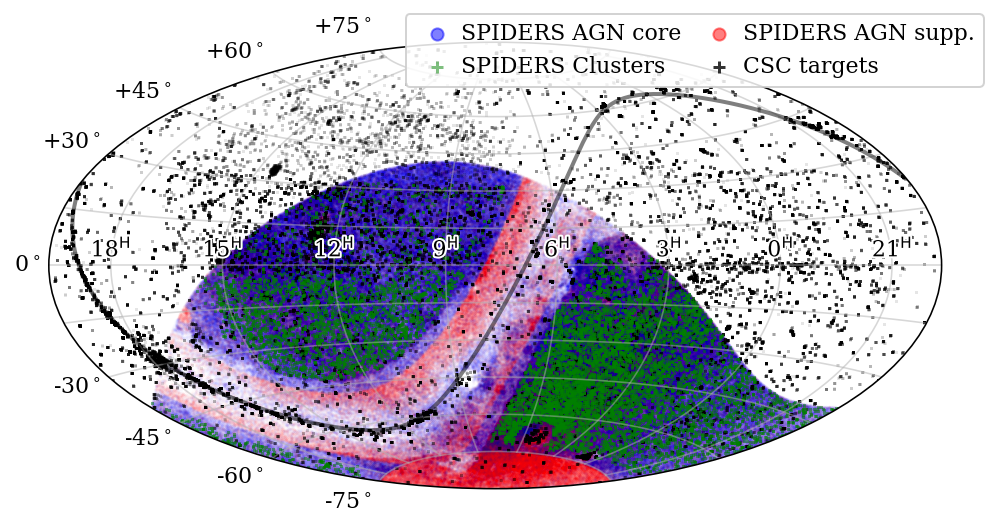}
\includegraphics[width=0.49\linewidth]{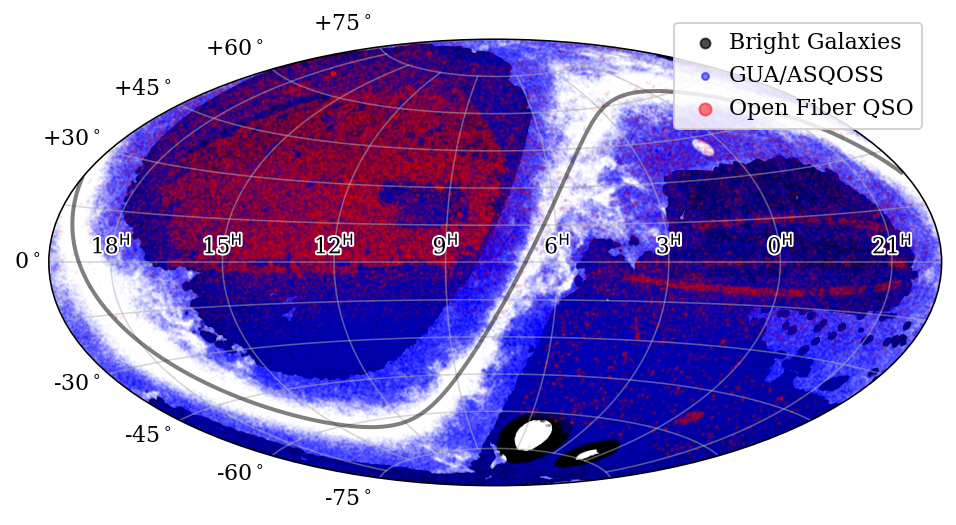}
\includegraphics[width=0.49\linewidth]{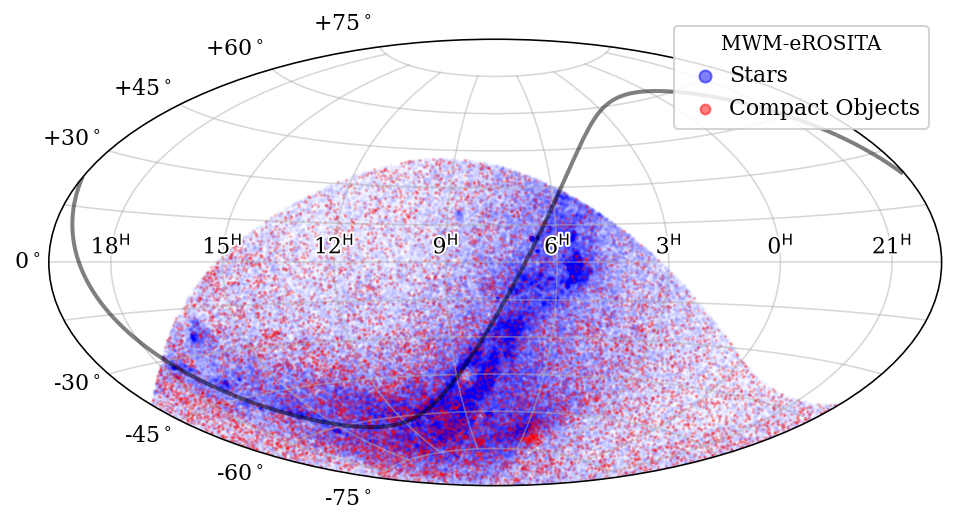}
 %\vspace{-0.2in}
  \caption{\footnotesize
{\bf Sky distribution of all BHM (and Galactic X-ray emitting) targets.} % (Equatorial coordinates).} 
{\it Top left panel:} QSO targets for BHM time-domain repeat BOSS optical spectroscopy. % (RM and AQMES). %The red plus magenta 
The gray background depicts the sky-density of bright ($i_{\mathrm{psf}}<19.1$) QSOs taken from the SDSS DR16 QSO catalog \citep{Lyke_etal_2020}, all of which have one or more earlier epochs of SDSS I-IV spectroscopy. 
Colored symbols indicate fields selected for new epochs of optical spectroscopy in SDSS-V; one or two new epochs for 425 AQMES-wide fields (green, c.~20k QSOs) and 10 new epochs for 36 AQMES-med fields (blue, c.~2k QSOs). 
The red symbols show the BHM reverberation mapping (RM) field locations; %which are both North and South, and 
these four fields (each with hundreds of QSOs) will receive $>$100 epochs of SDSS-V higher-cadence repeat spectroscopy. 
{\bf Top right panel:} Distribution of X-ray selected BHM optical spectroscopic targets across the SDSS-V survey. 
Candidate AGN targets selected from eROSITA eRASS:3 are indicated by the shaded blue (core sample) and red (supplementary sample) areas, 
eROSITA galaxy cluster targets are shown with green points, and Chandra Source Catalog (CSC) targets are shown with black points. 
Note that SDSS-V targeting of eROSITA sources is restricted to the western Galactic Hemisphere (approx. $180 < l < 360$\,deg).
{\bf Bottom left:} BHM-led ancillary targets, including bright well-resolved galaxies selected from the DESI Legacy Imaging Surveys \citep{Dey_2019_DESIsurveys}, 
QSO candidates selected from the Gaia unWISE AGN \citet[][GUA]{Shu2019} and \citet[][ASQOSS]{Yang+Shen2023} catalogs, 
additional QSO related open-fiber targets selected for their spectral, hard X-ray and optical variability properties.
{\bf Bottom right:} X-ray emitting Galactic targets selected from the eROSITA eRASS:3 survey, 
including candidate coronally active stars, and compact objects. Maps are shown in Equatorial coordinates. %, on a Mollweide projection.
}
\label{fig:bhmtargs}
\end{figure}

\begin{figure}[!ht]
 \centering
  \includegraphics[width=1.0\linewidth] 
  %{bhm_updated_spiders_plus_td-v1.pdf}
  {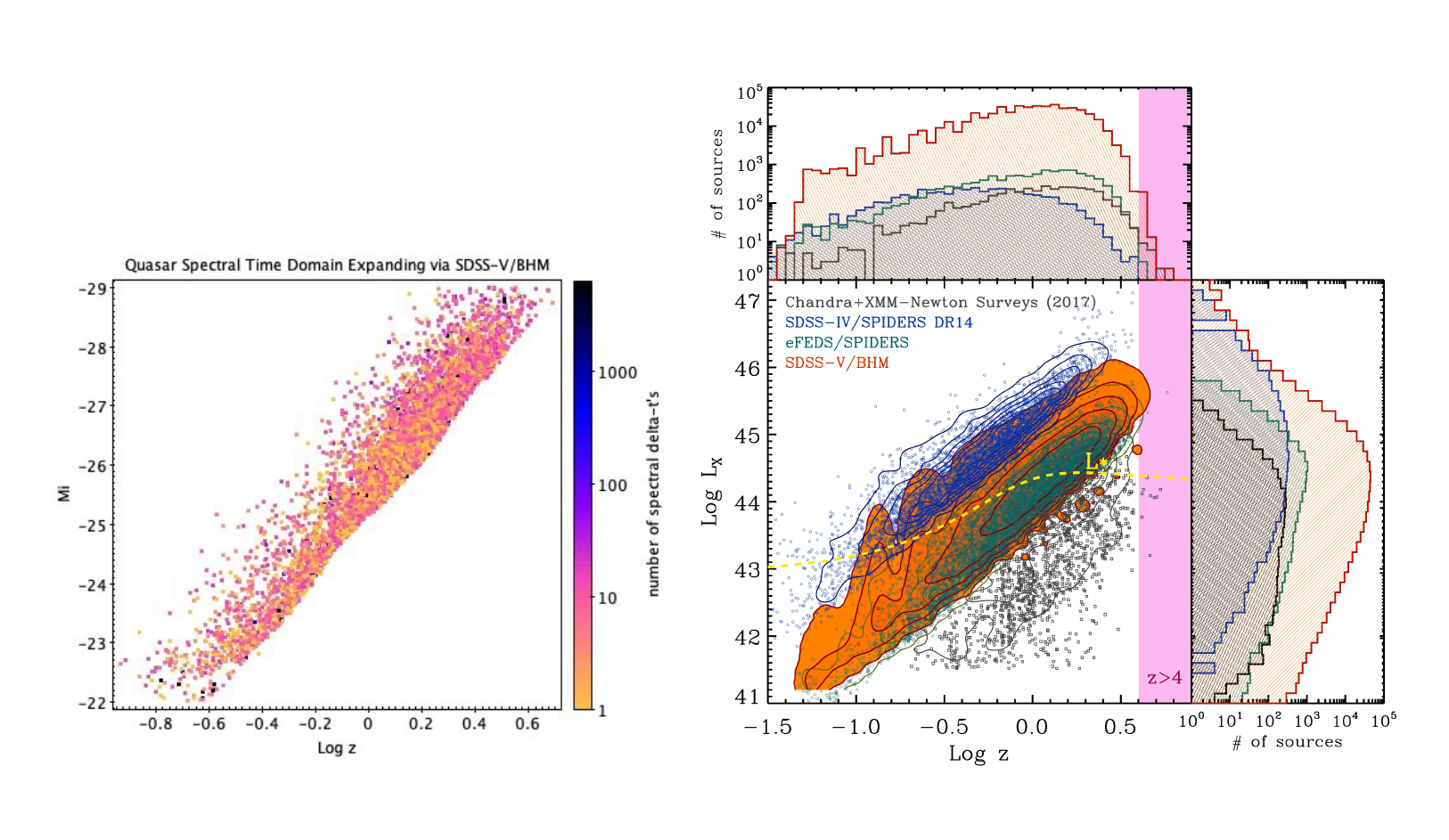}
  %{contours_l_z_log_chandra_xmm_spiders2_no500.pdf}
 %{contours_l_z_log_chandra_xmm_dr14_sdssv.png}
 %{contours_l_z_log_chandra_xmm_dr14_AS4.pdf}
  %\vspace{-0.4in}
  \caption{\footnotesize {\bf   
  %
%swap left-right panel option: 
%
Representative luminosity and redshift coverage of core BHM time-domain (left) and X-ray selected (right) surveys of AGN:} {\it Left panel:} The distribution of absolute magnitude vs. redshift for BHM core time-domain spectra of SDSS quasars (here, highlighting a subset of about $\sim$6000  with $i<19.1$) enabled from the first year of SDSS-V. Both AQMES and RM targets are included, but AQMES predominates in this figure: the number of RM targets (especially at the plotted limit of $i<19.1$) is comparatively small, though such RM targets are observed at a comparatively high cadence. The colors in this panel encode the number of time differences $\Delta t$'s between various pairs of optical spectral epochs (archival SDSS spectra, plus BHM spectra taken early with plates from APO). Building on the existing SDSS DR16 quasar survey allows BHM to efficiently expand spectral time-domain studies: the number of $\Delta t$'s for a quasar scales as $n(n-1)/2$, where $n$ is the number of distinct SDSS spectral epochs; the number of SDSS quasar $\Delta t$'s realized by BHM (to $i<19.1$) via its first year alone roughly doubles that previously available for the entire DR16 quasar sample.~~{\it Right panel:} For the SPIDERS X-ray/optical survey, the larger and denser the coverage in this parameter space, the more stringent the constraints that can be placed on the history of accretion onto SMBHs in the Universe. The central plot in this panel compares SDSS-V's {\it eROSITA} follow-up program (orange contours; $\sim$392,000 targets) and other current state-of-the art AGN surveys: SDSS-IV/SPIDERS (blue; $\sim$4600 targets); a compilation of deep {\it Chandra} and {\it XMM-Newton} fields (black; CDFS, CDFN, COSMOS, Lockman Hole, XMM-XXL; $\sim$4000 targets); SDSS-IV/V mini-survey (green; $\sim$ 10,000 targets) . The top and right histograms in this panel show the number of AGN expected in each of these comparison X-ray/optical samples. Note the logarithmic y-axes in both histograms: the {\it eROSITA} X-ray and SDSS-V optical sample will be about 100$\times$ larger than any existing sample, spanning a wide range of redshift and luminosities. The pink shading at $z>4$ highlights the $\sim$10$\times$ improvement in the X-ray selected SPIDERS AGN sample size at high redshift.
 }
%\label{fig:SPIDERS-Lz}
\label{fig:SPIDERS-Lz}
\end{figure}

All of the AQMES targets are drawn from the SDSS DR16 Quasar catalog \citep{Lyke_etal_2020}, as early epoch spectra are important for probing longer temporal baselines. Consequently, most of the AQMES targets are in the north. RM fields for BHM are both North and South. The upper left panel of Figure~\ref{fig:bhmtargs} shows the anticipated sky coverage of the BHM time-domain spectral components, with color coding highlighting AQMES and RM fields, also depicting the number of potential added epochs in SDSS-V. 
As a representative example, the left 
%right 
%[SA: **probably should switch plot panels up some to make consistent--yes switched this after some reported confusion in proofing**] 
panel of Figure~\ref{fig:SPIDERS-Lz} shows the distribution of absolute magnitude (optical $i$-band) versus redshift for the BHM repeat spectra of SDSS quasars with $i<19.1$ enabled from the first year of SDSS-V. The color coding in this latter figure signifies the
%anticipated end-of-survey 
number of spectral epoch differences ($\Delta t$) between various pairings of optical spectral epochs. These $\Delta t$'s reflect a combination of archival SDSS spectra and time-domain spectra taken early on in SDSS-V, during initial operations on plates from APO. Building on the existing SDSS DR16 quasar survey not only extends the epoch baseline; it also allows BHM to efficiently boost the number of $\Delta t$'s for any such quasar, since this number scales with the number of distinct spectral epochs as $\approx N_{epoch}(N_{epoch}-1)/2$. The $\Delta t$'s realized by BHM in the first year of SDSS-V alone already doubled those previously available for the entire DR16 quasar sample. 
The actual distribution of $\Delta t$'s anticipated for the end of the survey will be discussed in detail in upcoming  BHM publications, from the somewhat distinct perspectives of the RM and AQMES programs.

%\vspace{-0.9in}

\subsubsection{SPIDERS: BHM's Spectroscopy of eROSITA X-ray Sources}

BHM's program for the \emph{SP}ectroscopic \emph{ID}entification of
\emph{eR}OSITA \emph{S}ources (SPIDERS) is providing the first comprehensive set of spectra of eROSITA's unprecedented set of X-ray sources across the half of the sky that is accessible to the collaboration: the German (DE) half of the eROSITA sky at Galactic longitude $>$180 deg. Combining these spectra with the X-ray photometry allows BHM to peer deep into the AGN's central engine and probe SMBH evolution over large spans of cosmic time. 

Despite the high X-ray luminosity of nearly all AGN, we do not fully understand the physical origin of the tight coupling between the hot X-ray corona and the (comparatively) ``cold'' accretion disk. This is mostly due to the limited size of existing X-ray AGN samples, as X-ray telescopes with the necessary sensitivity ({\it Chandra}, {\it XMM-Newton}) have had relatively small FOVs.  Fortunately, the {\it eROSITA} satellite \citep{Merloni_etal_2024} launched in 2019 has both the sensitivity and  FOV needed to discover as many new X-ray sources in its first 12 months as was known from the previous more than 50 years of X-ray astronomy. SDSS-V is providing optical spectroscopic measurements, including identifications and redshifts, of $\sim$350,000 {\it eROSITA} X-ray sources detected in the first 1.5 years of the all sky survey, termed \emph{eRASS:3}. The SPIDERS targets were primarily selected to have a 0.5--2 keV flux $\gtrsim$$2 \times 10^{-14}$ erg~s$^{-1}$~cm$^{-2}$ 
\citep[see Figure~\ref{fig:SPIDERS-Lz} and][]{Merloni_2012}. 
SPIDERS spectroscopy will target sources as faint as $i_{AB}=21.5$, which is the approximate limit at which BOSS spectroscopy is able to provide reliable redshifts (in less than 1 hour, under typical conditions). This sample comprises foremost AGN at high Galactic latitude, but also contains some X-ray-emitting galaxy clusters and X-ray-bright stars in the MW and nearby galaxies. In addition, the BHM-SPIDERS program will inevitably characterize serendipitously discovered types of X-ray sources, extreme and rare objects, transients, and other peculiar variables found in the {\it eROSITA} survey 
\citep{Predehl_etal_2021, Merloni_etal_2024}. This combination of X-ray discovery and optical characterization provides a great leap forward in our description of the X-ray sky and will reveal the connections between large, statistical populations of X-ray sources and the cosmic structures in which they are embedded. 

As the German (DE) half of the eROSITA sky is predominantly in the South, most follow-up of eROSITA X-ray sources in the SPIDERS program is carried out from LCO (see Figure~\ref{fig:bhmtargs}; right panel).  The dramatic sample expansion resulting from SPIDERS is depicted in Figure~\ref{fig:SPIDERS-Lz} 
%(left 
(right panel), which compares the X-ray luminosity and redshift coverage of several AGN surveys. The larger and denser the coverage in this parameter space, the more stringent the constraints that can be placed on the history of accretion onto SMBHs in the universe. The central plot in the left panel highlights SDSS-V's {\it eROSITA} follow-up SPIDERS program (orange contours, representative of nearly 400k X-ray source candidate AGN targets planned for near-term BHM implementation), compared with other recent state-of-the art AGN surveys (the latter of which include thousands to tens of thousands of targets) including: the SDSS-IV/SPIDERS (blue symbols) survey; a compilation of deep {\it Chandra} and {\it XMM-Newton} fields (black; CDFS, CDFN, COSMOS, Lockman Hole, and XMM-XXL targets); and the very recent eFEDS mini-survey \citep{Brunner_etal_2022} for which optical spectroscopy started in SDSS-IV, was largely completed in SDSS-V and featured in SDSS-V DR18 \citep{sdss_dr18}. The top and right histograms (which are logarithmic) show that the {\it eROSITA} X-ray and SDSS-V-BHM SPIDERS sample will be about 100$\times$ larger than any
existing sample, covering a very wide range of redshift and luminosities.

BMH in SDSS-V has begun to produce new science across all of its main core science goals. Selected examples of BHM time-domain science publications include: 
\cite{Zeltyn_etal_2022}, %on a changing look quasar that may be traceable to a rare obscuration event; 
\cite{Fries_etal_2023}, \cite{Zeltyn_etal_2024}, \cite{Wheatley_etal_2024}, \cite{Fries_etal_2024}, and \cite{Stone_etal_2024}.
%on "breathing" behavior of the BLR of a quasar studied in SDSS reverberation mapping programs over a decade; 
Examples of early BHM science related to the characterization of X-ray source counterparts include:
\cite{Waddell_etal_2023},
%on X-ray/optical studies related to accretion-dependence (e.g., Eddington ratio) of X-ray AGN in the eROSITA eFEDS mini-survey region
%(Brunner et al. 2022)* 
%with warm-absorbers and soft X-ray excesses; 
\cite{Comparat_etal_2023}, \cite{Nandra_etal_2024}, and \cite{Saxena_etal_2024}.
%in which eROSITA/eFEDS AGN provide further evidence of clustering and lensing characteristics indicating large-scale structure halo occupation and bias measures depend on AGN luminosity.

\subsubsection{BHM Survey Strategy}
\label{sec:bhm:strategy}
BHM's observational strategy is embedded in the overall SDSS-V observing model in which focal plane resources (fibers, robots) are usually shared amongst more than one program with 15 minute exposure ``quanta''. In addition to its primary core science programs (AQMES, RM, and SPIDERS) described in the previous two subsections, BHM also takes (non-core) BOSS spectra via a significant number of otherwise spare or unused BOSS fibers often available, as indicated in the lower left panel of
Figure~\ref{fig:bhmtargs}; for BHM, these open-fiber and other ancillary targets are predominantly extragalactic objects for which classification spectra and/or redshifts are broadly useful, e.g., ranging from
%BHM open-fiber and ancillary targets include: 
wide-area catalogs of bright well-resolved galaxies 
\citep{Dey_2019_DESIsurveys}
%, optical/IR counterparts of X-ray sources from serendipitous X-ray catalogs (the Chandra Source Catalog \citealt{Evans2024}; the 4XMM serendipitous source catalog \citealt{Webb2020}; the 2SXPS Swift X-Ray Telescope Point-source Catalog \citealt{Evans2020}),
%repeat spectroscopy of carefully selected sub-samples of known QSOs \citep[from e.g. SDSS DR16Q,][]{Lyke_etal_2020},
%and 
to large all-sky quasar candidate samples
%, selected on the basis of their photometric/astrometric/variability properties 
\citep{Shu2019,Yang+Shen2023}.
Spectroscopic fibers/robots are shared not only among the multiple BHM programs but also routinely with MWM programs (and usually including a mix of targets that require either BOSS or APOGEE spectra). However, in contrast to most MWM programs, the majority of BHM targets require dark time. In order to obtain adequate SNR spectra of optically fainter BHM targets, we co-add multiple 15\,minute exposures, typically totaling 1--2 hours (per epoch).
%, rather than the ``fields'' and ``designs'' of earlier SDSS observations. [[Expand on this, where needed as it deviates from MWM]]
As noted above, BHM observations rely almost exclusively on the BOSS spectrographs and benefit from the existing reduction and analysis pipelines inherited from earlier generations of SDSS. Nevertheless, a number of modifications have been made to accommodate SDSS-V's particular setup and to improve the analysis. These changes are described in detail in the DR18 paper \citep{sdss_dr18}.

The primary data products of BHM will include the following: extracted 1D BOSS spectra (both for individual exposures, and different co-added versions) along with redshifts and spectral measurements (such as emission line fluxes) derived from the BOSS spectra. BOSS spectra are co-added to probe several time-scales: per night, per `epoch' (which may span up to two weeks), and for selected targets, over all of SDSS-V. The standard BOSS pipeline products also include summary files and additional stellar parameters for MWM targets. A more detailed description of standard BOSS pipeline products is provided in 
Section~\ref{sec:software_pipelines}. Data products beyond the standard pipeline outputs, such as vetted catalogs, improved systemic quasar redshifts, refined spectral measurements, etc., will be presented in value-added catalogs produced by the SDSS-V science teams. 

\subsection{The Milky Way Mapper}
\label{sec:mwm}
The ecosystem of stars, gas, dust, and dark matter in large galaxies like the MW has been shaped over billions of years by a variety of physical processes that operate across an enormous range of physical scales. Despite this complexity in galaxy formation, we now observe a population of galaxies that is ordered across galaxy masses spanning from hundreds of stars to hundreds of billion stars.  Explaining how this regularity emerges in a cosmological context from complex and varied physics, while wrestling with age-old questions about stellar structure and evolution, is a central challenge of modern astrophysics.  The Milky Way Mapper (MWM) survey will exploit our unique perspective within our Galaxy to address this issue by creating a global Galactic map that spans the H-R diagram and encompasses the evolutionary record contained in its stars and interstellar medium (ISM).  At the same time, the time-domain nature of the MWM program and its synergy with current missions will provide a critical component of understanding the physics of stars and stellar systems. 

\subsubsection{The Making of the Milky Way: Galactic Chronology}\label{sec:Milky Way}

From the earliest cosmological maps\footnote{Many historic records have been lost to the ravages of time and human savagery.  However, historical records ranging from the Dunhuang Charts found preserved within the Mogoa Caves dates from the late 7th century, as well as the star chart decoded on the ceiling of the tomb of Senmut, Queen Hatsheput's Vizier circa 1534  \citep{vonspaeth}, underscore both the universal nature of astronomy as a scientific practice, as well as the critical need for high-quality, temperature-controlled data archives. 
 (See S. Little, ``Mapping the Infinite: Cosmologies Across Cultures", in prep, for a historical sampling.)}  , the march of cosmological surveys has steadily proceeded outward in spacetime from the Milky Way moving out to the era of cosmic dawn.  The ability to survey galaxies at an industrial scale \citep[e.g.,][]{Hill1988, York2000, Gilmore2012,DeSilva2015, Cui2012, Gaia2016} simultaneously provided the opportunity to dissect and extract cosmological information locally, using the Milky Way galaxy as a cosmological end-point -- the practice currently referred to as ``near-field cosmology".  
 
 For this purpose, the Milky Way's magnificent mediocrity\footnote{This is meant with love, referring to the Milky Way's role as a typical, and therefore, mediocre galaxy -- in the mass, M$_{\star}$,  and luminosity, L$_{\star}$ function sense.} is critical. Although it is only one Galaxy, it is the one we can study most precisely, and model most completely.  It is therefore exceedingly fortunate, that we live in a rather typical galaxy, and not an outlier of the galaxy distribution. To truly utilize our Galaxy as a ``galactic model specimen", surveys must probe the entire hierarchy of space-time structure within the Galaxy using tracers that have memory of the Milky Way's past and that can faithfully recount the current Milky Way conditions.  These maps should be contiguous, and densely sampled throughout the (largely dust-obscured) disk- and bulge-dominated regions of the stellar MW and uniformly sampled in the halo. High quality spectra of stars are rich with information about their basic physical parameters, including their ages, chemistry, and kinematics. Interstellar spectral lines probe the composition and dynamics of the MW’s gas and dust, from which new stars are forming. These properties provide the best means to quantitatively test models of the most uncertain galaxy formation physics \citep[e.g.,][]{Bland-Gerhard2016}. 

Within the MWM's Galactic Chronology program, Galactic Genesis (GG) will produce the first spectroscopic stellar map that is contiguous, densely sampled, and all-sky but focused on the low Galactic latitudes where most stars lie and includes detailed information on {\it each} star and its foreground ISM. This mapping of the Galaxy spans a huge range in spatial and dynamical scales; from individual stars and their ecosystems to galactic-scale structure from the bulge to outer disk. With its near-IR and optical, multi-epoch spectroscopy through the entire Galactic plane, SDSS-V will also significantly expand the spectroscopic census of young stars in the MW, characterizing their masses, ages, multiplicity, etc., thus painting a global picture of the ``recent Galaxy.''

GG provides the best view of the evolution of the Galaxy using lower mass stars with ages $>$ 100 Myr. However, its view is hidden for the youngest stars and for older higher-mass stars. To compensate for this, programs in MWM observe the youngest stars, e.g. O and B-type stars and young stellar objects (YSOs) or pre-main-sequence stars. The deeper archaeological record for stars with intermediate masses up to $\sim$8~M$_{\odot}$ is probed by MWM observations of white dwarfs. Finally, by concentrating its fibers on the disk where the stellar concentration is highest, GG is not observing a high fraction of halo stars. However, the power of the facility is such that we are able to target halo stars in addition to our GG targets, separately, selecting halo stars through kinematic, metallicity, or distance tracers.

\begin{figure}[htbp]
    \centering
   \includegraphics[width=0.45\textwidth]{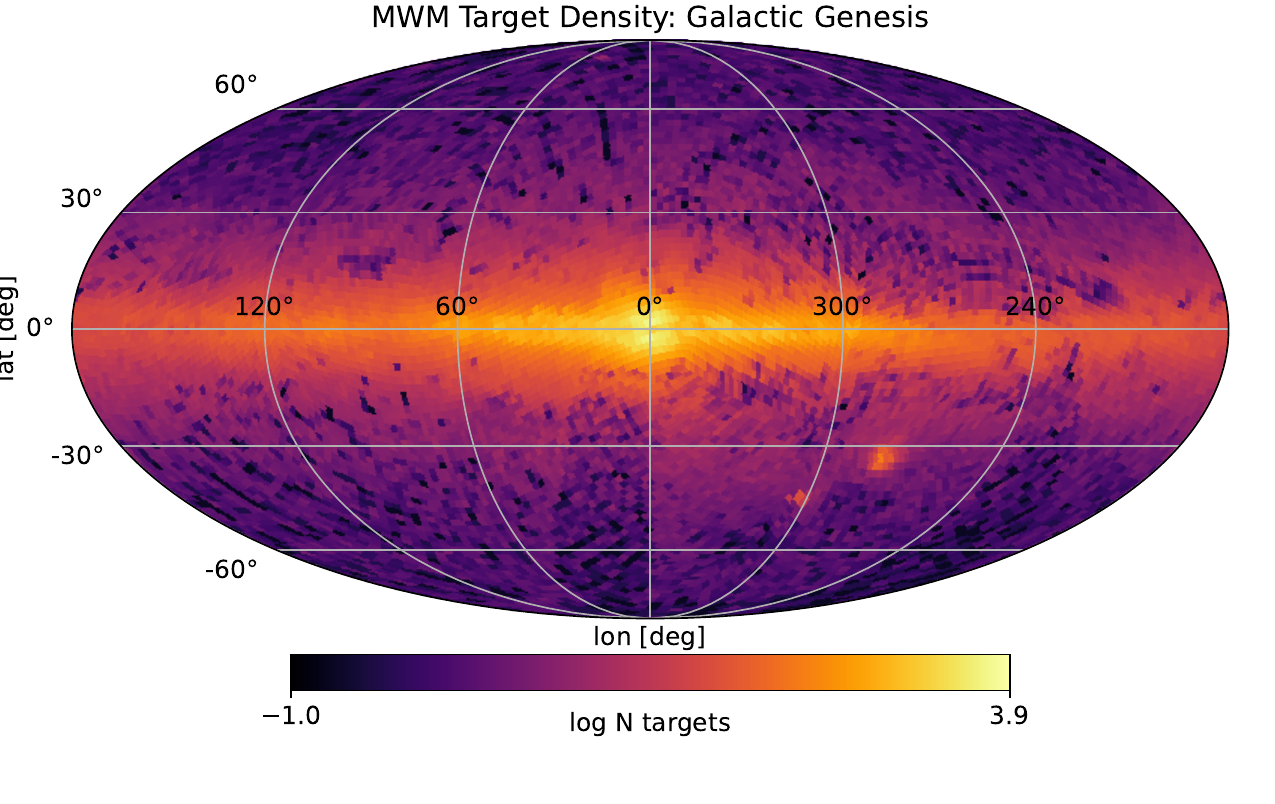}
    \includegraphics[width=0.45\textwidth]{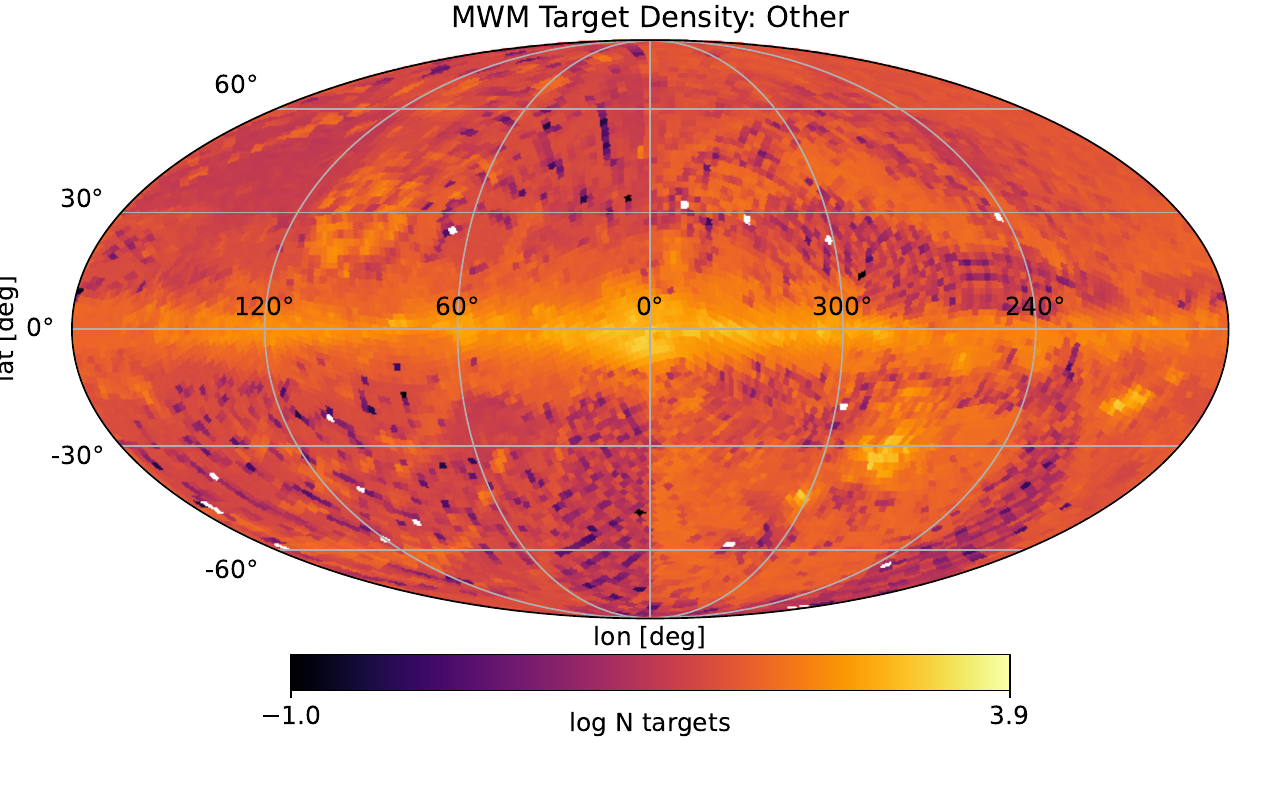}
  \caption{\footnotesize\linespread{1.2}\selectfont{} {\bf MWM on-sky target density.} These plots show the on-sky distribution of targets expected to be observed for MWM for a representative survey simulation.  The Galactic Genesis  survey (left) dominates the total number of stars observed under MWM, particularly in the Galactic plane. Because of the prevalence of \textit{potential} targets everywhere on the sky, some of the over-densities of observed targets are driven by observing pressures from other MWM and BHM science targets. The remaining MWM science targets (right) are more evenly distributed across the sky.
  %This is a tremendous leap forward in mapping galaxies, including our own Milky Way. 
}
  \label{fig:SDSSV_programs}
\end{figure}

\begin{figure}[ht!]
    \centering
  \includegraphics[width=1.0\linewidth]{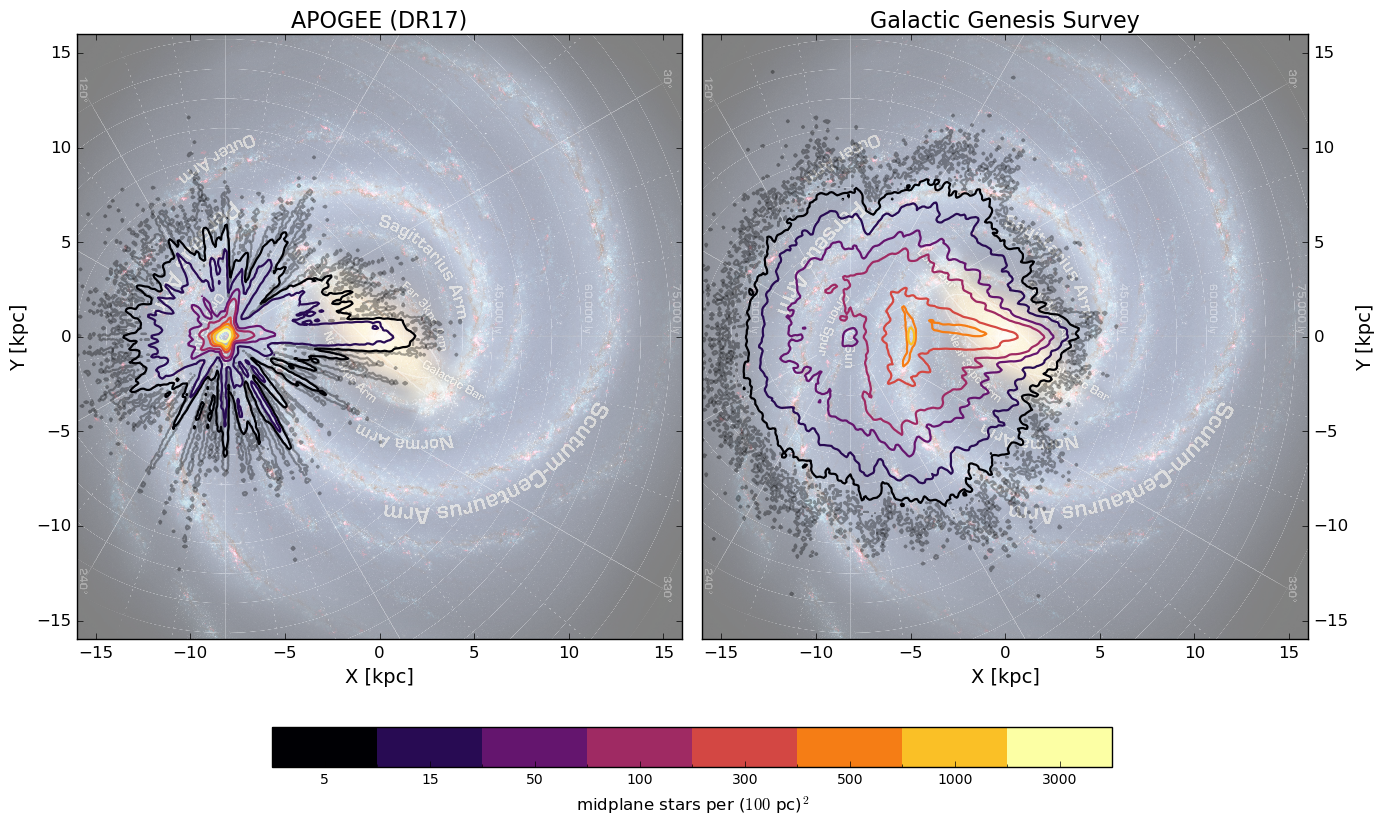}
  \caption{\footnotesize\linespread{1.2}
  {\bf Evolution of SDSS in-plane Galactic target density:} Target surface density of the APOGEE DR17 data release (left) and MWM's Galactic Genesis Survey (GG; right) for stars within $500$ parsec of the Galactic midplane. The maps show a face-on schematic of the Milky Way ({\it credit: NASA/JPL-Caltech/R. Hurt}) beneath target density contours. The Sun is located 8~kpc from the center of the Galaxy, at ($X, Y = -8.0, 0.0$). Light gray contours show areas with anticipated stars at surface densities 1 target per (100 pc)$^2$; colored contours follow the colorbar. These contours only contain stars within 500~pc of the midplane, comprising $4.7 \times 10^5$ in APOGEE DR17 and $1.9 \times 10^6$ stars in GG. Galactocentric target positions calculated using {\it Gaia} DR3 photo-geometric distances from \citet{BailerJones2021}. 
  %NMSU group distances for DR14
  }
  \label{fig:midplane_GG}
\end{figure}

 GG's rapid survey mode across the whole sky is enabled by its focus on bright ($H<11$) yet intrinsically luminous (and thus distant) sources. Its immediate product will be a Galactic census of stellar orbits, ages, and detailed abundances as a function of three-dimensional position across the entire Milky Way disk and bulge. GG alone will collect spectra from more than 3 million stars across the full sky, most of them from a contiguous area of $\gtrsim$3,000 deg$^2$ in the Galactic midplane (Figures~\ref{fig:SDSSV_programs} and \ref{fig:midplane_GG}; Table~\ref{tab:ggsa_classes}). These data will provide the means to address numerous long-standing questions, including the dominant formation mechanisms of the MW, hierarchical accretion, radial migration, and the place of the MW in a cosmological context.

The rigorous interpretation of the Galactic Chronology Survey's data will ultimately rely on knowing and understanding the complete lifecycle of stars from birth to death, including multi-star systems \citep[e.g.,][]{Iben1991,Maeder2000,Yakut2005,Zinnecker2007,Smartt2009,Marov2015}.  However, many questions remain about these lifecycles: What determines the mass and multiplicity of stellar systems?  How does multiplicity affect stellar evolution? What is the relationship between stellar and planetary properties \citep[e.g.,][]{Wang2015a,Wang2015b,Mulders2018}?  How does nucleosynthesis proceed throughout the lifetimes and death throes of different kinds of stars? What is the luminosity function of ancient dead dwarfs (a.k.a. white dwarfs)? What is their contribution to reionization and early metal enrichment, star formation histories, and early accretion history of the Milky Way? What is the spatial and temporal structure of star-forming regions? How structured/coeval are star-forming regions? How do populations evolve with time? How is the structure of the stars and gas influenced by the ionizing radiation from O and B stars? What is the spiral arm structure of the Milky Way as traced by young stars? What are the kinematic and dynamic properties of the young Milky Way disk? SDSS-V therefore also places special experimental emphasis on answers to these questions.

\begin{figure}[ht!]
\centering

  \includegraphics[width=0.45\linewidth]{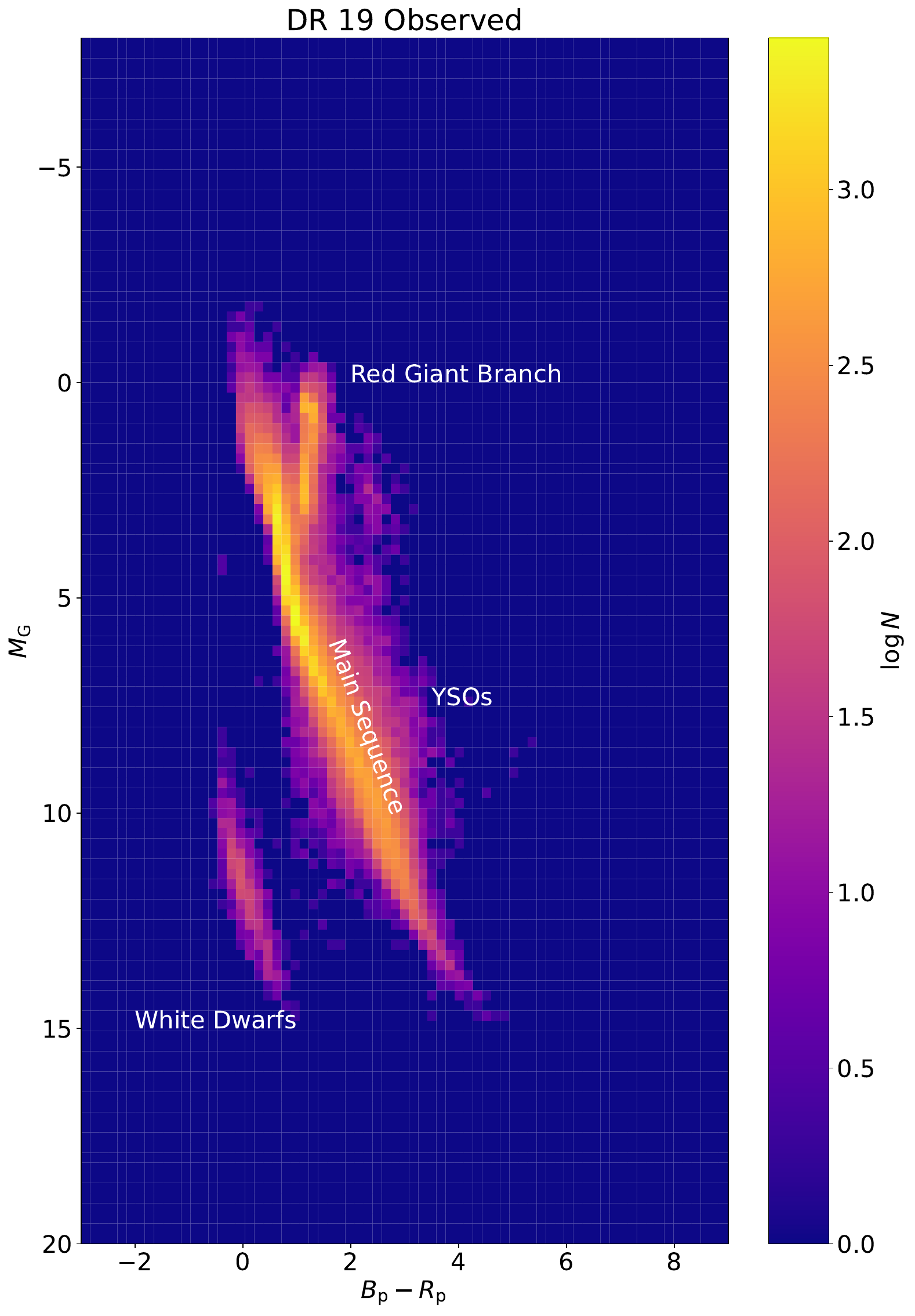}
  \includegraphics[width=0.45\linewidth]{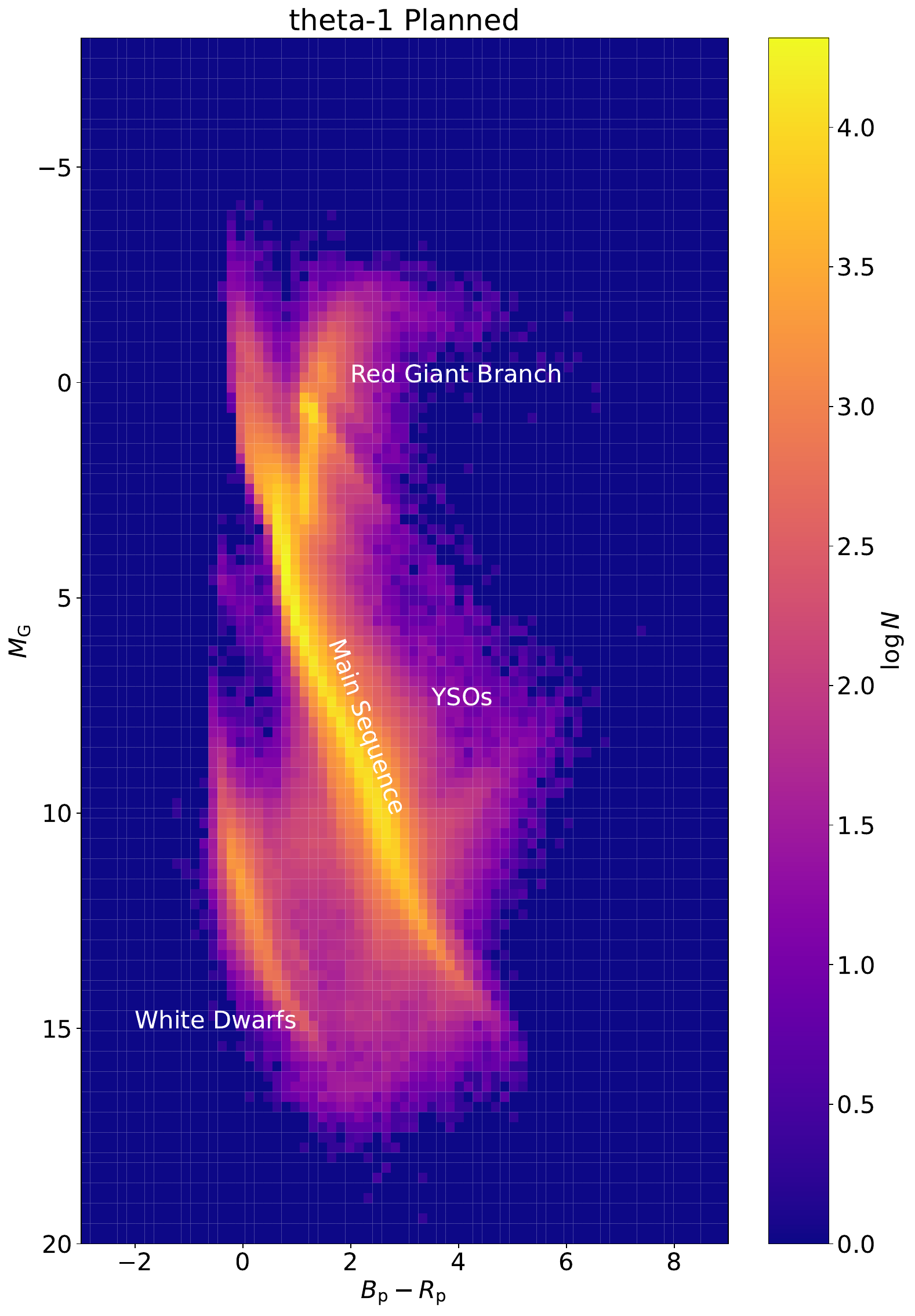}
%  \vspace{-1cm}
\caption{\footnotesize\linespread{1.2}\selectfont{} {\bf Stellar astrophysical targets in the MWM:} The {\it Gaia} $(B_p-R_p)$ color and absolute {\it Gaia} $G$ magnitude of MWM targets within 1~kpc, color coded by number density. The left panel includes all stars observed with either APOGEE or BOSS that will be included in the DR 19 data release.  The right panel shows stars expected to be observed by the end of the survey (based on a representative survey simulation, ``theta-1").  The diversity of spectral types allows the wide range of science cases of MWM  (Section~\ref{sec:mwm}). The luminous hot stars in the upper left ionize the gas seen by the LVM in the Milky Way (Section~\ref{sec:lvm}), and the cool dwarfs on the lower right yield prime hunting ground for rocky planets in the habitable zone, whose host stars must be carefully characterized.  The Galactic Genesis program will probe the Milky Way with luminous red giants, while the Young Stellar Object and Solar Neighborhood Census programs will probe stellar formation with stars along and near the main sequence. Thanks to {\it Gaia}'s measurements of distances to faint white dwarfs, MWM is well poised to characterize a large sample of these stellar ``cinders''.}
\label{fig:hr}
\end{figure}

\subsubsection{Stellar Astrophysics}\label{sec:mwm_sa}

The key data for understanding such outstanding questions of stellar astrophysics are long-duration, high-precision, time-series photometry and spectroscopy of large, cleanly selected samples of single- and multiple-star systems.  In particular, asteroseismology and {\it absolute} flux measurements \citep[e.g., from {\it Kepler} and TESS, and {\it Gaia}, respectively; see][]{Hekker2017,Aerts2019a,Basu2020,Aerts2021a,Creevey2023,GaiaDR3} have recently emerged as game-changers in terms of deepening our understanding of stellar astrophysics---literally, by letting us peer past the previous limit of stellar photospheres.  SDSS-V will use its panoptic spectral capabilities for a comprehensive investigation of Stellar Astrophysics (SA) over a range of 10$^4$ in the masses of stars, evolutionary stages from pre-main-sequence to white dwarfs, and a few pc to $>$15 kpc in distance from the Sun. The SA program will include observations necessary for precise age measurements of giant stars with asteroseismic detections; observations of massive stars to constrain the true relationships between masses, radii, rotation, and internal mixing; observations of  $\simeq65,000$ white dwarfs identified by \textit{Gaia} to improve our understanding of their evolution and mass return to the ISM; observations of deeply embedded stellar clusters, caught in the act of forming stars at numerous stages; and observations of a volume-limited sample of stars within $\sim$100~pc. The overarching goal of the SA program is to consistently and comprehensively measure mass, age, chemical composition, internal structure, and rotation for vast samples of stars across the color-magnitude diagram (Figure~\ref{fig:hr}).

\begin{figure}[ht!]
\centering
\includegraphics[width=\linewidth]{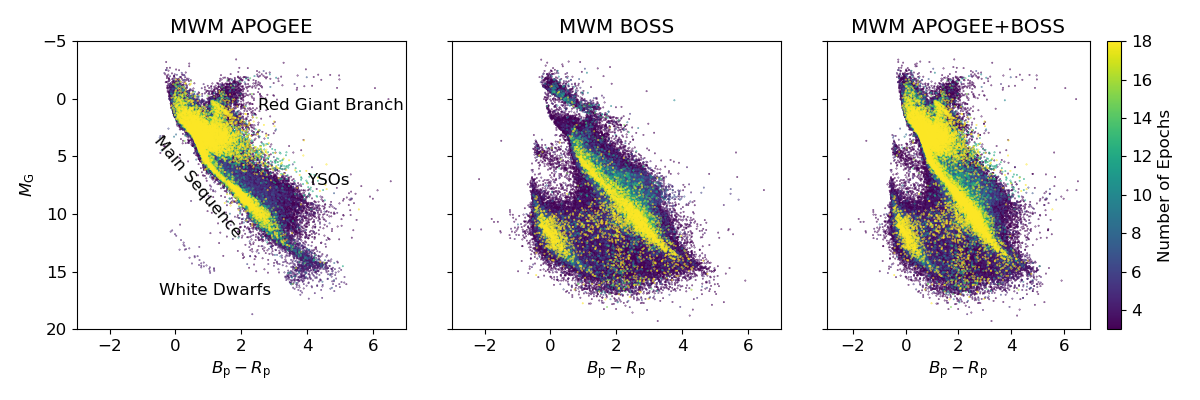}
\caption{\footnotesize\linespread{1.2}\selectfont{}  {\bf SDSS-V's MWM planned epochs shown on the H-R diagram.} Each of the three panels show targets drawn from Milky Way Mapper (MWM) cartons in the Theta-1 version of robostrategy survey simulation that received at least 3 epochs. {\bf Left:} All targets that receive 3 or more observations with the APOGEE spectrograph. {\bf Middle:} All the targets that receive 3 or more observation with the BOSS spectrograph. {\bf Right:} All of the targets from the first two panels combined. Note this does not include legacy SDSS data from BOSS and APOGEE from prior SDSS surveys which increase the number of epochs for some sources.} 
\label{fig:hr_bin}
\end{figure}

While MWM targets stars with companions across the HRD to provide detailed understanding of stellar system architectures (see Section~\ref{sec:mwm_ssa}), its SA program exploits binarity to quantify robustness of state-of-the-art stellar structure and evolution (SSE) models and provide the means for their improvement. Detached eclipsing and astrometric binary stars are a prime source of precise and accurate fundamental stellar parameters \citep[masses and radii, see e.g.][]{Torres2010,Serenelli2021}, provided their respectively photometric and astrometric data is complemented with orbital phase resolved spectroscopic observations. In tandem with asteroseismology, these precise and accurate measurements of fundamental stellar properties allow the SA program to calibrate poorly understood physical phenomena such as interior mixing, rotation, and chemical element transport in stars across the HRD \citep[e.g.,][]{Aerts_Tkachenko2023}.  The time-domain nature of SDSS-V spectroscopy allows our program to probe both the architecture of stellar systems and the physics of stellar components.  In Figure~\ref{fig:hr_bin} we show the multi-epoch nature of SDSS-observations across the HR diagram.

The SA observations of white dwarfs will provide the largest sample of homogeneously measured spectroscopic effective temperatures, surface gravities, primary atmosphere compositions (hydrogen vs helium), and magnetic field strengths. These parameters will be used to advance our understanding of the evolution of the white dwarf atmospheres \citep[e.g.][]{bedard2024} and convection zones \citep[e.g.][]{cunninghametal2019}as a function of their cooling ages, provide important constraints on the atomic and molecular physics under quasi-liquid like conditions encountered among the coolest white dwarfs, and yield insight into the origin of strong ($\simeq1-1000$\,MG) magnetic fields among a fraction of these stellar remnants.

\begin{figure}
\resizebox{0.49\hsize}{!}{\includegraphics{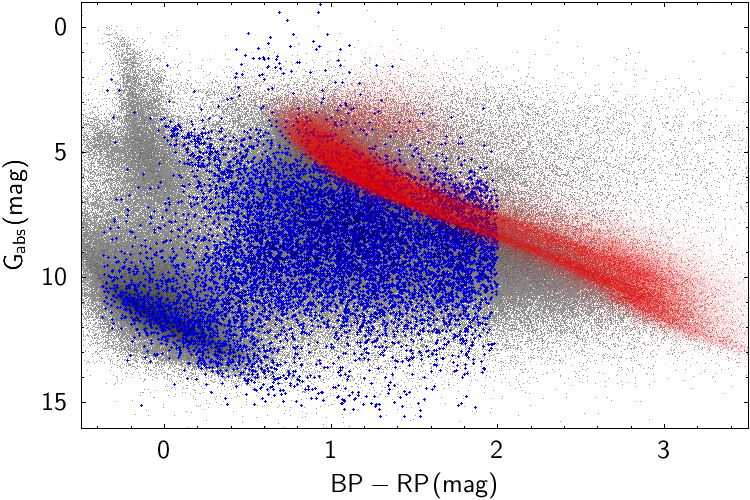}}
\hfill
\resizebox{0.49\hsize}{!}{\includegraphics{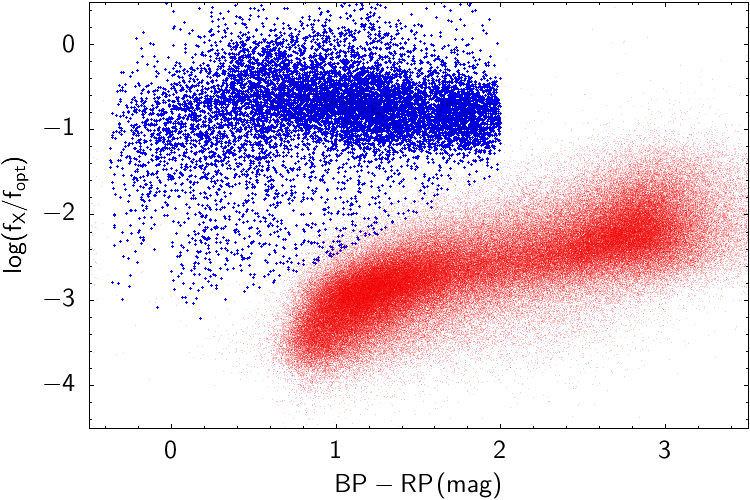}}
\caption{{\bf Color-magnitude (CMD) and color-color diagrams of {\it eROSITA}-selected Galactic targets}. The CMD is based on \gai magnitudes and distances from \cite{bailer-jones+21}. The CMD shows the X-ray to optical flux ratio versus the \gai color. Shown in red are likely coronal emitters, in blue likely accreting compact white-dwarf binaries. Background objects in the left panel shown in grey are UV excess objects, candidates for optically selected compact white-dwarf binaries. 
 \label{f:erogal}}
\end{figure}

\ero has been surveying the sky multiple times. The summed signal of the first three surveys, eRASS:3, was the basis to select targets for follow-up and identification spectroscopy with BOSS. SA will deliver a detailed and comprehensive high-energy view of the Milky Way, both for X-ray active stars (coronal emitters) and for accreting compact white-dwarf binaries. Brightness, color and flux ratio ranges of the selected targets are depicted in Fig.~\ref{f:erogal}. Coronal emitters will be studied from the
bottom of the main sequence up to the A-stars and into the giant branch, thus addressing questions about X-ray to optical activity dependent of age and environment. The space densities of the various classes of compact white-dwarf binaries have large uncertainties due to small memberships and strong selection bias \citep[e.g.][]{schwope18,pala+20}. So far this was limiting the predictive power of observed samples to constrain close binary evolution. SA will uncover compact white-dwarf binaries over ten orders of (absolute) magnitudes, thus being able for the first time to generate meaningful samples of all sub-classes (with magnetic or non-magnetic white-dwarfs, per- vs.~post period bounce objects, with main-sequence or giant or degenerate donor). The comprehensive samples being assembled will provide the input to synthesize the Galactic Ridge X-ray emission \citep{worrall+82}  and to constrain close binary evolution. The X-ray selected samples will be complemented by a comprehensive survey among galactic UV-excess objects to overcome any remaining selection bias. The success of the strategy was demonstrated recently \citep[e.g.][]{inight+23a, inight+23b, schwope+24}.

%and of Stellar System Architecture (SSA)   that belong to binaries, 0.5 hours to $>$12 years in orbital period, and the presence of companions 

\subsubsection{Stellar System Architectures}\label{sec:mwm_ssa}
\begin{figure}[ht!]
    \centering
  \includegraphics[width=\linewidth]{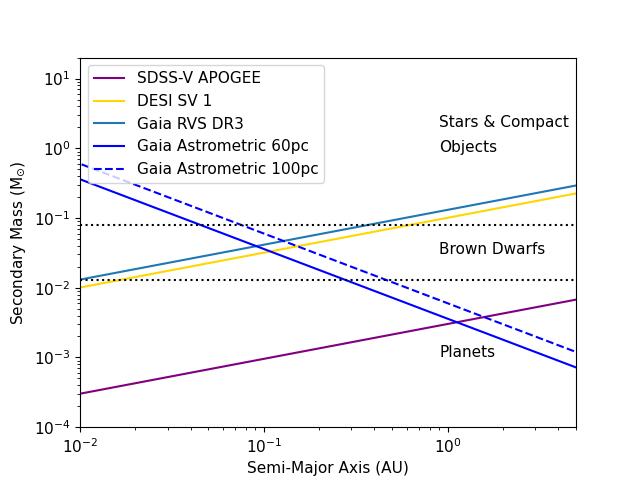}
\caption{\footnotesize\linespread{1.2}\selectfont{}  {\bf SDSS-V's stellar companion mass sensitivity.} The $3\sigma$ detection limit as a function of semi-major axis and secondary mass. The detection limit assumes a 1 solar mass primary and circular orbits. For the RV limits, $\sin i = 1$. The SDSS-V APOGEE instrument has an RV precision of 30 m/s. The DESI SV 1 survey has an RV precision of 1 km/s \citep{Cooper2023} and Gaia RVS DR3 has a RV precision of 1.3 km/s \citep{Katz2023}. The Gaia astrometric precision is 0.02 mas \citep{Lindegren2021} and the $3\sigma$ sensitivity is calculated for a system at 60pc and 100pc in distance from the Earth. The dotted horizontal lines are at 13 solar masses and 80 solar masses respectively.}
  \label{fig:binary}
\end{figure}

It is well established observationally that more than 30 percent of Sun-like and lower mass stars are found in binaries. The binary fraction grows rapidly with the stellar mass and reaches some 80 percent for the most massive stars \citep[e.g.][]{Sana2012,Sana2013,Almeida2017,Bodensteiner2021,Banyard2022}.
Due to the advent of large planet surveys using both transit and radial-velocity methods, we have entered an era where we can study solar systems around other stars with a variety of different architectures. For example, the wide variety of stellar architectures found by the \emph{Kepler} \citep{Borucki2010} and the Transiting Exoplanet Survey Satellite  \citep[\emph{TESS};][]{Ricker2015} missions suggests a complex stellar companion (planets, brown dwarfs, and binary stars) formation mechanism. The Stellar System Architectures (SSA) program will target hundreds of thousands of multiple-body systems with either single or multi-epoch observations  across a diverse range of Galactic environments, with a wide range of dynamical configurations as seen in Figure~\ref{fig:binary}.  SSA seeks to explain the dependence of multiplicity on stellar mass and environment, the frequency and properties of binary systems with compact objects that give rise to explosive events and gravitational waves \citep[e.g.,][]{LIGO_2017_GW17081}, and the effect of host system composition on exoplanet frequency and habitability.  

Although many planet discovery papers include follow-up, high-resolution spectroscopy to characterize the host star characteristics and compositions, these follow-ups are done on a variety of instruments and analyzed with a multitude of techniques, which results in a very inhomogeneous astronomical literature. SSA is utilizing the multiplexing capabilities of the SDSS-V FPS to create a sample of uniformly observed and analyzed stellar systems ranging from the planetary to wide binary star systems. Having all these stellar systems on a uniform system will allow us to do a detailed comparative analysis of elemental abundances to understand how they influence companion formation \citep{Brewer2018}.

Another way that SSA seeks to understand the formation of stellar companions is by detecting and characterizing brown dwarf stars. Brown dwarfs stars exist in a critical part of the companion mass spectrum where the formation mechanism for companions is transitioning from a disk formation model (i.e. planets) to a cloud collapse model (i.e. binary stars). Studying brown dwarfs can put limits on the size of planetary disks around different stellar hosts, which is essential for putting constraints on what can be formed in them \citep{Grether2006}. Brown dwarf companions are relatively rare around sun-like stars. This ``brown dwarf desert'' \citep{Kraus2011} means that large search programs are necessary to find them. The SSA program will again take advantage of the multiplexing abilities of the FPS to extend the search for companion brown dwarfs to other types of stars. In particular, we target M dwarfs, red giants, and subgiant stars to determine the overall abundance and characteristics of brown dwarfs around these hosts.

The SSA program will, for the first time, provide an unbiased census of compact binaries, both detached systems that emerged from a common envelope, and their later-stage interacting cousins. Both the common envelope phase, and the physics of accretion discs and accretion in the presence of strong magnetic fields are complex, observationally under-constrained, and of significant importance in a wide range of contexts, from the progenitors of Type Ia supernovae to young stellar objects and super-massive black holes. The SSA observations will provide detailed information on the physical parameters and the evolutionary states of the observed dwarf binaries, and the statistical analysis of the resulting sample will significantly advance our understanding of compact binary evolution and accretion physics \citep{Inight2023}.

See Table~\ref{tab:ggsa_classes} for more details on targeting for the SA and SSA programs.

\begin{table}[ht!]
%\scalebox{0.85}{
\begin{tabular}{ |l|l|l|r|l|  }
\hline
 \multicolumn{5}{|c|}{\bf Milky Way Mapper Targeting Summary} \\
\hline
{\bf Science Area} & {\bf Program} & {\bf Instrument(s)} & \multicolumn{1}{|l|}{\bf N$_{\rm Targets}$} & {\bf N$_{\rm Epochs}$} \\
\hline
\multirow{6}{12em}{\textcolor{violet}{Galactic Chronology } Section \ref{sec:Milky Way}} & mwm\_galactic & APOGEE & 2,674,808 & 1 \\
\cline{2-5}
& mwm\_dust &APOGEE & 16,722 & 1\\ 
\cline{2-5}
& mwm\_halo &APOGEE \& BOSS & 647,368  & 1-2\\ 
\cline{2-5}
& mwm\_magcloud &APOGEE & 122,680 & 1-4\\ 
\cline{2-5}
& mwm\_ob &BOSS & 323,912 & 1-3\\ 
\cline{2-5}
& mwm\_yso &APOGEE \& BOSS & 186,134 & 1-6\\ 
\hline 
\multirow{4}{12em}{\textcolor{violet}{Stellar Astrophysics } Section \ref{sec:mwm_sa}} & 
mwm\_snc &APOGEE \& BOSS & 739,461 & 1-2\\
\cline{2-5}
& mwm\_tessob & APOGEE & 20 & 8 \\
\cline{2-5}
& mwm\_tessrgb &APOGEE & 1,660,776 & 1-6\\ 
\cline{2-5}
& mwm\_wd &BOSS & 56,609 & 2\\ 
\hline 

\multirow{4}{13em}{\textcolor{violet}{Stellar System Architecture } Section \ref{sec:mwm_ssa}} & mwm\_bin &APOGEE \& BOSS & 131,754 & 1-18 \\
\cline{2-5}
& mwm\_planet &APOGEE & 172,078 & 1-6\\  
\cline{2-5}
& mwm\_cb &APOGEE \& BOSS & 105,405 & 1-2 \\
\cline{2-5}
& mwm\_erosita &BOSS & 127,863 & 1-2\\
\hline 
%\multirow{2}{10em}{\textcolor{violet}{Compact Binaries } Section \ref{sec:mwm_ssa}} & mwm\_cb &APOGEE \& BOSS & 251,446 & 1-2 \\
%\cline{2-5}
%& mwm\_erosita &BOSS & 136,192 & 1-2\\  
%\hline 
\end{tabular}
%}
\caption{\footnotesize\linespread{1.2}{{\bf Summary of primary MWM Science Targets}. Current targeting strategy (based on a representative survey simulation, \emph{theta-1}) for the major science programs of MWM. (Section~\ref{sec:mwm}).}}
\label{tab:ggsa_classes}
\end{table}

MWM can take advantage of several factors to produce this transformative data set: {\it Gaia} photometry and astrometry, the all-sky coverage of SDSS-V, the rapid target allocation enabled by the robotic fiber positioner (Section~\ref{sec:FPS} ), the SDSS-APOGEE spectrographs' IR wavelength coverage and resolution, the large FOV of the SDSS (APO) and duPont (LCO) telescopes (Section~\ref{sec:instrument_mos}), and novel spectral analysis techniques.
\newpage
\subsection{The Local Volume Mapper}
\label{sec:lvm}

The driving science goal of the Local Volume Mapper (LVM) is to constrain and understand the physics of the interstellar medium (ISM) and stellar feedback, connecting small ($<$pc) and large ($>$kpc) scales.  The LVM is designed to provide the missing physical understanding required to constrain empirically why star formation on galaxy scales is so inefficient, and how energy and heavy chemical elements are injected into and subsequently distributed across the ISM. A overview of the LVM project can be found in \citet{Drory2024}.

The LVM is mapping the warm ionized ISM through 3D spectroscopy of emission lines at a resolution of $R\sim 4000$ across the optical wavelength regime (3600-9800\AA). The data directly measure the physical conditions (densities, temperatures, abundances, and ionization states) and kinematics of the gas, which  set the conditions for star formation. It will do this in a regime where we have both high spatial resolution (from 0.1 to 10~pc) and complementary information on the hot, young (hence ionizing) stars, along with information on the neighboring cold molecular and atomic gas. 
LVM aims to resolve individual sources of feedback (HII regions associated with massive stars and young clusters) while at the same time covering contiguously large portions of the Milky Way and of nearby galaxies. In this way, the LVM can build a comprehensive and spatially contiguous view of the star formation and feedback cycle within the ISM.  This has never been done within the Milky Way on these scales and to this extent.  We are creating an optical spectral line atlas of the Milky Way, which previously existed only for H-$\alpha$ \citep{Drew2005} and without kinematic information.  

The key science themes that will be addressed include the connection between ionized gas, star formation, and feedback on multiple physical scales; the extraction of maximum information from the union of resolved and integrated stellar population data; and the topological structure of the ionized and dusty ISM to better understand chemical abundances and enrichment.  

Within these themes, there are numerous outstanding science questions that can only be answered with LVM-like data.  These range from understanding the energetics, sinks, and sources of stellar feedback; the dependence of star formation on the local galactic environment (e.g., gas density and local shear); the life cycle of GMCs and star-forming complexes; the co-evolution of stellar populations and the surrounding ISM; and the distribution of interstellar metals at high spatial resolution, including the mixing of metals produced in supernovae and AGB stars.  We briefly and qualitatively describe two of the primary drivers of the LVM below, with a more comprehensive description of the LVM science cases provided in \cite{Drory2024}. 

\subsubsection{Galaxy Formation and Evolution Physics}

Over the past five decades, we have come to understand that the transformation of cosmic gas into stars into stars is surprisingly inefficient. This cosmic inefficiency -- and the extent to which it reflects material expulsion -- has been a central question in observational galaxy formation.  
Models of galaxy formation have been heavily shaped by empirical relations --  observed in the local universe and at earlier cosmic epochs -- that relate different galaxy-integrated quantities (total luminosity, spectral energy distribution, size, environment, etc.).  However, the processes that actually regulate star-formation efficiency must occur on scales much smaller than whole galaxies \citep[e.g.][]{Schinnerer_Leroy_2024}.  As a result, our theoretical framework has been beset, on all sides\footnote{This challenging situation is, in some ways, reminiscent of that powerfully described by Samuel L. Jackson in {\it Pulp Fiction} \citep{pulpfiction1994}}, by a manifold of ``subgrid" prescriptions for star formation and feedback, with still insufficient constraints on the relevant scales.

\begin{figure}[tbh!]
\centering
\includegraphics[width=\textwidth]{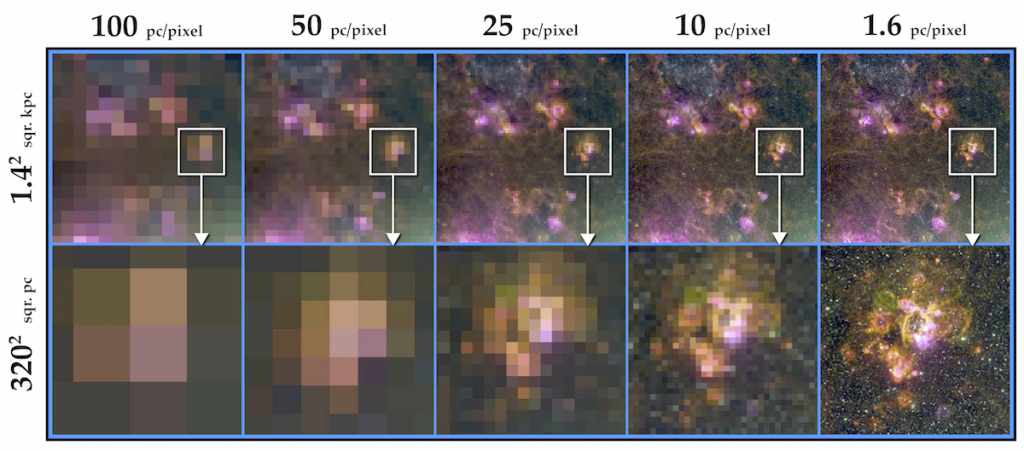}
\caption{
\footnotesize {\bf Resolving nebular ISM structures with LVM.}  
Optical (V-band) and narrow band ([O III], H$\alpha$, [S II]) imaging of the LMC reveal a wealth of nebular structures that are only apparent at the $\sim$10~pc resolution that LVM will achieve in the LMC/SMC and $\sim$pc resolution in the Milky Way. Providing quantitative measurements of the qualitative change observed at $\sim$25~pc scales, when ionization fronts and networks of shocks appear and become resolved, will uniquely enable LVM to directly constrain the physics happening at this ``energy injection scale'' within the ISM. SDSS-V's LVM program revolutionizes our view of the ISM, pushing beyond the extensive $\sim$kpc scale optical IFU surveys (e.g. MaNGA; \citealt{Bundy_2015_manga}) and $\sim$100pc scale local galaxy surveys (e.g., PHANGS; \citealt{Emsellem2022}), to uncover the smaller-scale physics regulating the star formation cycle. 
\label{fig:LVM_zoom}
}
\end{figure}

SDSS-V's Local Volume Mapper (LVM) addresses this problem by making comprehensive integral-field spectroscopy maps of the warm ionized ISM in the Milky Way and in other Local Group galaxies, with a resolution down to the physical scales from which these global correlations must arise.  Figure~\ref{fig:LVM_zoom} illustrates how the visible structure in the ISM changes \emph{qualitatively} starting at scales of 25~pc, below which the bubbles, shells and filaments that form the foundation of the ISM become apparent, and can be separated from surrounding diffuse ionized gas. Very few IFU studies are available of MW star-forming regions, and while some of the largest IFU mosaics cover $\sim$25~arcmin$^{-2}$ on the sky, this still only corresponds to $\sim$pc scales \citep[e.g. Orion,][]{Sanchez2007,Weilbacher2015,McLeod2015}. 

Connecting the internal structure of individual nebulae to galaxy-wide \emph{kpc} scales is critical to track the flow of energy and matter through the baryon cycle, and achieve a comprehensive understanding of the physics that regulate star formation.  %SDSS-V's LVM will provide comprehensive spectral mapping (at a resolution of $R\sim 4000$ across the optical wavelength regime) at high physical resolution over large regions of the Galactic disk, and of nearby galaxies, sampling the ISM across a wide range of local galactic environment.
The total area coverage of LVM will reach about one full steradian ($\sim$3,300~deg$^2$) of the sky. 
For context, the entire SDSS-IV MaNGA survey \citep{Bundy_2015_manga} of 10$^4$ low-redshift galaxies covers approximately 0.5~deg$^2$, and the total area covered by all single-fiber SDSS spectroscopy to date amounts to only a few deg$^2$ of the sky. On the LMC alone, the LVM program will observe $\sim$10$^6$ spectra, approximately as many spectra as were observed in all of MaNGA. These orders of magnitude improvement in terms of information content and physical resolution are essential for probing the ``energy injection scale'', quantifying the links between local and global stellar feedback processes, and placing novel constraints on galaxy evolution.   

\subsubsection{Star Formation and ISM Physics}

\begin{figure}[t!]
\centering

\includegraphics[width=1.0\textwidth]{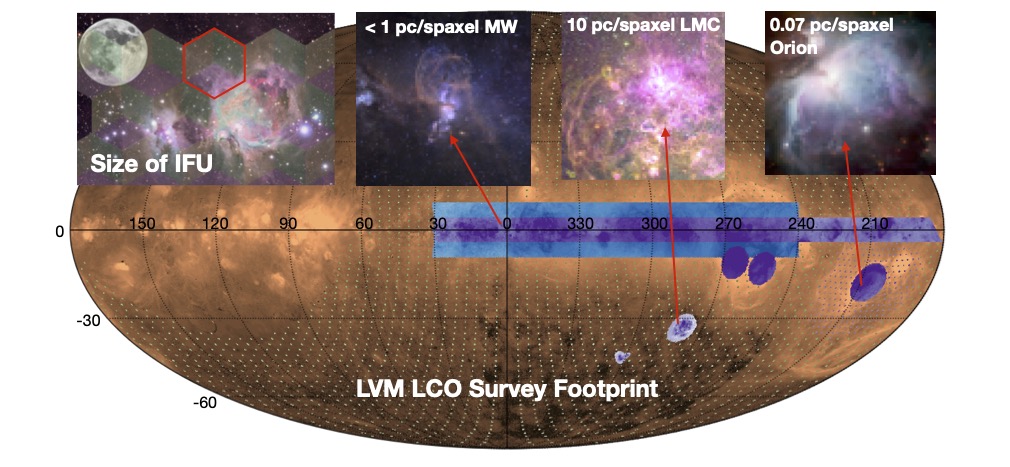}
\caption{{\footnotesize\linespread{1.0}\selectfont{} {\bf The LVM footprint.} Overlaid on a low-resolution all sky H$\alpha$ map (\citealt{Finkbeiner+2003}), the LVM will cover a $\pm$9 degree band along the Milky Way plane, a large area across the Orion and Gum Nebulas, and the full disk of the LMC and SMC. Higher priority zones are colored in purple, lower priority zones in blue. We supplement these core targets with a sparse grid tracking large-scale diffuse ionized gas above the Milky Way plane, and a sample of nearby (D$<$20~Mpc) galaxies. This remarkable survey area of $\sim$4300 deg$^2$ is only possible with our large 0.165~deg$^2$ IFU field of view, which is roughly the size of the full moon (as shown in the inlay on the upper left) and enables efficient tiling of wide survey areas, connecting small scale physics and large scale galaxy evolution. Samples of the scales probed by LVM are shown in the inlaid images.  Zooming into the Orion region (upper right), we see an image of ionized emission sampled at the LVM spaxel size ($\sim$0.07~pc). Within the Milky Way, at $\sim$3 kpc distances we still recover filamentary emission features (center left) at 0.6~pc spaxel sizes. Within the LMC (center right), we see a continuum $+$ ionized emission image of the 30 Doradus star forming region, sampled at the 10~pc spaxel size LVM will deliver on the Magellanic Clouds. Figure from \citet{Drory2024}}} 
\label{fig:lvm-overview}
\end{figure}

The LVM will survey the MW disk at 0.1--1~pc resolution, mapping the internal structure of individual ionized nebulae such as HII regions, planetary nebulae, and supernova remnants (see e.g. \citet{kreckel2024}), and will survey the full disk of the LMC and SMC at 10~pc resolution (see Figure~\ref{fig:lvm-overview}). LVM can and will also map \emph{Local Volume galaxies} at $<1$~ kpc resolution out to $\sim5$~Mpc and $\sim$3~kpc resolution out to 20~Mpc.  These are the critical scales for linking molecular clouds, young stars, and stellar feedback across all phases of the star-formation cycle.

%For comparison, the SDSS-IV MaNGA survey \citep{Bundy+2015} spans about 0.5$^\circ$~deg$^2$ across 10$^4$ low redshift galaxies, and {\it all} of the single-fiber SDSS spectroscopy to date sums to only a few deg$^2$ of the sky. On the LMC alone, the LVM program will provide as many spectra as all of MaNGA's objects taken together ($\sim$10$^6$).  
LVM's survey footprint is designed to align with the rich existing multi-wavelength datasets, and complete our view of this entire matter cycle within the ISM of galaxies. It is well matched in angular resolution and spatial coverage to far-IR, submillimeter and radio surveys that map dust along with the atomic and molecular gas phases, tracing the raw materials of star formation. Stellar spectroscopy from SDSS-V MWM and SDSS-APOGEE spectroscopy (Fig.~\ref{fig:lvm-overview} and Section~\ref{sec:mwm}) provide spectral typing and abundances for millions of stars, while photometry and CMDs in the LMC, SMC and other nearby galaxies provide direct constraints on {\em individual} sources of stellar feedback.   X-ray maps (e.g. from {\it eROSITA}; Section~\ref{sec:bhm}) constrain the often neglected hot ionized gas phase, and pinpoint individual X-ray binaries and other sources of interstellar ionizing radiation.  As LVM probes the ionized gas phase, at the interfaces between stars and denser gas, it will serve as a bridge to fill in the missing information on how the balance in the star formation cycle is maintained. 
%sky coverage will overlap with datasets that provide complementary information at a matched spatial resolution. Stellar spectroscopy with accurate typing and abundances from SDSS-V itself and earlier SDSS-APOGEE spectroscopy (Fig.~\ref{fig:lvm-overview} and Section~\ref{sec:mwm}) as well as resolved stellar photometry and CMDs (in the Magellanic Clouds, \& nearby galaxies) will allow us to connect the structures in the ISM to the radiation field and to {\em individual} sources of feedback. Far-IR, submillimeter and radio surveys probing the dust and H$_2$/HI phases of the ISM connect our observations to molecular clouds and cold gas. X-ray catalogs (e.g. from {\it eROSITA}; Section~\ref{sec:bhm}) indicate the location of X-ray binaries and other additional sources of interstellar ionization.

%\subsubsection{ISM Structure}

The combination of high physical resolution and line sensitivity provided by LVM further allows us to dissect the metallicity, ionization structure and kinematics of gas within individual HII regions. This will provide systematic constraints on the homogeneity of physical conditions in the ISM, including the study of electron temperature fluctuations (via direct detection of intrinsically weaker \emph{auroral} and \emph{recombination} lines) and density inhomogeneities. The IFU coverage and contiguous mapping facilitates a direct comparison of resolved and unresolved studies, to definitively determine their impact on chemical abundance measurements, a key parameter for regulating many physical processes in the ISM. %Sub-parsec resolution while covering kpc to global galactic scales allows for feedback effects to be traced starting from the individual sources through ionization fronts and shocks to global effects such as gas flows and galactic fountains. In its Milky Way Survey, LVM can avoid the trade between coverage and resolution that has beset all earlier optical 3D spectroscopy efforts.

%There is a rich historical data set on the ISM of the Milky Way ISM  and LVM's extensive sky coverage will overlap with datasets providing complementary information at matched spatial resolution. Within the Local volume, stellar spectroscopy with accurate stellar parameters from SDSS-V itself and previous APOGEE spectroscopy (Fig.~\ref{fig:lvm-overview} and Section~\ref{sec:mwm}) as well as resolved stellar photometry and CMDs that will allow us to connect the structures in the ISM to the radiation field and to {\em individual} sources of feedback. Far-IR, submillimeter and radio surveys probing the dust and H$_2$/HI phases of the ISM connect our observations to molecular clouds and cold gas. X-ray catalogs (e.g., from {\it eROSITA}; Section~\ref{sec:bhm}) indicate the location of X-ray binaries and other additional sources of interstellar ionization. 

%LVM's extremely wide-field optical spectral mapping adds critical information about the ionized ISM and integrated stellar populations to these complementary datasets, enabling a synthesis of the local $\leftrightarrow$ global physics in galaxy disks.  LVM's uniquely well-sampled 2D maps of the gas-phase properties of the ISM will provide empirical constraints at the ``energy injection scale''---the physical scale where stars return energy to their surroundings.  These measurements are critical for constraining theoretical models of star formation and feedback, as well as radiative transfer modeling of ISM structures.

\subsubsection{Ultra-wide field IFU mapping}

To address the ambitious goals described above requires a wholly new set of observational requirements, and thus hardware, than the BHM or MWM.  LVM's primary observational requirements of LVM relate to measuring precise and accurate emission line fluxes for an extensive set of optical emission lines that extend across the full optical wavelength range from [OII]$\lambda$3727~\AA\ to [SIII]$\lambda$9532~\AA. At the same time, LVM will record integrated stellar population spectroscopy, which can be related and calibrated to the resolved color-magnitude diagram photometry available in the Milky Way and the Magellanic Clouds.
To enable this ultra-wide field IFS mapping (Figure \ref{fig:lvm-overview}), SDSS-V has successfully implemented a highly innovative system of small (16 cm) telescopes at Las Campanas Observatory that feed a set of spectrographs with fibers that subtend 35.3$\arcsec$ on the sky at 37\arcsec\ pitch. The science IFU is comprised of 1801 lenslet-coupled fibers arranged in a hexagon of 30.2\arcmin\ (1813\arcsec) diameter,  accounting for the extremely-wide 0.165 deg$^2$ FOV. This stand-alone and robotic observing system is described in Section~\ref{sec:lvmi}.

In the initial science implementation of LVM, and to address the broad questions described above, it will map three sets of targets. First, there will be a \emph{Milky Way Survey} that will chart the bulk of the MW disk accesible from LCO, covering the vast majority of known optically selected H II regions and their interface with the diffuse ionized ISM. Second, in the \emph{Magellanic System Survey} LVM  will cover the nearby Small and Large Magellanic Clouds (SMC/LMC) to a line sensitivity limit allowing for the detection of strong emission lines over the full area and auroral lines on the majority of H II regions, sampling a lower metal abundance regime than is available in the MW while still resolving the inner structures of H II regions. Third, and living up to its name of \emph{local volume} mapper,  the \emph{Nearby Galaxies Survey}  will cover a set of $\sim$500 Galaxies within 20 Mpc of the Milky Way at kpc scales. This will provide a link to the MANGA survey, and can be anchored to ancillary observations to anchor how energy cascades to the largest global scales, measured across a diverse galaxy population. 
\section{MOS Instrumentation}
\label{sec:instrument_mos}
SDSS-V operates multiple telescopes and spectrographs in two hemispheres as a single unified survey across the whole sky. In this section, we describe the hardware required to carry out the multiobject spectroscopic programs of the Milky Way Mapper and the Black Hole Mapper.  At Apache Point Observatory (APO) in New Mexico SDSS-V continues to use the 2.5m Sloan Foundation Telescope \citep{Gunn_2006_sloantelescope}, the workhorse of past SDSS generations that remains dedicated to SDSS-V. At Las Campanas Observatory (LCO) in Chile, the Carnegie Observatories' 2.5m du~Pont telescope \citep{Bowen_1973_duPontTelescope} is dedicated to SDSS-V for the duration of the survey. The new LVM facility is described in Section~\ref{sec:lvmi}.

\subsection{Legacy Spectrographs \& Upgrades}
Both APO and LCO each house a set of well-established survey instruments for SDSS-V: a near-IR APOGEE spectrograph \citep[298 fibers, R$\sim$22,000, $\lambda=1.5-1.7\mu$m;][]{Wilson_2019_apogeespectrographs}; a large optical spectrograph \citep[500 fibers, R$\sim$2000, $\lambda=360-1000$nm;][]{Smee_2013_bossspectrographs}.  At LCO, we have added a new facility that includes three medium-resolution optical spectrographs feeding a large IFU (2000 fibers, R$\sim$4,000, $\lambda=360-1000$nm). See Figure~\ref{fig:SDSSV_schematic} for a schematic layout of these instruments. For the multi-object spectroscopy used by the MWM and BHM, each 2.5m telescope is able to observe up to 500 targets at a time, drawing on both the optical and IR multifiber spectrographs simultaneously. SDSS-V relies heavily on the established SDSS observatory infrastructure, instrumentation, and operations model, including the use of the near-IR APOGEE and optical BOSS spectrographs for the MWM and BHM.  SDSS-V improves and expands this infrastructure in important ways, foremost through new robotic fiber positioning (RFP) systems for both MOSs, and an ultra-wide field IFS.

\subsubsection{APOGEE}
\label{sec:APOGEE}
Both APOGEE spectrographs have been upgraded in three respects to improve the radial velocity precision from 100 -- $200\,\rm{m/sec}$ to $\lesssim 30\,\rm{m/sec}$, as described in \citep{Wilson_2022_rvupgrades}.  First, segments of octagonal core fiber were incorporated in the APOGEE fiber train of both robotic Focal Plane System units.  Octagonal core fibers improve radial scrambling and thus positional stability of the fiber illumination at the end faces of each fiber terminated within the spectrographs despite variations in the input illumination of the fibers at the telescope focal plane due the effects of seeing and telescope pointing and fiber positioning errors.  As the fiber end-faces, which form the spectrograph pseudo-slit, are reimaged by the spectrograph onto the detectors, this improved illumination stability directly leads to improved radial-velocity stability.  Second, back pressure regulation systems were added to control the boil-off pressure of the internal spectrograph liquid nitrogen tanks.  Controlling this pressure, rather than letting it drift with changes in ambient atmospheric pressure, stabilizes the liquid nitrogen boiling temperature. This, in turn, minimizes internal expansion and contraction of the instrument's opto-mechanical modules.  Lastly, Fabry-Perot Interferometric (FPI) calibration sources were developed to provide a picket fence of spectral lines with uniform spacing and illumination across the APOGEE wavelength range to improve wavelength calibration.  Wavelength calibration images with FPI light that illuminates all 300 fibers are taken before and after every night of observing.  During on-sky observations, two fibers are illuminated with FPI light to provide real-time wavelength calibration.  All three of these upgrades were pioneered by the Penn State Radial Velocity Group for their Habitable Zone Planet Finder \citep[HPF][]{Mahadevan_2012_HPF,Mahadevan_2014_HPF} and NEID \citep{Schwab_2016_HPF} instruments.

The APOGEE fiber system relies on high-fiber-count MTP connectors, manufactured by \emph{US Conec}, to enable the rapid connection and disconnection of the 300 fibers emanating from the instruments at either the FPS or calibration ports.  Ten MTP connectors, each containing 30 fibers, are grouped within gang connector assemblies which enable all the fibers to be cycled simultaneously.  This system had been used during the SDSS-III APOGEE-1 and SDSS-IV APOGEE-2 surveys to enable rapid changes of pre-plugged cartridges throughout the night. Consequently, the most heavily used connectors had over 17,000 cycles, far exceeding the 500 cycles that the manufacturer typically used for durability testing and throughput.  In advance of starting SDSS-V operations with the FPS at APO, the most heavily used MTP fiber connections were re-terminated in the field.  The gang connector assemblies were also replaced.  \citet{Wilson_2022_mtpreterminations} describes the re-termination process and the lessons learned.  No re-terminations have yet occurred at the LCO fiber system, which had been introduced only in SDSS-IV and hence has had less usage.

\subsubsection{BOSS}
\label{sec:BOSS}
In SDSS-V, the twin BOSS optical spectrographs retain their fundamental capabilities described in \cite{Smee_2013_bossspectrographs}. One BOSS spectrograph was moved to LCO, refurbished, and mounted on the du Pont telescope, in order to enable SDSS-V's all-sky MOS capability in the optical.

Several BOSS spectrograph modifications were required to assure availability of spare parts and ease of maintenance at LCO, as many of the original spectrograph parts are no longer manufactured. Unavailable parts were replaced with equivalent items. In particular, the electronics systems required refurbishment. Additionally, the collimator mirror position motors, collimator mirror motor controllers, the primary motion control computer, and the CCD readout controller were replaced with new items.

In addition, a standalone autofill system for liquid nitrogen was deployed with a 15 liter buffer dewar. The original system (still used at Apache Point Observatory) used the CCD controller to handle the autofill duties from a 10 liter buffer dewar.

A new slithead was built from a monolithic glass part and has been integrated into the mechanical structure of the FPS. This replaced the former slithead that was assembled from multiple steel V-groove blocks.

The BOSS CCDs remained the same for SDSS-V, as they have retained their efficiency. But the original Princeton CCD controller was replaced with a commercial Archon controller, designed and manufactured for SDSS-V by Semiconductor Technology Associates (STA). The controller is fitted with dual readout channels, so the two CCDs can be read out simultaneously on the same clock train.  This series of updates, while leaving the BOSS performance unchanged, extends the spectrograph lifecycle and allows SDSS-V to make efficient and economical use of high-quality heritage hardware, to focus ressources on constructing the new hardware required for SDSS-V's ambitious science goals and ``success-oriented" schedule.
%\footnote{The SDSS-V schedule was called many names.  This was the most favorable and most accurate one according to the SDSS-V Director, but the reader should consult with the sub-system leads for their perspectives, which may differ.} schedule.

\subsection{Robotic Focal Plane System}
\label{sec:FPS}

Qualitatively new hardware requirements arose from SDSS-V's high target densities and all-sky survey scope that implied rapid exposure 
sequences: SDSS's long-standing plug-plate fiber system had to be replaced with robotic Focal Plane System (FPS) units on both telescopes.
Each FPS unit has 500 fiber positioners in a fixed hexagonal array in the focal plane of the telescope, where each positioner can reach a surrounding zone. Each positioner carries three 120 micron diameter fibers: one back-illuminated metrology fiber and two science fibers, one for feeding the BOSS spectrograph, and one for APOGEE. However, only 298 fibers are connected to the APOGEE spectrograph at one 
time. Due to differing telescope plate scales at each site, the on sky fiber diameter is 1.3 arcseconds at LCO and 2 arcseconds at APO.  In addition to the 500 fiber positioners, the focal plane carries 60 fiber-illuminated fiducials (FIFs). These are distributed 
throughout the hexagonal array and around its periphery. They
establish a fixed reference frame against which the robot 
fiber positions are measured using a telescope-mounted 
Fiber Viewing Camera (FVC) system. Six CCD cameras 
mounted around the periphery of the robot array provide 
guiding, acquisition, and focus monitoring functions, 
completing the configuration of the focal plane.

The focal plane elements are mounted on a precision-curved mounting plate that is shaped to put the fiber faces and
guide camera CCDs in the focal plane of the telescope. The
fully integrated focal plane assembly is installed in an 
optomechanical mount and thermal enclosure that mates to 
the telescopes with the same 3-point kinematic latching 
system that had been used by the heritage SDSS plug-plate cartridges. The FPS 
enclosure carries the BOSS slit head and the APOGEE fiber 
connector, along with the metrology fiber back-
illumination system and the glycol cooling manifolds, which 
remove waste heat from the 500 robotic fiber positioners 
and CCD guide cameras. A suite of temperature, humidity, 
and glycol coolant pressure and flow sensors complete
the FPS system. An outboard instrument electronics box, 
mounted on the instrument rotator with the FPS, carries 
all of the control electronics and distributes AC power 
and fiber ethernet services to the FPS.

\begin{figure}[tbh!]
\centering
\includegraphics[width=\textwidth]{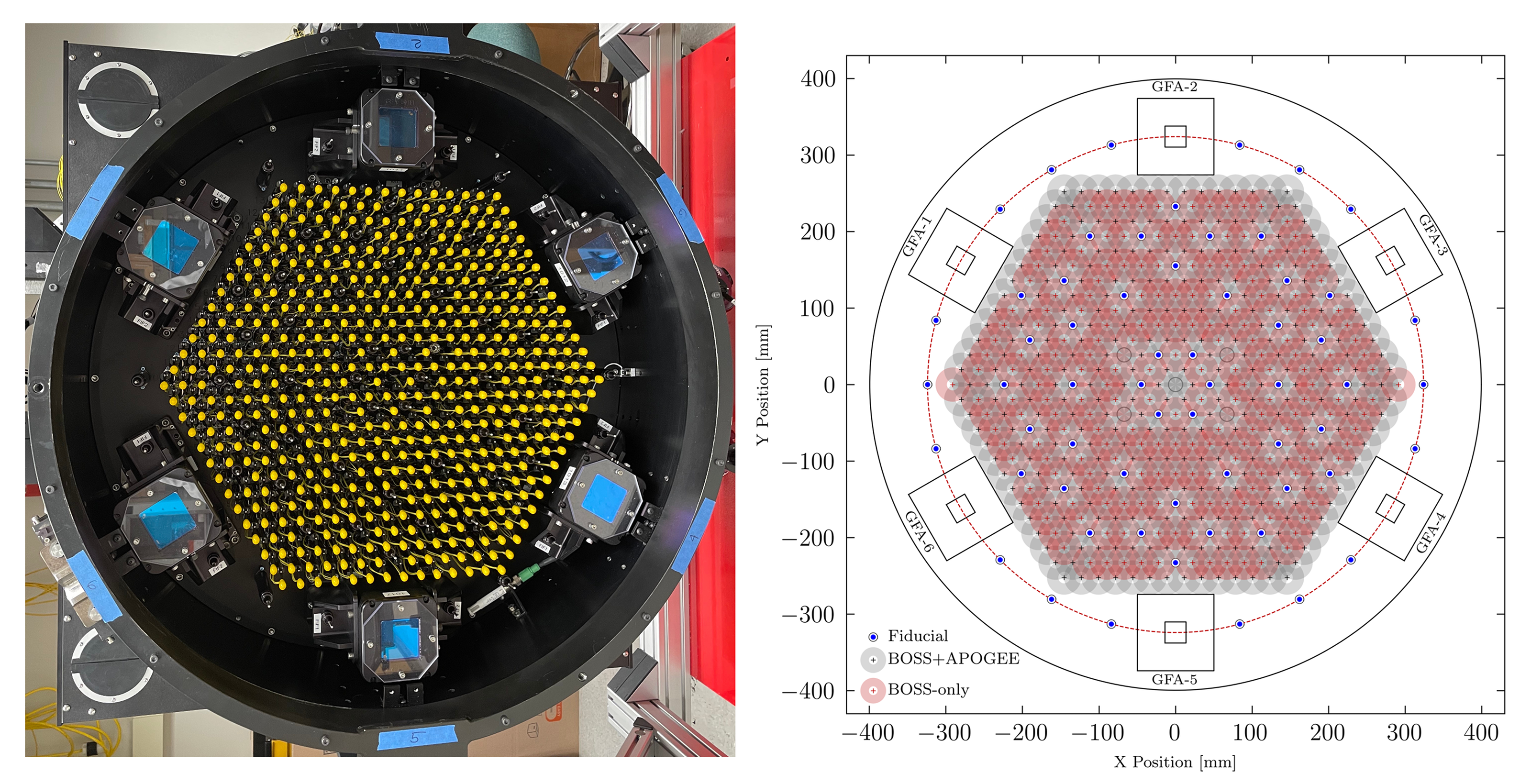}
\caption{
\footnotesize \textbf{FPS focal plane configuration}. 
Left: Sloan FPS unit, with yellow dust caps installed on 
the fiber positioners; Right: layout showing fiber 
positioner patrol fields (colored +s and annuli), fiber-
illuminated fiducials (blue and white circles), and 
guide/focus/acquisition camera locations (squares).
\label{fig:FPS_FocalPlane}
}
\end{figure}

The layout of the FPS focal plane had been designed to maximize 
coverage of the telescope corrector fields of view with 
500 BOSS fibers and 300 APOGEE fibers.  A photo of the 
Sloan Telescope FPS unit is shown in Figure\,
\ref{fig:FPS_FocalPlane}. This Figure also shows a schematic of the 
focal plane layout with the patrol fields of the 500 BOSS fibers and of the 298 robots that have active APOGEE fibers. It also shows the 
locations of the fixed fiber-illuminated fiducials and the 
locations of the guide/focus/acquisition CCD cameras.  This layout ensures full-field coverage in both the optical and infrared consistent with the science requirements of SDSS-V.

The FPS units reduce the field-to-field reconfiguration time from
the previous $\sim$15 minutes needed to swap plug plate to $\lesssim$3 minutes at fixed field position for the new FPS robotic reconfiguration. The previous plug-plate system was also limited to at most twelve field changes per night, and nine more typically planned, which capped the survey cadence. The new FPS system now allows $\gtrsim 30$ fields (or fiber configurations) to be observed in a given night. However, the robotic FPS system enables also a qualitatively far more dramatic improvement. Classic SDSS survey plates needed to be drilled and delivered to the site months in advance of operation and be planned for operation on specific nights and at a specific airmass. {\it The total lead time to settle on target coordinates  has therefore shrunk from months to literally minutes.} Fundamentally, the SDSS-V hardware system can now tackle transient science spectroscopically.

Now the fiber positioners can be configured to account for atmospheric refraction more easily than drilled plates, greatly 
increasing the available observing window for any field 
and boosting the survey efficiency. Targets can also 
be incorporated on short timescales to allow observations of 
transients and other targets of opportunity.

The first FPS unit arrived at Apache Point in November 2021 and was installed and commissioned on the 
2.5-m Sloan Telescope in December 2021.  The second unit was installed and commissioned 8 months later, in August 2022, on the 2.5-m Ir\'en\'ee du Pont Telescope at 
Las Campanas Observatory.  This hardware implementation and comissioning (which included extensive international travel) was successfully carried out despite the severe impediments that the COVID-19 pandemic caused.  More details about the FPS can be found in Pogge et al. (2025, {\it in prep.}).

\subsection{Wide Field Corrector}
The new robotic Fiber Positioning System placed different requirements on the focal plane characteristics of the telescope.
This necessitated the design and implementation of a new three-element wide field corrector for the SDSS 2.5m telescope at APO. The original SDSS two-element corrector produced a focal surface that was non-telecentric and suffered from issues with axial color, throughput and image quality, when used in the H-band with the APOGEE spectrograph. The combination of the new optical prescription with broadband anti-reflection coatings now addresses the telescope's previous optical shortcomings. The optomechanical design of this new corrector required very minimal changes to the telescope interfaces. It also allowed for in-situ axial adjustment of one lens element in order to fine-tune the as-built spherical radius of the focal surface to match the FPS' nominal design value.

The SDSS-V corrector design consists of three elements, all of which are fused silica \citep{2022SPIE12182E..3OB}. The center thickness of L1 and L3 are 50~mm while the thickness of L2 is 40~mm.  These are thick when compared to the original SDSS corrector lenses.  Fabricated by SESO\citep{2022SPIE12188E..0MG}, the extra thickness allowed these lenses to be constructed without bonding a support to one side while grinding the lens to near-final thickness; reducing fabrication risk.  The design has two aspherical surfaces, one on the front surface of L1, the other on the rear surface of L2.   

The coatings for the original SDSS survey were optimized for the visible spectrum, with no consideration given to transmission in the infrared. Previous measurements have shown that these coatings do not perform very well in the H-band.  For SDSS-V, a broadband anti-reflection coating was used, with low reflectance in the visible and H-band, and with relaxed performance in the deadband region (1–1.5 um).  This compromise provided slightly better performance, at the level of 0.5\% per surface, over a continuous broadband design.  The overall corrector transmission is 90\% based material transmission data and coating performance predictions from SESO. 

The optomechanical design of the SDSS-V corrector borrows extensively from the previous SDSS
corrector design, which served the project well for many years. SDSS-V lenses are mounted in
steel cells with a room-temperature-vulcanizing silicone (RTV) annular layer between the cell-internal diameter and the lens' outer diameter. The RTV layer accommodates differential contraction between the cell and glass; the radial thickness was sized such that the net expansion of the RTV internal diameter tracks
the expansion of the lens. Steel (C1020) tube sections support the lens cell assemblies and meter the spacings
from lens to lens. Given the addition of a third lens with the SDSS-V corrector, some new features were added,
such as vent ports between L1 and L2.  Mechanically, the main difference between the old and new designs is the overall mass.  Three thick lenses compared to two thin lenses in the SDSS design added not just the mass of the additional glass but additional metal as well; an overall mass increase of 177~kg.  

\begin{figure}[!ht]
 \centering
  \includegraphics[width=1.0\linewidth] 
    {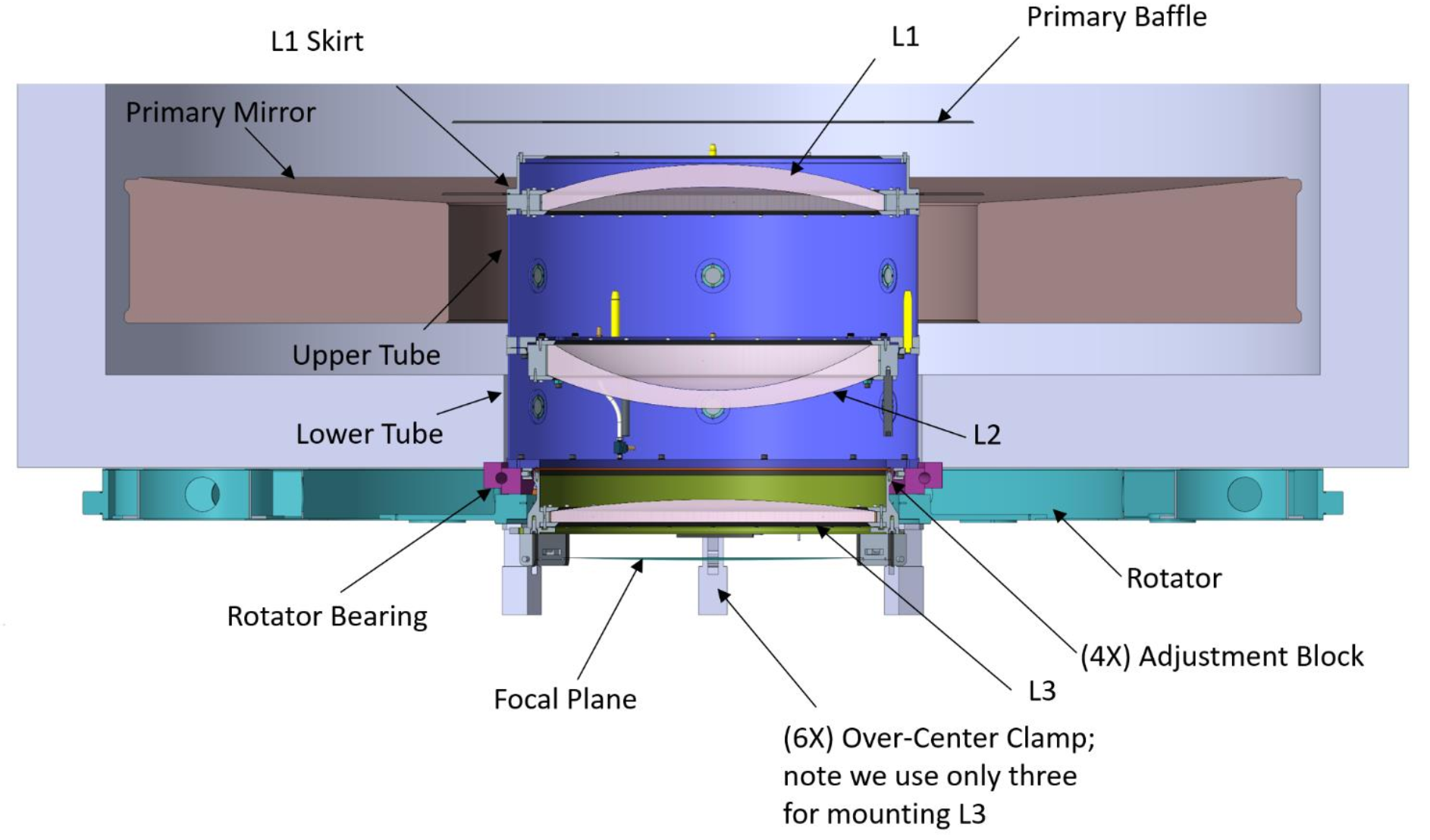}
 % \vspace{-0.4in}
  \caption{{\bf SDSS-V Wide-Field Corrector for the 2.5m at APO.} A new corrector was required for the SDSS Telescope at APO to enable our robotic instrument.  This figure shows a schematic layout of the new corrector.  The three new optical elements (L1, L2, L3) are shown along with details of the optomechanical assembly. %while the bottom image shows Photographs of the SDSS-V wide field corrector in the cleanroom at JHU, just prior to shipping. Both corrector subassemblies are shown; the L3 subassembly is shown in the foreground, and the main subassembly containing lenses L1 and L2 is shown in the background.
  }
  \label{fig:SDSS-V_photo}
\end{figure}

The SDSS-V wide field corrector was installed and commissioned on-sky between late July and early August of 2021. The design drawing of the corrector and a photograph of the corrector in the clean room at Johns Hopkins University just before shipment are shown in Figure~\ref{fig:SDSS-V_photo} Imaging performance and focal surface radius of curvature were measured using a custom-built \emph{focal surface camera} (FSC); a CCD camera mounted to a three axis (R/$\theta$/Z) stage housed in a re-purposed fiber cartridge.  With the FSC mounted to the telescope Cassegrain rotator, the CCD camera was commanded to an array of field locations, with precise through-focus scans taken at each field point.  A filter wheel containing SDSS u, g, r, and i band filters attached to the camera allowed data to be taken in select SDSS imaging filter bands.  A detailed description of the focal surface camera and the procedure used to collect focal surface and performance data is described in \cite{2022SPIE12184E..62G}. The processed through-focus data provide a measure of the sag in the focal surface at each field point, while individual images at best focus provide a measure of the imaging performance at that field location.

Commissioning of the system started at Apache Point in very early hours of August 7 and 8, 2021 in fortuitously good seeing.  Using the i-band filter, data was collected with the FSC to measure the focal surface radius of curvature for comparison against the prescribed value.  Through focus sweeps measured the image point spread function (PSF) size as a function of FSC camera piston at a multitude of points intermixed with repeated sweeps at the center of the field to monitor for focus drift.  In the very early hours of August 7, during a routine focus sweep, a minimum PSF of 0.68" FWHM (Full-Width at Half Maximum) was recorded in i-band at the center of the field.  A lower FWHM of 0.635" was recorded later in the observing run.  The following day, in the early hours of August 8, a FWHM of 0.585" was logged, also at the center of the field in i-band. To put this performance in context, the best i-band imaging ever recorded in 2400 hours of SDSS imaging was 0.698" FWHM.  The new SDSS-V wide-field corrector bested that number three times during the first two nights on sky. More details about the SDSS-V corrector can be found in Smee et al. (2025, \emph{in prep.}.

\section{Local Volume Mapper-instrument (LVM-i)}
\label{sec:lvmi}

The Local Volume Mapper Instrument (LVM-I) is the instrument that was designed and built to execute the LVM survey. It will create a spectral map of over 3500 square degrees of the Southern Galactic plane with $37^{\prime\prime}$ spatial resolution and R\textasciitilde4000 spectral resolution over the wavelength range 3600 - 9800 \AA ; and it will map the Magellanic Clouds. LVM-I consists of five subsystems: the telescopes, the Integral Field Unit (IFU) and fiber system, spectrographs, the enclosure, and software. Each of them has been designed to work seamlessly with the others. LVM-I represents the first completely new end-to-end facility since the first incarnation of the Sloan Digital Sky Survey more than 25 years ago.  An overview of the LVM instrumentation can be found in \citet{Herbst2024}.

\begin{figure}[!ht]
 \centering
  \includegraphics[width=1.0\linewidth] 
    {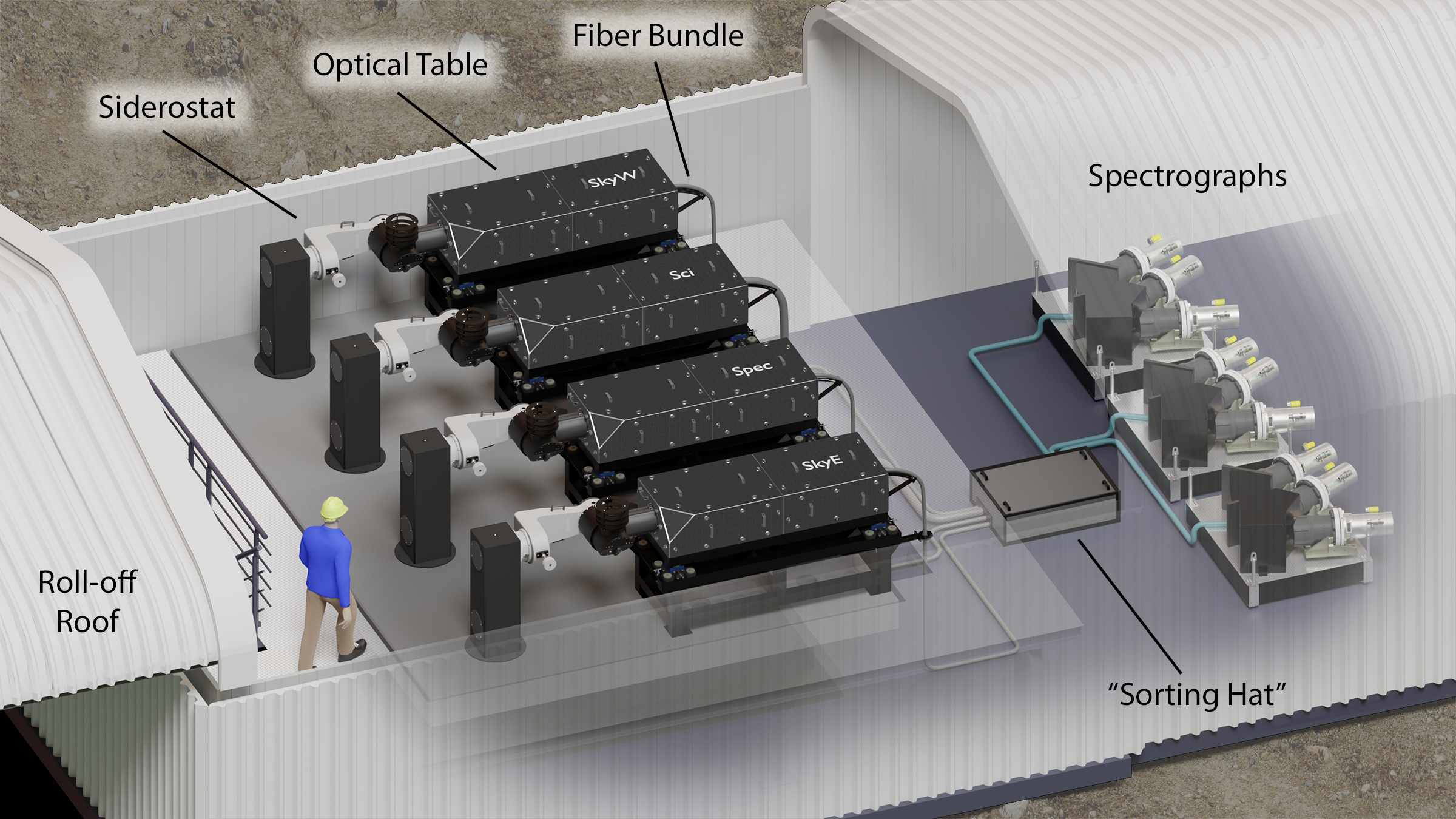}
 % \vspace{-0.4in}
  \caption{{\bf Overview of the LVM Instrument.} Each of the four  telescopes uses a siderostat in alt-alt configuration feeding components on an optical table. The IFUs in the focal plane convey the light to an environmentally-controlled spectrograph chamber containing a “sorting hat” to distribute the fibers to three identical spectrographs.
  }
  \label{fig:LVMI1}
\end{figure}

Figure~\ref{fig:LVMI1} shows an overview of the LVM Instrument. Each of the four telescopes consists of a two-mirror siderostat in horizontal, altitude-altitude configuration relaying the light to components on an optical breadboard protected by a dust-proof cover. This produces a fixed and stable focal plane for the microlens-based IFU's, which are connected to optical fiber bundles that convey the light to three spectrographs housed in an adjacent temperature-controlled chamber. This dual-walled chamber lies at the heart of the custom-built enclosure, which also incorporates a roll-off roof to shelter the telescopes, a utility area, and a small control room. The software subsystem controls all of these elements, currently allowing fully robotic and/or remote observations.

%% Figure LVMI1 here

LVM-I was implemented as a truly a global project, with contributions
from five continents: North America, South America, Europe, Asia, and Australia. The following sections describe each of the LVM-I subsystems in detail, and Konidaris et al.(2025 in preparation) provides more information on the instrument as a whole.

\subsection{Telescope Subsystem}\label{telescope-subsystem}

\begin{figure}[!ht]
 \centering
  \includegraphics[width=1.0\linewidth] 
    {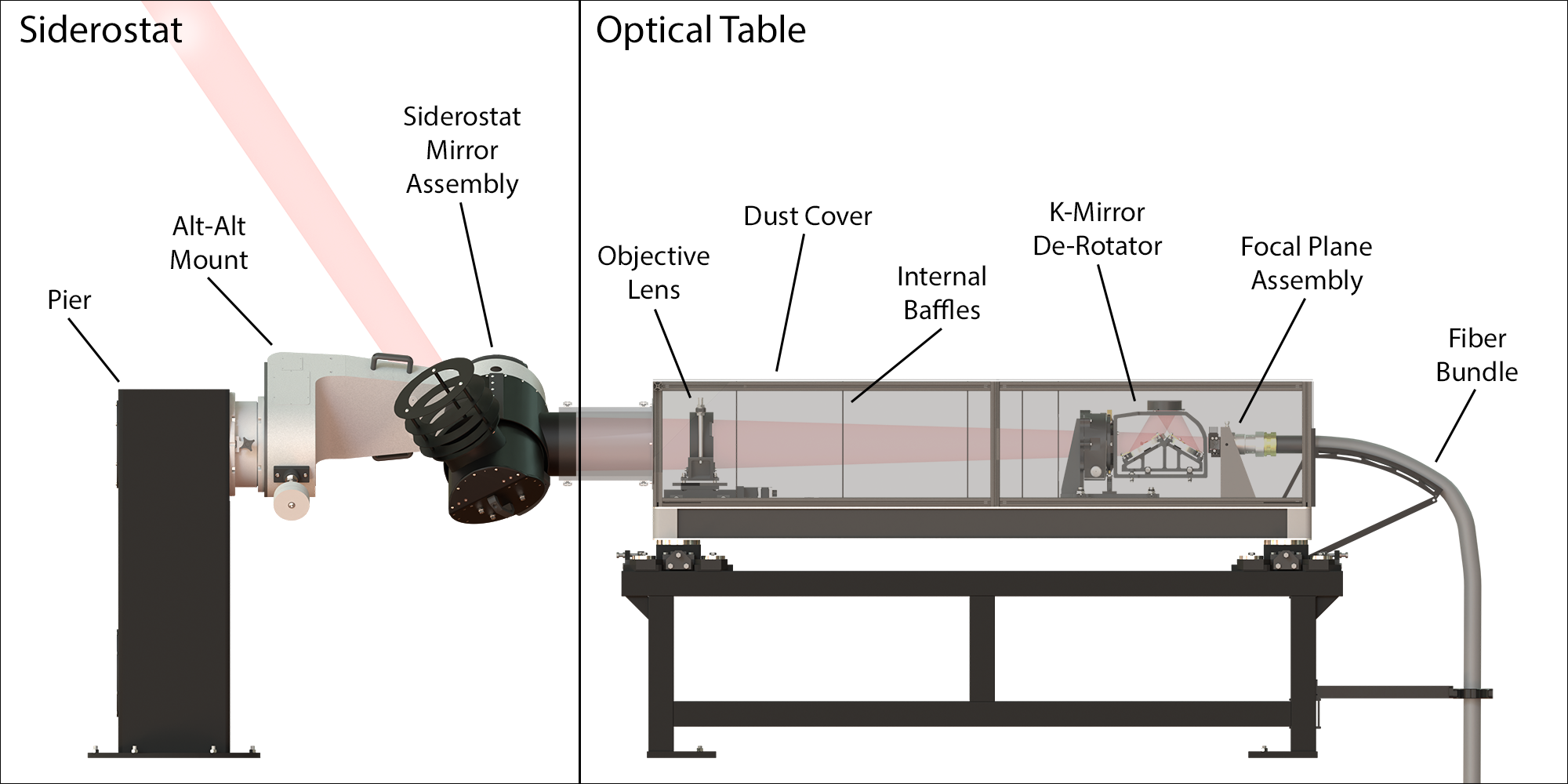}
  \vspace{-0.04in}
  \caption{{\bf Architecture of the LVM telescopes.} Note that the Science and two Sky telescopes contain K-mirror de-rotators as shown, while the fourth, Spectrophotometric telescope does not. It hosts a rotating fiber selector mask in its focal plane for isolating the flux of individual bright stars.}
  \label{fig:LVMI2}
\end{figure}

Figure~\ref{fig:LVMI2} shows a single telescope and labels the major components. Formally, the LVM telescope subsystem consists of all opto-mechanical elements between the sky and the microlenses of the fiber bundle. This includes the siderostats, objective lenses, K-mirror de-rotators, the focal plane assemblies (FPA), and optical tables, as well as the associated electronics and low-level control software. The design and production of all of the telescope hardware was overseen by the Max Planck Institute for Astronomy (MPIA) in Heidelberg, except for the objective lenses that came from the Carnegie Observatories in Pasadena. Herbst et al. (2020, 2022)
report on the design and production of the telescopes, while Lanz et al. (2022) describe the lens design, manufacturing, and testing.

As shown in Figure~\ref{fig:LVMI1}, there are four telescopes in total. One telescope hosts the Science IFU, while the two ``Sky'' telescopes observe adjacent fields to calibrate geocoronal emission. The fourth ``Spectrophotometric'' telescope makes rapid observations of bright stars to compensate telluric absorption.

As with previous Sloan surveys, the Local Volume Mapper depends on consistency in terms of both measurement and calibration to achieve its science goals, and this consistency has been a guiding principle in the design of the LVM telescopes. The overall telescope architecture was explicitly motivated by the need for simplicity and robustness in a multi-year, automated survey such as LVM. For example, with the exception of the siderostat, all elements are in a fixed, gravity invariant environment enclosed within a protective dust cover. This means that the crucial Integral Field Units (IFU's) and fiber bundles never move, in principle eliminating concerns about time variable Focal Ratio Degradation (FRD) and throughput.

\subsection{Fiber IFU Subsystem}\label{fiber-ifu-subsystem}
The Fiber IFU subsystem comprises the four microlens-fed Integral Field Units, the three spectrograph slit heads, and the ``Sorting Hat'' which inter-connects the IFU and slit fibers. It also includes the fiber bundles between the telescopes, Sorting Hat, and spectrographs, along with flat-field and test slits and related hardware for laboratory integration. Australian Astronomical Optics (AAO) in Sydney produced the IFU subsystem, and Feger et al. (2020) provide additional details.

\begin{figure}[!ht]
 \centering
  \includegraphics[width=1.0\linewidth] 
    {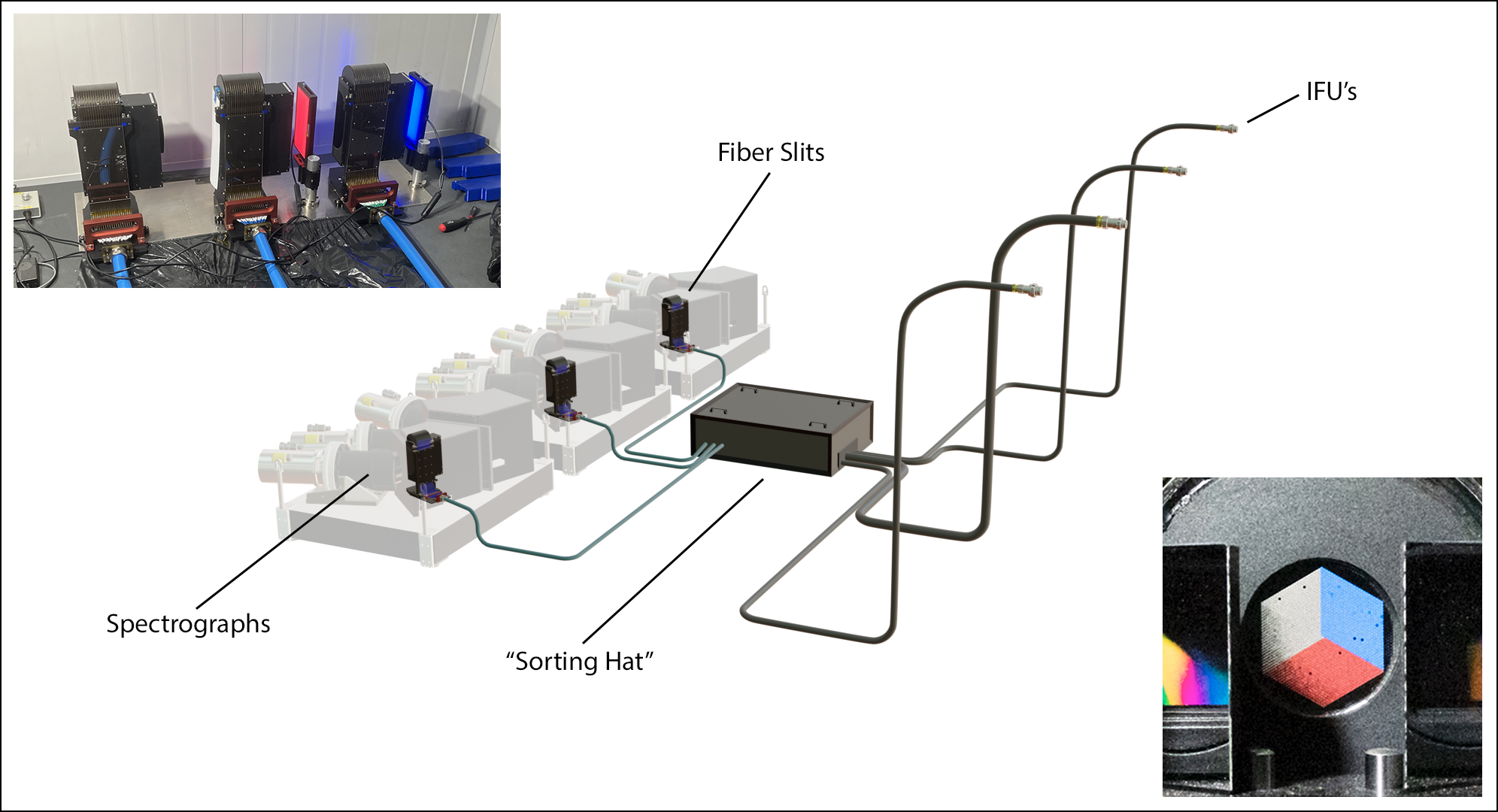}
  \vspace{-0.04in}
  \caption{{\bf The LVM Fiber System.} The four fiber bundles coming from the IFU’s (right) enter the “Sorting Hat”, where almost 2000 fiber splices redistribute them to the fiber bundles leading to the entrance slits of the spectrographs (left). The inset photos show a back illumination test performed at LCO during installation in February 2023. Three different coloured lamps illuminate the slit heads (upper left). The science telescope IFU then reveals the distribution of fibers by spectrograph (lower right). See also Figure 19 and note the very small number of broken or damaged links.}
  \label{fig:LVMI3}
\end{figure}

Figure \ref{fig:LVMI3} shows the architecture of the Fiber IFU subsystem. The IFU's employ a pair of microlens arrays to produce an image of the sky on the tips of the fibers. The light is then conveyed via four fiber bundles to the ``Sorting Hat'', which contains 1944 fusion splices in an enclosure. The fibers emerge from the Sorting Hat in three bundles, which are connected to the individual entrance slit head mechanisms of the three spectrographs.

\begin{figure}[!ht]
 \centering
  \includegraphics[width=1.0\linewidth] 
    {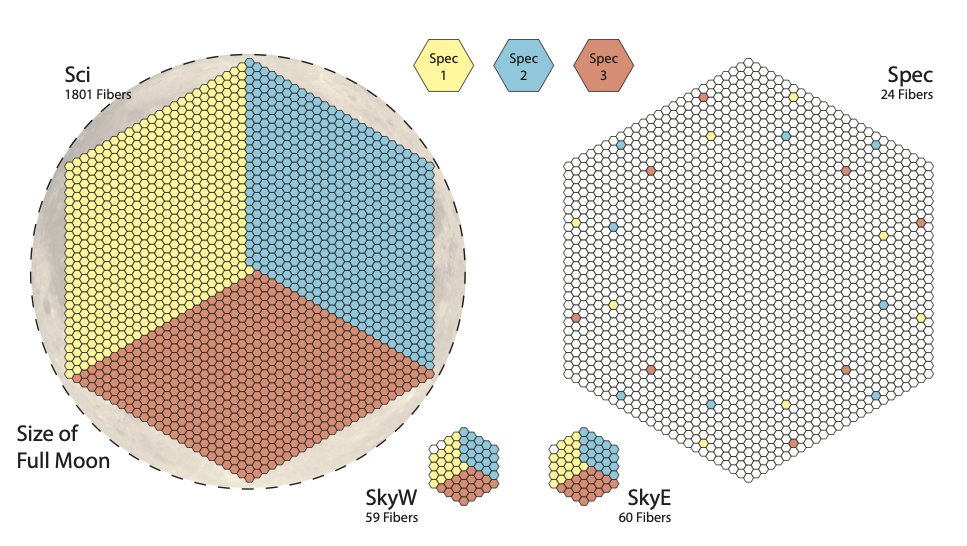}
%  \vspace{-0.4in}
  \caption{{\bf IFU layout showing the allocation to each of the three spectrographs.} The larger Science and Spectrophotometric IFU's have a 30 arcminute footprint on the sky, essentially the diameter of the full moon (shown for scale).  The Sci, SkyW, SkyE, are bundles of 1801, 59 and 60 fibers respectively with 24 fibers for calibration interspersed (see text).}
  \label{fig:LVMI4}
\end{figure}

Calibration is a core aspect of the LVM survey, in particular ensuring that the information from the Sky and Spectrophotometric telescopes can be used to remove terrestrial geo-coronal emission features, as well as telluric absorption. To accomplish this, the Sky and Spectrophotometric fields are close to the Science target on the sky, and the fibers from the three calibration telescopes are interspersed in the focal plane slits with those from the Science telescope. Figure~\ref{fig:LVMI4} shows the exact focal plane arrangement, with the three colors corresponding to the respective destination spectrographs.

\subsection{Spectrograph Subsystem}\label{spectrograph-subsystem}

The spectrograph subsystem consists of three spectrographs, nine CCDs and cryostats, three Archon CCD controllers, a spectrograph controller, and the automated liquid nitrogen (LN\textsubscript{2}) refill system. Bertin Winlight, located in Pertuis (Southern France), manufactured the spectrographs, which are very similar to those used in the Dark Energy Spectroscopic Instrument (DESI). The Imaging Technology Laboratory at the University of Arizona produced the cryostats and integrated the STA4850 CCD's from Semiconductor Technology Associates (STA). Carnegie Observatories in Pasadena led the spectrograph subsystem effort and performed the lab integration and testing. Further information appears in Konidaris et al. (2020).

\begin{figure}[!ht]
 \centering
  \includegraphics[width=1.0\linewidth] 
    {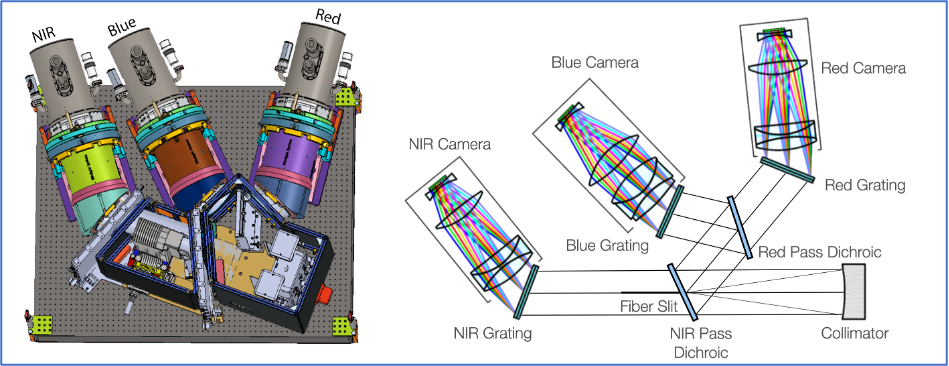}
%  \vspace{-0.4in}
  \caption{{\bf Three-dimensional rendering of the opto-mechanics of one spectrograph (left) and schematic optical path (right).} See text for details.}
  \label{fig:LVMI5}
\end{figure}

Figure \ref{fig:LVMI5} shows the configuration of the spectrographs, along with a simplified schematic of the optical path. A spherical mirror collimates the light from the fiber slit head and sends it toward a Volume Phase Holographic (VPH) grating, which disperses the radiation for capture by the Near Infrared (NIR) channel camera. A visible-reflecting dichroic mirror placed upstream directs radiation blueward of 760~nm to the remaining two channels, where a similar arrangement of a blue-reflecting dichroic and two channel-specific VPH gratings produces spectra in the red and blue channels split at 580~nm.

Each of the nine cryostats contains a five-element lens system to focus the light on the detector. The first of these elements also serves as the cryostat window. Each CCD's is 61 mm square, containing 4k x 4k, 15 µm pixels. These STA devices exhibit excellent photometric properties: high Quantum Efficiency (QE\textasciitilde95\%) and low noise (read noise \textless{} 3 electrons and dark current \textless{} 3 electrons per hour). To achieve this performance, an automatic LN\textsubscript{2} refill system from Cryoworks, coupled with a thermal control loop in the Archon controllers, keeps the devices at optimum operating temperature (\emph{ca.} --180 C)  Each 15-minute survey exposure with the LVM-I spectrograph system  produces nine, 4k frames, totaling roughly 150 megapixels.

\subsection{Enclosure Subsystem}\label{enclosure-subsystem} 

The enclosure subsystem protects the telescopes, the IFU fiber system, and the spectrographs. The overall structure has a 32 x 8 m footprint containing the telescope platform with its roll-off roof, a double-walled thermally controlled clean spectrograph chamber with its ante-room, a utility / server area, and a small control room. The enclosure subsystem also provides remote sensing and telemetry (video /audio feed and environmental sensors) and includes the support infrastructure needed for the survey, including spectral and flat-field calibration lamps and screen, the Heating Ventilation and Air Conditioning (HVAC) system, network infrastructure, and uninterruptible electrical power units. 

The LVM-I enclosure was designed by the Carnegie Observatories in close collaboration with FRISO, an architectural firm in La Serena, Chile. Construction at Las Campanas' ``Robot Ridge" took place between 2021 and 2023, in the midst of the COVID-19 global pandemic. The building was delivered in March 2023.  The facility control systems enables local, remote, and robotic operations.

%\begin{figure}[!ht]
% \centering
%  \includegraphics[width=1.0\linewidth] 
%    {LVMI6.png}
%  \vspace{-0.4in}
%  \caption{The LVM-I enclosure subsystem}
%  \label{fig:LVMI6}
%\end{figure}

\subsection{LVM-I Design, Installation, and
Commissioning}\label{lvm-i-design-installation-and-commissioning}

The Local Volume Mapper Instrument represents a completely new hardware development, from the enclosure to the telescopes to the IFU's and fiber bundles to the CCD sensors of the spectrographs. The bulk of this development work took place during the period 2019-2022, overlapping the COVID-19 pandemic. In fact, many of the collaborators on the geographically-dispersed teams had never met face to face nor visited the observatory site prior to final deployment. Needless to say, this presented unique challenges.

Installation of the various subsystems in the enclosure took place during February and March of 2023. This was followed by 2-3 months of component testing, some on-sky observation, and addressing remaining issues. %Figure 8 contains a selection of photographs of the completed LVM Instrument hardware. 
Handover of LVM-I to the science team took place on 10 July 2023 for science commissioning.  The LVM-I was ready for SDSS-V LVM operations on November 1, 2023.

\section{MOS Survey Planning and Execution}
\label{sec:SurveyOperations_MOS}

Multi-object spectroscopic data collection for SDSS-V began at APO in October 2020 using plug plates to feed the APOGEE and BOSS spectrographs.
The 2020 plug plate program (described in the following section) was used to mitigate telescope downtime and FPS instrument delays related to the COVID pandemic.  In November 2021, the first FPS unit was delivered to Apache Point Observatory (APO), and the second FPS unit was delivered to Las Campanas Observatory (LCO) in August 2022.

\subsection{Early Plug Plate Observations From APO}

In the initial SDSS-V project plan, the FPS unit at LCO was scheduled to be commissioned first, and the FPS unit at APO second.  To avoid telescope downtime at APO during the LCO FPS commissioning phase, SDSS-V had planned an early ``Plug Plate'' period at APO.  SDSS-V did not, however, plan for a global pandemic.  Disruptions to supply chains, nullified contract timelines by ``force majeur,'' created extensive restrictions on staffing and travel. In the end, these safety considerations across our global collaboration delayed (by 1 year) and complicated the completion of our FPS units and the associated telescope modifications.

The MOS mapper teams responded by extending our original plans to jump-start SDSS-V data collection using plug plates at APO and by swapping our deployment plan to lead with FPS commissioning at APO (ahead of LCO).  APO could operate on lean staffing that made it possible to operate it safely, responsibly, and in compliance with all local ordinances during that period.  An interim plug plate program at LCO was not feasible, due to travel restrictions to LCO, uncertainty surrounding international shipping of plug plates and staffing regulations.

SDSS-V started observations at APO using the existing plug plate system on October 23, 2020, after the northern SDSS-IV observing program had been extended by a few months to mitigate observing time losses that SDSS-IV experienced due to COVID-19 shutdowns and staffing restrictions.

As discussed perviously, the overheads between different fields (20 minutes) and limited number of at-the-ready plug plates per night, meant a lower sustainable field cadence during the plug plate operations, with exposure times of typically 60 minutes and a 20 minute overhead for cartridge switching.  Even if plug plate observation times were shortened, the total number of fields available for observations each night was limited by the plug plate cartridges available on site (typically 9 BOSS cartridges and 9 APOGEE cartridges although 10-12 were possible long winter nights). If plate and cartridge limitations were not present, half the night would be spent performing plug plate cartridge changes as a result of shortened observation times.

Consequently, SDSS-V focused during that period on target categories that were more compatible with plate-based observations than others.  These were faint objects that required $>$ 15 minutes of observation to reach sufficient S/N, or targets slated for long baseline time monitoring (e.g., reverberation mapping and radial velocity monitoring).  A special focus of the SDSS-V plug plate program included the eFEDS mini-survey, which covered a $\sim$140 deg$^2$ field for optical spectroscopy follow-up of X-ray sources from eROSITA.

 A total of 265 plates were observed as part of the SDSS-V plate program. Depending on the science goals, each SDSS-V plate was scheduled for $\sim$30--120 minutes of open shutter exposure time (in contrast to the typical SDSS-IV exposure time of $\sim$60 minutes of open shutter time per plate). Modifications were made to the cartridges that held the plates to enable joint observations with the APOGEE and the one BOSS spectrograph remaining at APO, while the second BOSS spectrograph was being readied for shipment to LCO.  These cartridge conversions were carried out primarily between October 2020 and January 2021 with earlier testing in spare cartridges in 2019\footnote{We could not convert the majority of the cartridges earlier than the end of SDSS-IV operations which were extended owing to COVID-19 interruptions.}. During this period, some plates with only APOGEE targets were observed using unconverted cartridges. This mitigated the difficulty of filling long winter nights with a limited number of APOGEE+BOSS cartridges and reduced staffing for daytime plate changes and plugging, due to COVID-19 protocols.

During the plate program, each plate was typically observed two to four times, with each distinct observation identified as a single plate \emph{epoch}. Plates containing BOSS targets were observed for a total of 569 plate epochs,  acquiring 199,295 BOSS spectra of 89,201 science targets.  Plates containing APOGEE targets were observed for 945 plate-epochs, acquiring 246,631 APOGEE spectra of 56,021 distinct targets.

Plate observations concluded on June 27th, 2021, at which point engineering work to prepare for the installation of the SDSS-V focal plane system began, to be completed during APO's summer monsoon shut-down. SDSS-V was unable to conduct a similar plate program from LCO, given mountain restrictions at Las Campanas Observatory during this period owing to COVID-19. With the delivery of the FPS units, the SDSS Plug Plate system, and its two decade legacy was retired.%\footnote{Don't worry.  We kept all that stuff if you are feeling nostalgic.  SDSS is for hoarders.}.

\subsection{``Mow the Sky": From Concept to Reality\footnote{This is meant in the agrarian sense, however turned skyward;  Cover the full-sky field with spectroscopy leaving no untended patch see e.g. Tolstoy's Konstantin Levin for more detail on the benefits of mowing.}}

SDSS-V's adoption of a robotic focal plane for its two MOS mapper programs has impacted nearly every pre-existing piece of SDSS survey hardware, infrastructure, and operation, including the survey planning and execution, which is largely implemented in software.  Most of SDSS-V operations software is written in Python (3.7+) or Python-wrapped C/C++.  SDSS-V has made an effort to write well tested and documented code, adhering to a fleshed-out coding standard\footnote{\url{https://sdss-python-template.readthedocs.io/en/latest/standards.html}}.  The majority of SDSS-V software products are open source and available on SDSS's GitHub space\footnote{\url{https://github.com/sdss/}}.  Several SDSS-V software products depend on specific PostgreSQL database connections and schemas used to store spectroscopic targeting data.

For computationally intensive work (e.g., target selection, data reduction), SDSS-V uses a dedicated cluster of Linux servers at the University of Utah's Center for High Performance Computing (CHPC).  For nightly operations and data collection, each observatory (APO and LCO) houses a set of SDSS-dedicated Linux servers that run software for telescope/instrument control and data quality assurance.  Data are synced between the two observatories and the CHPC daily.

The subsections that follow provide a broad overview of the organization, logic, and flow of SDSS-V FPS operations from target selection to data collection, providing descriptions of various important software products.  Additional references relevant to FPS operations topics can be found in \citet{SanchezGallego2020} and \citet{SanchezGallego2022}.

Broadly speaking, the MOS survey execution entails the steps of target selection, optimizing the robotic fiber assignment in each field, and casting these fields into a global observing plan, both for the global optimization of the survey strategy and its actual execution.

\subsection{MOS Target Selection}

\subsubsection{Underlying Cross-Matched Target Catalogs}
\label{sec:crossmatch} 

The first stage of survey operations involves identifying the desired targets for MOS observations.  These lists are built from a number of published catalogs from astronomical surveys including Gaia \citep{GaiaDR3}, TESS \citep{tessInputCatalog}, the DESI Legacy Surveys \citep{Dey_2019_DESIsurveys}, 2MASS \citep{2MASS}, SDSS \citep{sdss_dr13}, and about 50 others.  See \citet{sdss_dr18} for target selection details relevant to SDSS DR18.  Each catalog used in target selection is stored in database tables on the CHPC servers in a schema called {\tt catalogdb}.  

Common targets between catalogs are linked by a cross-match strategy that uses a sequential approach to link targets between catalogs. We begin with an initial, base catalog (TESS Input Catalog v8 for previous versions of the cross-match, Gaia DR3 for current ones) and assign a unique identifier, {\tt catalogid}, to each target within. A relational table {\tt catalog\_to\_X} (where X is the name of the parent catalog) is created in {\tt catalogdb} to link catalogids with their associated targets in the parent catalog.  For each successive catalog we apply three cross-matching phases.  In phase one, we consider known relationships between the new targets and other parent catalogs already processed (for example, the Legacy Survey DR10 catalog includes a column linking targets to Gaia DR2 and DR3). A new {\tt catalog\_to\_X} is created and populated with the associations between matched {\tt catalogid} and the catalog targets.  In phase two we perform a spatial match between existing {\tt catalogid} and targets not matched in phase one, using a resolution-matched search radius (usually 1 arcsec, which matches our on-sky fiber radius at APO). All associations are registered in the {\tt catalog\_to\_X} table and the closest match is marked as ``best".  In phase three, any target not associated with a {\tt catalogid} in phases one or two is considered a new unique target and assigned a new {\tt catalogid}. \textcolor{red}{}{Thus, at the end of the process there is a {\tt catalogid} associated with every entry in each parent catalog.}

This approach ensures that each {\tt catalogid} is associated with only one best target in each cross-matched catalog (the opposite, each target being associated with only one {\tt catalogid}, is not enforced). The resulting cross-match is sensitive to the order in which parent catalogs are processed, and in general we arrange catalogs in order of better to worse spatial resolution.  The cross-match process is computationally intensive, and it is highly dependent on the fine tuning of the Postgresql server and the cross-match SQL queries. A full cross-match run requires two to three weeks to run.

While a {\tt catalogid} identifies a unique target in a given crossmatch, each crossmatch rerun will assign a new {\tt catalogid} to a given object. To simplify the identification of targets and track observations across different cross-matches, each object that {\it could} potentially be observed in SDSS-V is also assigned a unique {\tt sdss\_id} integer identifier. Each {\tt sdss\_id} is associated with one or more {\tt catalogid} and is used in publicly released SDSS data as the main target identifier. Such {\tt sdss\_id}s have also been assigned to observed targets in legacy SDSS I-IV data.

\subsubsection{Target ``Cartons" for the Mapper's Different Science Goals}

Each scientific subprogram of SDSS-V's MWM or BHM is required to provide a target selection function in the form of a well-defined SQL query operating on {\tt catalogdb}.  The results from each subprogram's target selection constitute a single target \emph{carton} in SDSS-V parlance.  A carton contains the set of targets available for observations to achieve one of SDSS-V's science goals. Note that any one target may belong to more than one carton.  SDSS-V has defined over 200 cartons for the FPS survey.  Additionally, each target selection carton requires definitions for several observational criteria: maximum acceptable sky brightness (lunation), spectrograph (APOGEE or BOSS), observation epoch/cadence (given a fixed observation time of 15 minutes), airmass, value, and priority.

In addition to \emph{core science cartons} that drive the survey program at highest priority, SDSS-V has periodically solicited calls across the collaboration for \emph{open fiber cartons}.  Targets from open fiber cartons are used to fill any unused science fibers after the initial allocation of core science targets has been made.  Any member of the SDSS-V collaboration may propose an open fiber carton.  Unlike core science cartons, open fiber cartons can be defined not only algorithmically, but also through a (scientifically justified) ad-hoc list of targets, as long as they are already part of the current cross-match and have a valid {\tt catalogid} \textcolor{red}{}{(i.e. are in one of the parent catalogs described in Section \ref{sec:crossmatch}, such as Gaia, TESS, DESI Legacy Surveys, etc.)}. 

Target selection is periodically updated, usually instigated by either the availability of new or updated imaging catalogs or a sanctioned decision to change survey strategy.  All target selection cartons are stored in a database schema called {\tt targetdb}.  SDSS-V's {\tt target\_selection}\footnote{\url{https://github.com/sdss/target_selection}} git repository contains a version controlled set of scripts that perform catalog crossmatching and target selection for the SDSS-V FPS.  As target selection evolves throughout the survey, its versioning is carefully tracked for book-keeping purposes.  {\it This is especially important for scientific analyses in which selection effects must considered.}

The {\tt target\_selection} scripts use a standalone Python package {\tt sdssdb}\footnote{\url{https://github.com/sdss/sdssdb}} that provides a programmatic interface to catalog, targeting, and other databases within the SDSS Linux environments.  This tool is an important component in the SDSS-V operations code, and provides convenient access for individual users to explore, verify, or analyze various stages of survey planning and progress.

\subsection{FPS Survey Strategy: Time Allocation and Fiber Assignment}

The number of MOS targets and the variation of the observing requirements among carton types  present a challenge for planning an efficient fixed-duration survey. The SDSS-V MOS cartons vary greatly in the number of objects in each carton and the 
observing requirements.  For example, MWM's White Dwarf carton contains $\sim$200,000 
objects that require two exposures per object under dark-sky lunation conditions 
using the BOSS spectrograph with at least 1 day separating each exposure.  MWM's 
Galactic Genesis carton contains $>$4 million objects each requiring only a single 
APOGEE exposure under any sky brightness condition.  Some cartons have very demanding 
epoch requirements, such as cartons selected to study radial velocity variations in  
binary stars, or to monitor quasar time-variation through frequent repeat observations 
over the full duration of the survey.

We developed the {\tt robostrategy}\footnote{\url{https://github.com/sdss/robostrategy}} package
to plan the survey. We start with a set of around 12,000 fixed positions uniformly 
distributed across the sky (except with a higher density in the region assigned to LCO
to account for its smaller field-of-view).
This software decides what set of fields to observe, how many observations
to conduct in each and with what timing, and which objects to target in each observation.
The first two tasks comprise the time allocation problem, and the third comprises the 
fiber assignment problem. For each field we will choose a cadence of observations, 
and the plan for each observation is expressed as a ``design,'' for each of which there 
is a fiber assignment of each APOGEE or BOSS fiber on a robot to a target. 
Each design is intended to receive only a single observation, so any target that 
requires repeat observations will appear in two or more designs.

The time allocation proceeds through an approximate method of maximizing a metric 
of value over the possible cadence choices for each field.
The process needs to respect several boundary conditions: (1) the duration
of the survey, (2) the chosen exposure times, (3) the estimated overheads, and (4)
a realistic model for the open dome fraction as a function of time of year. These
boundary conditions lead to a certain number of observations that can be performed
in bins of Local Sidereal Time (LST).
It must also account for efficiency as a function of airmass in the optical, which
depends on the LST at which a given field is observed.
Conceptually, {\tt robostrategy} performs fiber assignment {\bf for} all possible cadence choices 
for each field, and estimates a value for each choice. Then it assigns a set 
of cadence choices that maximizes the total value over all fields, while 
respecting the overall boundary conditions, using a linear programming package
(Google's OR-Tools\footnote{\url{https://developers.google.com/optimization}} Glop solver; \citealt{ortools}). 
The software uses a number of simplifications and approximations in estimating
the value to make this process tractable. It results in a cadence choice for
each field and a recommendation for the hour angle distribution of the observations
of that field.
{\tt robostrategy} does not provide a rigid survey schedule (i.e. when a specific 
design should be observed during a given night), which is instead chosen 
in real time based on the hour angle recommendations.

Given a cadence choice for each field, {\tt robostrategy} then performs a 
final fiber assignment for each design. For each design, all cartons are 
considered, though some targets are assigned higher priorities than others.
In addition, there are requirements for standard star and sky background
observations that {\tt robostrategy} accounts for. Furthermore, in the 
assignment:
\begin{itemize}
\item A target must land on the focal plane within the patrol zone of its intended fiber.
\item A target-to-fiber assignment must not collide or interfere with a neighboring robot's target-to-fiber assignment.
\item 500 robots may be assigned targets per design.  Any robot may be assigned a BOSS target, while a subset of 298 robots may instead be assigned an APOGEE target.
\item Targets with higher priority (usually core science) take precedence over targets from lower priority (usually shell science) cartons.
\item All targets sharing a design must be valid for the same observing conditions (airmass, sky brightness, target cadence, time allocation constraints).
\end{itemize}

The final stage of target selection involves loading the resulting {\tt robostrategy} designs into {\tt targetdb} which makes them available for observation at the telescopes.  A detailed description of {\tt robostrategy} and the survey planning logic is given in Blanton et al. \emph{(in preparation)}.

\subsection{Robotic Reconfiguration and Field Acquisition}
Each {\tt robostrategy} design requires a specific mechanical configuration for each robot in the positioner array.  When the telescope is pointed and guided at a design's intended (RA,Dec) field center and each robot's arm angles are precisely set to the design's specifications, photons from targets will land on the expected APOGEE or BOSS fibers that feed the spectrographs.  {\tt robostrategy} ensures that each design is technically feasible in that no two robot arms may occupy the same physical space in the focal plane after being placed in their desired positions for a science exposure.  

Finding efficient and collision-free paths for each robot to follow while navigating between configurations is an important and complex problem.  A solution to this problem was critical for the successful use of SDSS-V's heavily overlapping array of fiber positioners. The software package {\tt kaiju}\footnote{\url{https://github.com/sdss/kaiju}} was developed specifically to compute collision-free trajectories for robots during reconfiguration.  Navigation algorithms for robot ``swarms" are a subject of general interest in the robotics community (e.g. \citealt{multi_agent_control}) and are a necessary component in today's astronomical survey instruments that use overlapping arrays of robotic fiber positioners (e.g. \citealt{laleh}, \citealt{moons_sw}).  {\tt kaiju} employs a unique ``reverse-solve" heuristic to determine safe trajectories for all robots during a reconfiguration.  This heuristic algorithm results in a fast runtime suitable for on-the-fly path generation.  
The total robot motion time during reconfiguration averages to 70 seconds, and the {\tt kaiju} code achieves near-perfect target acquisition efficiency.  In survey operations thus far, we have measured {\tt kaiju} to be $>99.99$\% efficient.  This means less than 1 target per 10,000 is lost due to path generation constraints.  Details of the {\tt kaiju} implementation and its performance are described in \citet{Sayres_2021_kaiju}.

The accuracy of fiber placement after a robotic reconfiguration has a direct impact on the overall throughput of the FPS.  The ``blind move" fiber-positioning accuracy for SDSS-V robots is approximately 50 microns (RMS) in the focal plane.  To improve upon these blind-move positioning errors, feedback from the Fiber View Camera (FVC) is used after a {\tt kaiju}-directed reconfiguration.  The FVC images the robotic focal plane and measures centroids of back-lit metrology fibers that are carried by each robot, relative to the fixed grid of back-lit fiducials. From this information one can determine and command small corrections in each robot's orientation.  After a single FVC measurement and commanded correction, fiber position errors are currently estimated at $\sim$20 microns (RMS).  The fiber positioning accuracy has continued to improve with time and we expect to reach a maximum fiber positioning error of $\leq$18 microns, which corresponds to 0.3$^{\prime\prime}$ at APO and 0.2$^{\prime\prime}$ at LCO.

In addition to accurate and precise fiber positioning, telescope guiding performance is critical to maximize the overall throughput of the FPS instrument.  We developed software named {\tt cherno}\footnote{\url{https://github.com/sdss/cherno}} that is responsible for field acquisition and guiding during science exposures.  The six peripheral guide cameras (Figure \ref{fig:FPS_FocalPlane}) are exposed simultaneously on a continuous loop with a typical exposure time of fifteen seconds and limiting magnitude of $\sim$18.5 in the SDSS $r$ band.  One camera is located slightly above best focus, and another camera is located slightly below best focus, allowing for a continuous monitoring of focus drift and direction.  Each guide loop yields an error measurement for RA, Dec, position angle, focus, and telescope plate scale.  Corrections in RA, Dec, position angle, and focus are controlled by a proportional-integral-derivative (PID) loop and continuously sent to the telescope during active guiding.  Telescope plate scale drifts very slowly with temperature variations, and it is not continuously corrected for during guiding.  Instead, this plate scale is provided as an input parameter at reconfiguration time between spectroscopic fields so that the robotically positioned fibers match the last-measured plate scale derived from recent guider exposures throughout the night.

During the guide loop, {\tt cherno} uses an offline installation of {\tt astrometry.net} \citep{Lang2010} to determine the WCS astrometric solutions for each exposure from the  stars that happen to have landed on the guide chips.  This strategy provides robust telescope pointing solutions in very crowded fields where guiding is typically difficult, and it also relaxes dependencies on the accuracy of the telescope pointing model for guide star identification.  The WCS solutions from {\tt astrometry.net} are then used to match guide chip centroids to Gaia DR2 sources from {\tt targetdb}.  Using every Gaia-matched guide centroid across all six cameras (10's to 1000's of matches depending on stellar density at telescope pointing), we update the telescope pointing solution.  These Gaia-matched pointing refinements are small ($\sim$0.1 arcsec) when compared to the initial {\tt astrometry.net} solutions, but even small pointing offsets will have a measurable effect on total system throughput.

A prerequisite for both guiding and fiber positioning is knowledge of where a guide star or spectroscopic target is located in the focal plane for a given (RA,Dec) field center, observatory location, and time of observation.  A package named {\tt coordio}\footnote{\url{https://github.com/sdss/coordio}} was developed to perform these types of computations, and it is a core piece of survey planning and execution infrastructure.  {\tt coordio} provides wavelength-dependent conversions between sky coordinates (RA,Dec), guide coordinates (pixels), robot coordinates (arm angles), and FVC coordinates (pixels) using a well-defined stack of coordinate systems.  The description of each coordinate system and the mathematics of each coordinate transformation are described in \citet{Sayres_coordio}.

The accuracy of {\tt coordio} is largely dependent on the quality of the FPS and FVC calibration and metrology.  Instrument calibration includes: fitting kinematic models for each robot, measuring the absolute location and orientation of each robot or guide camera relative to the fiducial field, knowing the relative offsets between metrology and science fibers carried by the robot arm, and characterizing the optical distortion in FVC images.  Calibration efforts began with extensive lab testing and measurement.  Calibration was further tuned on-sky using bright stars to measure flux responses under small star-to-fiber offset conditions using a method adapted from the DESI instrument team \citep{Schlafly_2024}.  The process of FPS calibration is described in detail by \citet{Sayres_thesis}, although calibration strategies and techniques have evolved and improved since that work.  We expect to continue improving fiber position accuracy, throughput, and efficiency over the course of the survey.

\subsection{FPS Nightly Operations}

Nightly operations with the SDSS-V FPS are planned using an observing queue and carried out by a team of observing specialists at each telescope site.  The observing queue contains an ordered list of {\tt robostrategy} designs to visit throughout the night.  A {\tt roboscheduler}\footnote{\url{https://github.com/sdss/roboscheduler}} process populates the nightly observing queue, and the queue can be regenerated at anytime during the night if observing conditions change.  {\tt Roboscheduler} determines which designs are observable for a given night (based on airmass, sky brightness, and cadence requirements), and prioritizes them based on factors such as the number of designs in a field or how much time is left in a field's cadence. These parameters are tuned using full survey simulations to maximize long-term survey outcomes. A full description of prioritization and {\tt roboscheduler} is given in Donor et al. \emph{(in prep)}.

The nightly objective is to move through the queue as quickly as possible, which means minimizing the overheads between spectroscopic fields.  A high-level software process named {\tt hal}\footnote{\url{https://github.com/sdss/hal}} maximizes data collection speed by automating the sequence of field setup steps, including:

\begin{itemize}
\item Load target-to-fiber assignments for the next queued design from {\tt targetdb}.
\item Compute and execute a robotic reconfiguration, accounting for current telescope plate scale and differential atmospheric refraction during slew to new field center.
\item Use FVC image feedback to correct small fiber position errors.
%\item Slew the telescope to the new field center.
\item Obtain any required flat field or arc calibration frames.
\item Acquire the field and begin the guide loop.
\item Begin a 12-15 minute science integration, exposing APOGEE and BOSS spectrographs simultaneously.
\end{itemize}

{\tt hal} is able to execute several of these steps in parallel, resulting in a typical field setup time of five minutes.

After each science exposure is completed, quick reduction pipelines for both APOGEE and BOSS instruments are run to estimate immediately the signal-to-noise accumulated for each observed design.  Each design is expected to require only a single science exposure, but if signal-to-noise criteria are not met, a design may receive one or more additional science exposures.  These quick reduction pipeline results are displayed for the observer alongside recent exposures in a webapp called {\tt kronos}\footnote{\url{https://github.com/sdss/kronos}}.  The observer uses {\tt kronos} throughout the night to monitor progress and make adjustments to the observing plan when necessary.  After each night, the raw data are transferred from each observatory to CHPC servers in Utah for full spectroscopic reduction pipeline processing and downstream analysis.

\section{LVM Survey Planning and Execution}
\label{sec:SurveyOperations_LVM}

\label{operating-principle-a-single-survey-observation}
In Section~\ref{sec:lvm} we have described the overall ISM science goals of the LVM program and the top-level science requirements on the data. Here, we sketch the specifics of the survey implementation, with a focus on the survey areas and depth.

\subsection{LVM Targets}
In its initial science implementation,  LVM will map the Milky Way ISM including two nearby HII regions (Orion and Gum), the Magellanic Cloud system, a number of nearby 
galaxies (D $<$ 20~Mpc) with large apparent sizes, and an all-sky grid of regions at high Galactic latitude. In this section, we provide an overview of the survey targeting and operations, with further details provided in \cite{Drory2024}.

Figure~\ref{fig:lvm-overview} shows the footprint of the LVM survey and Table~\ref{tab:lvm-targets} gives an overview of the survey areas (or targets), the number of spectra in each LVM sub-program, the spatial resolutions of a spaxel at the target distance, and the flux sensitivities.  All target areas are broken down into survey tiles, representing individual LVM pointings that combine to produce contiguous mosaics covering each target.
We lay out the rationale for the specific set of survey target regions in the subsequent paragraphs.

\begin{table*}
    \centering
    \begin{tabular}{cccccc}
    \hline
       Target & Area & \# Spectra & Spaxel & 5$\sigma$~depth\\
              & sq.\,deg. &  & pc & erg\,s$^{-1}$\,cm$^{-2}$\,arcsec$^{-2}$ & \\
        \hline\hline
        MW mid-plane    & 2700    & 35M & 0.1-1 & $6\times10^{-18}$ (H$\alpha$)&\\
        MW plane extension   & 1744    & 22M & 0.1-1 &''&\\
        Orion       & 132.7   & 1.7M & 0.07 & "&\\
        Gum & 190 & 2.2M & 0.1 & " &\\
        \hline
        MW high latitude & $\dots$ & 1M & ... & $6\times10^{-18}$ (H$\alpha$) & \\
        \hline
        LMC         & 78.5    & 1M & 10 & $2\times10^{-18}$ (H$\alpha$; 23.0 $V_{\mathrm{AB}}$) & \\
        SMC         & 16.4    & 200k & 10 & " & \\
        \hline
    \end{tabular}
    \caption{Overview of LVM targets, covered area, number of spectra and the spatial resolution, and finally, flux limits.}
    \label{tab:lvm-targets}
\end{table*}

\textbf{The Milky Way Survey} of the LVM program will chart the bulk of the MW disk covering the vast majority of known optically selected H II regions, along with their interfaces to the diffuse ionized ISM. LVM's target line depth will allow the detection of enough emission lines to map the metallicity and ionization structure of the ISM over the  surveyed area and characterize the feedback-induced kinematics of gas in and around H II regions. With LVM's sensitivity, we can also study electron temperature fluctuations via the intrinsically weaker \emph{auroral} lines, and their impact on chemical abundance measurements for a significant fraction of optically visible H II regions in the sky. LVM's sub-parsec resolution -- while covering the full set of Galactic scales -- allows for feedback effects to be traced starting from the individual sources through ionization fronts and shocks to global effects such as gas flows and galactic fountains. In this way, LVM's Milky Way Survey can avoid the trade-offs between coverage and resolution that have beset all earlier 3D spectroscopy efforts.  The Orion nebula and the Gum nebula are two of the closest and best studied and HII regions in the Milky Way. Due to their proximity they extend far above the disk mid-plane. We therefore target them individually. 

\textbf{The LVM Magellanic System Survey} will cover the nearby Small and Large Magellanic Clouds (SMC/LMC) with an emission line sensitivity that allows for the detection of strong emission lines over the full area, and of the much weaker auroral lines in the majority of H II regions. This part of the LVM program samples the ISM at lower metallicity than is available in the MW disk, while still resolving the inner structures of H II regions. At the same time, LVM's measurement of the stellar continuum for the luminous hot stars will enable an unprecedented integral picture of more this metal-poor stellar populations. LVM's forseen survey area will encompass after 4 years the entire area (above a surface brightness of 25th magnitude) of the SMC and the LMC. %And it will encompass 80\% of the analogous area in the LMC. 
At the LMV and SMC distance, LVM data will have a 10 pc physical resolution, enough to distinguish shocked vs. photoionized emission, resolved large-scale ionization gradients in star-forming regions. And it will fully sample all sites of optically visible star formation and large-scale diffuse gas globally.

\textbf{The Nearby Galaxies Survey}  builds towards the core idea of a \emph{local volume} mapper,  as this survey component will cover $\sim$500 galaxies within $\sim$20~Mpc of the Milky Way with a physical resolution of $\sim 1$kpc (see Drory et al. 2024, Kreckel et al. 2025 in prep). A Local Group sub-sample, consisting of 29 particularly nearby galaxies (Table \ref{tab:lvm-target-localgroup}), has been selected from highly complete catalogs \citep{Kennicutt2008, Karachentsev2013} to have declination below +15\degr\, distance less than 4~Mpc, and a total integrated H$\alpha$ + [NII] flux above 10$^{-13}$ erg/s/cm$^2$. This includes both dwarfs and more massive galaxies, with stellar masses ranging from 10$^{6.8}$-10$^{11}$ M$_\odot$. A Local 20~Mpc sub-sample draws from catalogs that are not necessarily complete \citep{Karachentsev2013, Leroy2019}, and we select galaxies below 15\degr\ declination, M$_* \gtrapprox 10^{9}$ M$_\odot$), and SFR $>$ 0.1 M$_\odot$/year, although these limits are not strict.  Both sub-samples are observed out to R25, and in most cases this is achieved with a single dithered LVM pointing. 

Taken together, these nearby galaxies will provide a link to the MANGA survey, and it constitutes the most immediately related use of the facilities when neither the Milky Way nor any of the Magellanic Clouds are sufficiently high in the sky.

\begin{table}[]
    \centering
    \begin{tabular}{l|c c c}
\hline
Name & RA  & Dec  & Dist    \\
  &  [hms] &  [dms] &  [Mpc]  \\
\hline
\hline
WLM  &  0:01:58.1 & -15:27:40  &  1.0  \\
NGC~55  &  0:14:53.6 & -39:11:48  &  2.1  \\
NGC~247  &  0:47:08.3 & -20:45:36  &  3.6  \\
NGC~253  &  0:47:34.3 & -25:17:32  &  3.9  \\
NGC~300  &  0:54:53.5 & -37:40:57  &  2.1  \\
IC~1613  &  1:04:47.8 & +2:08:00  &  0.7  \\
M33  &  1:33:50.9 & +30:39:37  &  0.7  \\
NGC~625  &  1:35:05.0 & -41:26:11  &  3.9  \\
NGC~1313  &  3:18:16.1 & -66:29:54  &  4.1  \\
NGC~2915  &  9:26:11.5 & -76:37:35  &  3.8  \\
Sextans~B  &  10:00:00.1 & +5:19:56  &  1.4  \\
NGC~3109  &  10:03:07.2 & -26:09:36  &  1.3  \\
UGC~05456  &  10:07:19.7 & +10:21:44  &  5.6 \\
Sextans~A  &  10:11:00.8 & -4:41:34  &  1.3  \\
IC~3104  &  12:18:46.1 & -79:43:34  &  2.3  \\
GR~8  &  12:58:40.4 & +14:13:03  &  2.1  \\
NGC~4945  &  13:05:26.1 & -49:28:16  &  3.8  \\
NGC~5102  &  13:21:57.8 & -36:37:47  &  3.4  \\
NGC~5128~(CenA)  &  13:25:28.9 & -43:01:00  &  3.8  \\
ESO~324-024  &  13:27:37.4 & -41:28:50  &  3.7  \\
NGC~5253  &  13:39:55.8 & -31:38:24  &  3.6  \\
ESO~325-011  &  13:45:00.8 & -41:51:32  &  3.4  \\
ESO~383-087  &  13:49:18.8 & -36:03:41  &  3.5  \\
Circinus  &  14:13:09.3 & -65:20:21  &  4.2  \\
ESO~274-001  &  15:14:13.5 & -46:48:45  &  3.1  \\
IC~4662  &  17:47:06.3 & -64:38:25  &  2.4  \\
NGC~6822  &  19:44:57.7 & -14:48:11  &  0.5  \\
IC~5152  &  22:02:41.9 & -51:17:43  &  2.0  \\
NGC~7793  &  23:57:49.4 & -32:35:24  &  3.9  \\
\hline
    \end{tabular}
    \caption{Local Group sub-sample from the Nearby Galaxy Survey (parameters taken from \citealt{Karachentsev2013})}
    \label{tab:lvm-target-localgroup}
\end{table}

To make optimal use of telescope time,  we supplement these galaxy targets with a sparse grid of all-sky pointing through the out-of-disk ISM to be observed when no other target is observable given the scheduling constraints. The spacing of the grid points is 4\degr.

%The Milky Way is observed at any lunation, with observing constraints of distance to the Moon larger than 60$\degr$, airmass below 1.75, and shadow height of $>1000$~km to lower the impact of geocoronal H$\alpha$\ on the MW signal. Due to the relative brightness of H$\alpha$\ in the Orion and Gum nebulae, the shadow height constraint is relaxed to $>500$~km, with all other parameters the same as for the rest of the Galaxy. 

\subsection{LVM Survey Execution}

The LVM-I was designed to operate in a fully robotic operations mode after an initial period of remote operations. As of late 2024, the telescope control software and automated scheduler (see Section \ref{software-subsystem}) serve to reduce the input needed from remote observers to weather monitoring and debugging. Initial commissioning and survey operations had been ongoing since November 2023, and in this phase they have been remotely carried out through the volunteer efforts of a cohort of over 30 astronomers. Taking advantage of the the international nature of the SDSS-V collaboration, each night is split into two halves, with North and South American team members observing in the first half, and European team members observing in the second half. On-site support in case of emergency is available from the neighboring SDSS-V DuPont observers. 

The LVM survey can be thought of as a sequence of single survey tile observations. At the beginning of such an observation, the four telescopes slew to and acquire their targets, and the K-mirror derotators move to their initial angles. Feedback from Acquisition and Guide cameras adjacent to the IFUs allows fine corrections for pointing, rotation, and focus, if necessary. A 15-minute survey exposure begins when all telescopes are ready. The ``Sci" and ``Sky" telescopes track and de-rotate on their targets for the full exposure, while the ``Spec" telescope points at 12 different bright calibration stars for approximately 1 minute each during the 15 minutes. The rotating mask in the Spec telescope prevents contamination during calibration star acquisition. At the end of the exposure, the telescopes move to and acquire their next target during readout of the CCDs in the spectrographs. The siderostat mounts can achieve 50$^\circ$ per second slew speeds, so there is very little dead time between exposures, leading to a total system overhead of $\sim 10\%$.

All observations have a nominal exposure time of 15~minutes, to reach a line sensitivity of $5\sigma=6\times10^{-18}$~~erg~s$^{-1}$~cm$^{-2}$~arcsec$^{-2}$ at 6563\AA. Survey tiles within the Milky Way and all-sky grid are only observed with a single 15~minute exposure. Survey tiles for the Magellanic Systems and Nearby Galaxies Survey are observed in a nine-point dither pattern to achieve improved emission line sensitivity and fill the gaps between fibers to recover the full flux.

\subsection{LVM-i Software}\label{software-subsystem}

The software subsystem provides control over the other four subsystems -- telescope, fiber/IFU, spectrograph, and enclosure --~and will eventually support fully robotic operation.  LVM-I software follows the general SDSS framework, simplifying maintenance during the survey. Sanchez-Gallego et al. (2022) describe the software in greater detail.

\begin{figure}[!ht]
 \centering
  \includegraphics[width=1.0\linewidth] 
    {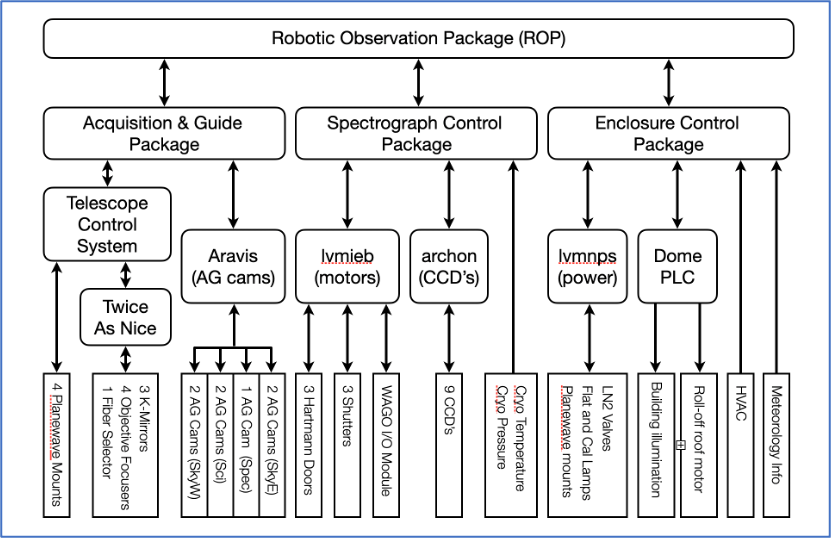}
 % \vspace{-0.4in}
  \caption{LVM-I software subsystem architecture. See Section~\ref{software-subsystem} for details.}
  \label{fig:LVMI7}
\end{figure}

Figure \ref{fig:LVMI7} shows the hierarchical structure of the LVM-I software subsystem. Lower-level components are largely hardware drivers, while the mid-level items integrate and control sets of low-level components. The highest-level packages for Acquisition / Guiding, Spectrograph Control, and Enclosure Control gather the elements of the middle-ware into functional blocks corresponding to specific survey operations. Finally, the Robotic Observation Package can automate all aspects of LVM-I.

\section{Data Reduction and Analysis Overview}
\label{sec:pipelines}
\label{sec:software_pipelines}
A hallmark of SDSS data products has been not only the use of state-of-the-art data reduction and analysis pipelines, but also to provide raw, reduced, and analyzed data to the community. SDSS has always released the raw data (``publish all the bits") so that anyone from the global community can improve on the collaborations' work, or make entirely new use of the data, not originally envisioned.  Enabling easy and far-reaching data use beyond the SDSS consortium has been core to SDSS's philosophy and success for over two decades. 
In this Section we describe the software pipelines that make this possible for SDSS-V.
This ``publish all the bits" is both a best-practice as well as the backbone philosophy of the survey.  We do this to enable members of the community to improve -- from the raw frames -- upon our work in the future.

\subsection{Data Reduction Pipelines}
In SDSS data \emph{reduction} pipelines take the raw instrumental and calibration data and produce analysis-ready spectra.  This is the first step after obtaining our raw observational data.

\subsubsection{BOSS Spectral Reduction}
\label{sec:boss-red}
Optical SDSS-V data taken with the BOSS spectrographs are processed with an updated version of \texttt{IDLSPEC2D} \citep[][Morrison et al.\,in prep]{Bolton2012,Dawson_2013_boss}. The \texttt{v6\_X\_X} versions of \texttt{IDLSPEC2D} build on the final (\texttt{v5\_13\_2}) SDSS-IV/eBOSS version of the pipeline.\footnote{All SDSS-V versions of \texttt{idlspec2d} are available for download from the SDSS GitHub at \url{https://github.com/sdss/idlspec2d/releases}.} These updates are outlined in \citet{sdss_dr18}, with a general description below.

Data from BOSS are processed through a quicklook pipeline (Son Of Spectro, SOS) during the night at each observatory. SOS provides an estimate of the signal-to-noise ratio as exposures are taken, providing data quality feedback to the observers, informing decisions on additional exposures to meet the required S/N for each program. It also produces a set of quality assurance plots that help identify failures within the pipeline, spectrograph, telescope, and fiber positioning systems. 

Following the transfer of the data from the observatories to the University of Utah's Center for High Performance Computing, the BOSS data are processed through the full reduction pipeline. This provides reduced spectra and a set of target spectral parameters to the collaboration on a daily basis. The pipeline uses an optimal extraction \citep{Horne1986} with profiles built on the traces flat lamp spectra to extract all science and calibration frames. In SDSS-V, different spectrophotometric requirements for different programs lead to different observing modes. But regardless of mode, the same basic procedure is followed. The extracted arc-lamp spectra are then used to provide an initial wavelength calibration in vacuum wavelengths, with a secondary calibration provided by the sky lines in the target fibers. The BOSS pipeline then uses the set of spectrophotometric standard stars to build a 2D flux distortion and calibration model across the field of view of the plate or field to calibrate each frame from each camera. This calibration step applies a Milky Way dust extinction correction from either the SFD model for dust extinction \citep{Schlegel1998, Schlafly2011} or the 3D Bayestar2015\footnote{This map will be replaced with a more complete coverage 3D map in the near future.} dust map \citep{Green2015} to properly model the standard star spectra in building the fluxing vectors. However, the final flux-calibrated spectra are uncorrected for Milky Way dust extinction. The choice of the two dust maps depends on the location of the field, with the 3D map used in the Galactic plane fields. The pipeline then first combines the extracted and calibrated frames from the red and blue cameras for single exposures, followed by the combination of individual red-blue-merged exposures to form \emph{epoch spectra} for any given target. In the plate era of SDSS, each epoch coadd was performed following a few simple rules, e.g., coadd two consecutive nights of RM plates with the same plugging, coadd all exposures of a plate (excluding RM plates) with the same plugging, and special coadds of all eFEDS plates for each Modified Julian Date (MJD). However, in SDSS-V, the epoch-coadding scheme is more complex, leading to the daily reductions providing coadds only from that night, with an additional set of epoch coadds (and other special coadds) produced for internal and external data releases. 

Following the reduction of the spectra, \texttt{IDLSPEC2D} uses template fitting to perform spectral classification and redshift determination. There is a separate set of templates for stars, galaxies, and quasars, with the final classification and redshift determined by the $\chi^2$ of the model fits. For stars, these templates are spectra for individual  stellar types, while for galaxies and quasars the models are supplied as a Principal Component Analysis (PCA) basis sets. In SDSS-V, with an increased number of stars being allocated BOSS fibers, \texttt{pyXCSAO} \citep{marina_kounkel_2022_6998993}, a python replication of the \texttt{xcsao} package \citep{Kurtz1992,Mink1998,Tonry1979}, was added, replacing the stellar parameters from the ELODIE models. The quasar PCAs were updated in SDSS-V, using 849 SDSS-IV RM quasars \citep{Shen2019} and the Weighted \texttt{empca} package \citep{Bailey2012,Bailey_2016_empca}. This new set of quasar PCA templates is implemented with 10 eigenvectors instead of the 4 used in SDSS-I through SDSS-IV. Following classification, the pipeline provides measurements of the emission-line flux and equivalent widthfor a number of major emission lines.

The standard BOSS pipeline products include the calibrated epoch-coadded spectra as well as individual, 900s-exposure spectra, and a FITS summary file ({\it spAll*}) for all epoch-coadded targets. The basic content and format of these spectroscopic products largely follow previous SDSS data releases. However, in detail there are many modifications in the SDSS-V FPS era that are documented on the SDSS-V Data Release pages and in Morrison et al. (\emph{in prep.}). 

\subsubsection{APOGEE Spectral Reduction}
The APOGEE data from both spectrographs are reduced with an updated version of the APOGEE Data Reduction Pipeline \citep[\texttt{apogeedrp}][Nidever et al., in prep.]{Nidever2015} \footnote{\url{https://github.com/sdss/apogeedrp}}.

Data are initially processed on the mountain using the APOGEE ``quicklook'' software,
\footnote{\url{https://github.com/sdss/apogee_mountain}}
to provide real-time S/N information as the exposure is being acquired and completed.  The information permits the observers to assess whether the S/N goals have been met.

Once the data have been transferred to the University of Utah's High Performance Computing Center, the full data APOGEE reduction pipeline is run on them.
The NIR detectors are read out non-destructively every 10.6 seconds, creating a datacube for each exposure.  These cubes are collapsed to 2-D images using up-the-ramp (UTR) sampling, where cosmic rays are flagged and removed (when possible).  The 300 fiber spectra are then extracted to 1-D using a model PSF that is generated for each exposure using trace offsets determined empirically from the exposure itself. Each column is extracted separately using an optimal extraction that takes into account the overlap of neighboring fibers as well as the broad trace profile wings. 

One important addition to the calibration data in SDSS-V comes from the Fabry-Perot Interferometer (FPI), which provides a very stable wavelength calibration source for APOGEE.  The afternoon daily calibration exposures of FPI and arc lamps (Thorium-Argon-Neon and Uranium-Neon) are used to determine accurate daily wavelength solutions.  Each science exposure has FPI light in two dedicated fibers that allow for the determination of small offsets of the daily wavelength throughout the night.  

For each field, sky subtraction and telluric correction is performed for each single exposure and then the spectra for each fiber are stacked or dither combined (if the data were spectrally dithered).  A final flux calibration is applied using the shape of hot stars and the 2MASS $H$-band photometry of each source.
For each star, the visit-level spectra (``apVisit'') are combined (when there are more than one epoch) and resampled onto a uniform logarithmic wavelength scale (``apStar'').  Radial velocities are determined with
\texttt{Doppler}\citep{Nidever2015}\footnote{\url{https://github.com/dnidever/doppler}} which forward-models the data and jointly fits all the visit-level spectra of a star simultaneously.

\subsubsection{LVM Spectral Reduction}
\label{sec:lvm-data}

For any exposure, the ``raw" data from the LVMI system consist of nine individual FITS files, one for each channel ($b$, $r$, $z$) in the three spectrographs. Each frame is read out through four amplifiers. The first step in the reduction is therefore to de-trend these frames by using amplifier-specific bias, overscan, gain, bad pixel masks and pixel flats. A combination of dome flats and twilight flats is used for tracing and building an image model for the individual fiber spectra. Arc-lamp exposures are used to create wavelength solutions. Throughout the data processing we track bad pixels and pixels affected by cosmic rays through a pixel mask extension that propagates through to the final data products. We also perform error propagation appropriate for CCD devices throughout the pipeline and store the resulting inverse variances.

We extract the the 1D spectra for each fiber using fiber-profile fitting extraction, which yields $\lesssim 1\%$ accuracy across all wavelengths. This algorithm accounts for variations in the fiber profiles across the CCDs and cross-talk between neighboring fibers, providing us with de-blended fiber spectra.

After the extraction step is run on all nine frames, we flatfield the spectra, combine all 3 spectrographs in each channel to end up with 3 frames for the blue, red, and infrared spectrograph channels. This step also provides a correction to any fiber-to-fiber, spectrograph-to-spectrograph and telescope-to-telescope variation in sensitivity. To do so, information is taken from twilight flats observed with all 4 telescopes pointing to the same region in the twilight sky. At the same time, the spectra are resampled onto a uniform 0.5$\AA$ grid spanning $3600 \le \lambda \le 9800\AA$

We perform flux calibration for each channel ($b,r,z$) independently. In this step the spectra are multiplied by a sensitivity curve to convert the signal from the instrumental units of electrons~s$^{-1}$~cm$^{-2}$~\AA$^{-1}$~spaxel$^{-1}$ to physical flux density units of erg~s$^{-1}$~cm$^{-2}$~\AA$^{-1}$~arcsec$^{-2}$. In this step we also remove the effects of atmospheric extinction specific to the exposure. We obtain a sensitivity curve for each of the 10-12 standard stars as the ratio between the instrumental spectrum and a reference spectrum at the top of the atmosphere. The final sensitivity curve for each exposure is given by a weighted average of the individual stars sensitivity curves. The weights are given by the fraction of the total science exposure time during which each standard star was exposed. A rough sky subtraction is sufficient for this purpose, since we choose bright stars that dominate the sky by large factors.

The reference spectrum of each star comes from a best-fit stellar atmosphere model to both the low-resolution Gaia BP/RP spectra and the higher-resolution continuum-normalized LVM-I absorption line spectrum. The wealth of spectroscopic information on the relative strength of absorption features in the LVM-I spectra, and the overall shape of the optical continuum of the star given by the Gaia BP/RP, are used jointly to accurately constrain the stellar parameters of the spectro-photometric standards. This, combined with our target pre-selection focused on F-type stars for which stellar atmosphere models are very robust, allows us to produce a reliable reference spectrum for each standard. The averaging of 10-12 calibration sources helps us to average out the systematic errors that arise from the uncertainties in typing individual stars. Once the sensitivity curves are applied to the fiber spectra of each channel, the overlap regions between them share a common flux level and can be adequately combined. 

The LVMI system uses two ``sky'' telescopes to observe two different patches of sky simultaneously with every science exposure. Our ultimate goal is to model the terrestrial night sky emission at the location of the science telescope, and subtract it off the observed astrophysical spectrum. In the case of the Milky Way, this is particularly challenging as line emission from the ISM is detectable over an area covering tens of degrees, forcing us to choose sky positions at relatively large angular separations. To mitigate this issue, we have chosen a strategy where one sky telescope always points to a patch of sky that is carefully selected to be free of Milky Way foreground contamination (but may be over 30~degrees away), and the other sky telescope always points to the darkest patch of sky  within $\sim$10~deg angular separation. Different components that make up the sky background and are modeled in different ways. 
The continuum components (from the moon, zodiacal light, airglow continuum, and atmosphere contributions) will be estimated by using the nearest sky patch and the ESO sky model \citep{Noll+2012} to extrapolate to the science field location. The more distant, darker sky patch will be used for secondary corrections. For airglow lines, we will use skycorr \citep{Noll+2014}, with any sky line residuals will corrected using a secondary principal component analysis (based on \citealt{Wild2005}). For geo-coronal emission, we will use the more distant, dark sky patch to estimate the intensity as a function of the H$\alpha$ line intensity and shadow height, in order to subtract any residual contribution in the science spectra.

The final product of the data reduction pipeline is a set of row-stacked calibrated spectra, errors, masks, sensitivities, sky spectra, and metadata for each fiber in our system. The LVM data reduction pipeline is described in detail in Mejía et al.,\emph{in prep.}

\subsection{Science Analysis Pipelines}

An integral part of the SDSS legacy is that a set of astrophysically interesting parameters (redshifts, velocities, stellar parameters, emission line fluxes, etc..) are extracted from all, or at least most , reduced spectra through a set of
\emph{scientific analysis pipelines}. These pipelines inevitably reflect to enable high-level scientific objectives at the ``catalog level".  We briefly describe these analysis pipelines and their choices here.

\subsubsection{MWM (ASTRA)}
\newcommand{\sw}[1]{\texttt{#1}}
\newcommand{\Astra}{\sw{Astra}}
MWM targets are exclusively stars, but span a wide range of spectral types (OBAFGKM). The detailed physics required to accurately model stellar photospheres in one extreme (e.g., O-type stars) is not relevant for stars in the other (e.g., M-type), and vice versa.  
Therefore, the diversity of MWM targets requires multiple analysis pipelines.  Multiple analysis pipelines present both challenges and opportunities to cross-check and improve the results.  If multiple pipelines analyze the same spectra and there is important disagreement in the estimated parameters, then it motivates where we should investigate. If there is sensible agreement then it builds confidence in the results.  A key advantage to this approach is that it explicitly motivates continuous improvements on all contributed pipelines, another long-standing tenet of data analysis in SDSS.  Prior to a data release, the MWM team will work with the results from a minimal set of pipelines necessary to derive parameters for all target types (i.e., we will not average results from these pipelines). In parallel, we will work to understand why there are remaining discrepancies.  The software responsible for coordinating these analyses is called \Astra.

\Astra\ includes wrappers to execute more than a dozen analysis codes in a consistent manner.  This includes spectral synthesis codes (\sw{Korg}, \sw{TurboSpectrum}), full spectrum fitting with pre-computed grids (\sw{FERRE}, \sw{ASPCAP}, \sw{SnowWhite}, \sw{The Payne}, \sw{MDwarfType}), data-driven techniques (\sw{The Cannon}), machine-learning methods (\sw{Classifier}, \sw{APOGEENet}, \sw{BOSSNet}, \sw{SLAM}), utilities that are common to many pipelines (e.g., for continuum modelling), as well as bespoke tasks to measure line ratios or radial velocities (e.g., \sw{corv}, \sw{LineForest}).  All of these codes have been re-factored (or rewritten entirely) to better interface with the SDSS-V data models.  Any function in these pipelines that returns a measurement given some input spectrum now has a consistent interface in \Astra\, and is wrapped (decorated) as a `task'.  Decorating these functions as tasks allows \Astra\ to control the execution of work and attach necessary metadata to the results.  For example, if a user executes an \Astra\ task in an interactive Python environment, \Astra\ could execute that work on a GPU or remote cluster, without significant user effort.  Immediately before results are sent back to the user's interactive session, \Astra\ writes those results to a database (efficiently, in batches) with additional metadata (e.g., time elapsed, task keywords for reproducibility, tagged software versions). 

We use \sw{Airflow} to orchestrate these \Astra tasks in an automated way.  One advantage of using \sw{Airflow} over a timed script is that different processing tasks can be paused, stopped, or started through a web interface.  Tasks that temporarily fail (e.g., due to some connectivity issue) are automatically rescheduled, and limited resources (e.g., compute clusters) are balanced appropriately across tasks.  This workflow helps scale and orchestrate the analysis of all reduced data products from the MWM program; it produces scientific validation figures that are used for quality assurance or to identify any discrepancies between pipelines. It also prepares summary data products for release to the collaboration. \Astra\ will be presented in more detail in Casey et al. (\emph{in prep}).

\subsubsection{BHM}

For the SPIDERS and AQMES parts of SDSS-V/BHM, the data products remain largely unchanged, compared to the previous data releases DR17 and DR18 \citep{Abdurrouf_2021_sdssDR17,sdss_dr18}, and remain based on the the standard BOSS pipeline. This includes coadded spectra where appropriate, and the very extensive {\it spAll*} summary file parameters (e.g., spectral class, redshift, line parameters, and many more). 

For BHM-RM, a custom process called \sw{PrepSpec} \citep{Shen_etal_2016} is applied to the 1D pipeline spectra of the BHM-RM targets to further improve spectrophotometry to facilitate RM lag measurements \citep{Shen_etal_2023}. \sw{PrepSpec}-calibrated spectra are not part of the standard SDSS-V public data releases, but will be released in science articles reporting RM lag measurements. To accompany SDSS-V monitoring spectroscopy, the BHM-RM collaboration also conducts dedicated photometric monitoring programs using different ground-based facilities to obtain continuum light curves for BHM-RM targets \citep[e.g.,][]{Zhuang_etal_2024} that are published separately. Photometric light curves from public surveys such as ZTF and LSST will also be used in the BHM-RM lag measurements.

\subsubsection{LVM}
\label{sec:lvm-dap}

The LVM survey was designed from the very beginning with the goal of providing not only the raw and reduced spectral data cubes but also high-level data products. Those include observational and/or physical parameters extracted from the observed spectra that allow the community to make use of the data in a convenient manner.

Beyond the Data Reduction Pipeline that produces reduced and calibrated row-stacked spectra, we have developed a data-analysis pipeline (DAP) that decomposes the observed spectra into the (stellar and nebular) continuum and the warm gas emission lines. The DAP then extracts a set of observational and physical quantities that characterize each component.

Individual LVM fibers in the Local Group and MW often contain single stars to few stars, or parts of single star clusters. For LVM-DAP, we developed a strategy dubbed \emph{resolved stellar populations} (RSP), described in \citet{mejia24b}. We do not model the stellar spectra using linear combinations of single stellar populations (SSPs), but rather a library of individual stellar spectra covering a wide range of stellar properties (T$_e$, log(g), [Fe/H] and [$\alpha$/Fe]) instead.

The current version of LVM-DAP (pilot DAP, or pDAP) uses the {\sc pyFIT3D} algorithms included in the {\sc pyPipe3D} package \citep{pypipe3d}. A complete description of the pipeline will be presented in \citet{sanchez24}. The pDAP analyzes the LVM-data pointing-by-pointing, using as inputs the spectra and errors provided by the data-reduction pipeline. For each individual spectrum, pDAPperforms the following steps.

We first estimate the systemic velocity, v$_\star$, velocity dispersion, $\sigma_\star$ and dust attenuation, A$_\mathrm{V,\star}$, of the stellar component by fitting the continuum with a linear combination of only four stellar spectra that are shifted in velocity, broadened, and dust attenuated, while masking the strongest emission lines.  Once these non-linear parameters are determined, a first model of the stellar component is subtracted, and a set of strong emission lines are fit using Gaussians.
We subtract these emission lines, and then again fit the remaining continuum, this time with a larger stellar template library covering a wide range of stellar properties, yielding our model of the stellar spectrum. This stellar spectrum is then subtracted from the original spectrum. The remaining ``gas-only'' spectrum is then analyzed using a weighted-moment nonparametric procedure to estimate the integrated flux, velocity, velocity dispersion, and EW of the a predefined set of 192 emission lines. The parameters derived for the stellar and ionized gas components in each of the previous steps are stored in a set of FITS tables that are packed into a single multi-extension FITS-file and the model spectra are stored in a row-stacked spectra (RSS) format. For details of the algorithms and their implementation, we refer the reader to \citet{sanchez24} and \citet{pypipe3d}\footnote{http://ifs.astroscu.unam.mx/pyPipe3D/}.

\section{Data Management and Archiving in SDSS-V}\label{sec:data}

Each phase of SDSS has added to the cumulative experience in managing and archiving SDSS data, adding sophisticated online access tools which have enabled thousands of papers by professional scientists, and creating educational resources that use real data to positively impact student learning.  The SDSS-V data team continues to build upon and expand existing SDSS data archive servers, such as the Utah-based Science Archive Server (SAS) and JHU-based Catalog Archive Server (CAS) in Baltimore.  The SDSS Data Management and Archiving (DMA) Team includes roles that span the SAS, CAS, Web, and Help Desk, and interfaces with the Control and Operations Software team, each of the SDSS-V Mappers, and Education and Public Outreach (EPO).

 The SAS serves file-based science data products including images and spectra in FITS format, hosted within the University of Utah's Center for High Performance Computing (CHPC) Tier-3 data center that provides \(>  99.982\%\) uptime availability,  a large team of network specialists, and cluster infrastructure. Since the beginning of SDSS-IV in 2014, SDSS science data have been reduced by pipelines running at CHPC that deliver fully processed data to the SAS\footnote{\url{https://data.sdss.org/sas/}}.  The current data volume of the SAS exceeds 1 Petabyte, with over 737 Terabytes in the cumulative data releases, DR8 -- DR18, and preparing for DR19 (MOS) and DR20 (LVM) there are 180 terabytes and growing as the data reduction pipelines are frozen for downstream analysis, such as the astra pipelines, value-added catalogs and a new generation of dynamic (containerized) web applications. Internally supported by a dedicated (postgreSQL) database with fully loaded Solid State Drive (SSD) bays for both high performance computing, to make the pipelines which run on 1500 nodes on the SDSS-V computation cluster as efficient as possible, and for the downstream visualization of the data.  This architecture has enabled a data visualization Working Group, to provide it's first round of value-added products, including a fastAPI to be made public in DR19  and a front-end built on modern platforms, providing an easier to maintain software stack and deployed as docker containers on a CHPC virtual machine, and portable to other servers for either mirroring or migration into the cloud.  See section \ref{data-viz} for the history and plans for data visualization in SDSS.
 
 The CAS provides performance-optimized SQL (Structured Query Language) based access to science catalog data stored in commercial database management systems (DBMS). These services continue to serve new data for SDSS-V, and are being modernized to make SDSS data more accessible to the larger community.  Since the original SDSS Early Data Release in 2000, all SDSS catalog data has been served by the CAS. The CAS database schema, data loading scripts and data access tools were all developed by the CAS team at the Johns Hopkins University (JHU). The primary public interface to CAS data is the SkyServer website\footnote{\url{https://skyserver.sdss.org}}, which was developed at JHU in 2001. More powerful and flexible data access tools for astronomers are provided by the CasJobs batch query workbench\footnote{\url{https://skyserver.sdss.org/casjobs}}, which was also developed at JHU and launched in 2004. Around the same time, the CAS team developed the sqlLoader pipeline to load new SDSS data releases into CAS~\cite{Szalay2008}. These software products still provide the bulk of the catalog data access for SDSS users worldwide, and in 2015 JHU's Institute for Data Intensive Engineering and Science (IDIES\footnote{\url{https://www.idies.jhu.edu/}}) launched the SciServer Science Platform\footnote{\url{https://www.sciserver.org/}}~\citep{Taghizadeh-Popp2020} that integrates the SDSS building-block services into a collaborative data-driven science framework.  SciServer was developed with NSF DIBBs\footnote{\url{https://new.nsf.gov/funding/opportunities/data-infrastructure-building-blocks-dibbs/504776/nsf17-500/solicitation}} funding and importantly added a server-side Jupyter Notebook \footnote{\url{https://jupyter.org}} analysis capability to the existing tools, so that CAS users can submit queries and entire workflows in the form of interactive and batch mode Jupyter Notebooks. 
 Users run Jupyter Notebooks in the SciServer Compute \footnote{\url{https://apps.sciserver.org/compute}} application, 
which leverages Docker Containers \footnote{\url{https://apps.sciserver.org/compute}} to provide isolated software environments for data analysis, as well as as direct access to a file system hosting data from multiple science domains, including astronomy and SDSS in particular. A dedicated ``SDSS" container image is also available for creating a Jupyter session with specialized software needed for the access and analysis of the SDSS data hosted on SciServer, such as the sdss-access, sdss-semaphore, astropy, and specutils python packages. This special SDSS image also facilitates local access to the SAS data that is copied to JHU along with the CAS. The DR19 tutorials available on the SDSS website include examples of how to access the SAS data via SciServer Compute, with account signup and container creation instructions available at \url{https://www.sdss.org/data/data_access/sciserver_compute/}.   
 
 The existing SDSS services SkyServer and CasJobs were re-engineered and integrated into SciServer so that they were upgraded in important ways without losing their essential functionalities. In addition to the Jupyter notebook access via the SciServer Compute subsystem, a new platform-independent version of CasJobs is also under development called SciQuery. SciQuery will ultimately replace CasJobs but this is happening incrementally in such a way that there is no break in service and functionality for the large CasJobs user base. SciServer's collaborative features are accessible via the Groups page, whereby users can share their resources - files, folders, Jupyter notebooks, data volumes and databases - with their collaborators or their students through layered access capabilities, which are especially suited to using SciServer in the classroom to enable students to work with the SDSS data products, as part of data science course curricula. As mentioned above, one unique capability that SciServer adds is the ability to run Jupyter notebooks and scripts in batch mode, just like queries in CasJobs.

 Parts of the SkyServer will be internally redesigned to work with the new SDSS V data model, being centered now in the 'sdss\_id' unique identifier for object searches, and including a revised Navigate tool for displaying and interacting with the targeted objects on the sky in places where the legacy SDSS photometric survey has no coverage. The SpecDash \footnote{\url{https://specdash.idies.jhu.edu}}~\citep{Taghizadeh-Popp2021} spectral visualization and analysis tool linked from SkyServer is now also able to search for and load SDSS-V targets which is mirrored at JHU and also made available in SciServer as a public data volume.

A new addition to SkyServer is a form-based interactive webpage
\footnote{\url{https://skyserver.sdss.org/dr19/CrossMatchTools/crossmatch}} (with a standalone version at \url{https://skyserver.sdss.org/public/xmatch})
that allows running interactive two-way on-the-fly 2-dimensional spatial cross-matches and cone searches across the co-located astronomical catalogs. These include SDSS DR19 and all previous cumulative data releases, as well as more than 50 other astronomical surveys. To support this, we have stored all the catalog data tables in the new `xmatch` public context in CasJobs, and integrated them with the new SQLxMATCH software \footnote{\url{https://github.com/sciserver/SQLxMatch}}, which allows running an in-database cross-match by means of a simple SQL query \citep{Taghizadeh-Popp2023}. Users can also upload their own catalogs into their MyDB context in CasJobs, using the Import feature, and cross-match them against the existing catalogs on-the-fly through the interactive web interface. More advanced users can customize the demo cross-match Jupyter Notebooks \footnote{\url{https://github.com/sciserver/SQLxMatch/blob/main/demo/SQLxMatch_CasJobs_demo.ipynb}}
in SciServer for programmatic analysis and storage.

\subsection{Data Processing}

The flow of data from the SDSS Observatories starts with an automated daily transfer~\citep{Weaver2015}, which runs multiple rsync transfer streams to download each MJD, followed by data verification by checksum match, and backup.

The data files generated on the SAS by the various pipelines depend upon:
\begin{itemize}
\item a software repository in GitHub for the SDSS Organization\footnote{\url{https://github.com/sdss}},
\item IDL software language,
\item Python software language,
\item Postgresql database containing targeting catalog information,
\end{itemize}
The SAS is setup with user configurable modules, allowing software branch management for testing, development, and production, including automated daily runs for each spectrograph to keep the data reduction and analysis pipeline's updated for members of the collaboration running analyses on proprietary observations. Each pipeline records data reduction and analysis information to a high performance database, which can then be used for data visualization and quality inspections.

Each mapper team uses the SAS to run their respective software pipelines, with BHM and MWM processing BOSS and APOGEE spectra with modernized versions of packages that were developed for previous SDSS generations, and LVM processing spectra from the DESI spectrographs using a custom version of the DESI Pipeline, as described in Section \ref{sec:software_pipelines}.

A dedicated 1500 core computer cluster, maintained by the University of Utah CHPC, provides the primary computational cluster for all SDSS data reductions and analysis, including multiple interactive nodes providing login access to members of the SDSS collaboration.  There are daily reductions, which process the recentmost MJD on a single computer node, and co-ordinated reductions, which use many nodes in order to process many MJDs, to either test new developments, or to generate an Internal Product Launch (IPL).

IPLs provide working groups and collaboration scientists with a common set of reductions, from which further analysis can be performed, leading to Value-added Catalogs (VACs) which are candidates for inclusion in a subsequent data release, such as those included in the most recent Data Release 18\footnote{\url{https://www.sdss.org/dr18/data_access/value-added-catalogs/}} After vetting and quality control, the IPL's data products are made publicly available as public data releases (DRs).

\subsection{Data Products}

The two broad types of SDSS data consist of flat-file based data and catalog data products, which continues to be the case since the beginning of SDSS. Within these broad categories, the SDSS archive servers support the following categories of data products:  SDSS imaging, targeting, optical spectra, infrared spectra, IFU spectra, and value-added catalogs as summarized in Table~\ref{tab:dataproducts}.  For each data product, a unique environmental variable provides a path location on the SAS, supporting multiple versions including those for testing and branch development.  This knowledge is organized in the SDSS tree product\footnote{\url{https://sdss-tree.readthedocs.io}} which includes separate configurations for each data release and internal product launch, allowing a flexible definition for each data product's output file paths on the SAS, supporting the same keywords used by the source code.

Access to data products via their flat-files on the SAS is available over HTTPS\footnote{\url{https://data.sdss.org/sas}}, rsync, globus, and python.  Whereas access over HTTPS is best suited for web browsing the directory structure and downloading individual files, rsync using a dedicated data transfer node\footnote{\url{rsync://dtn.sdss.org}} provides a more robust method which synchronizes downloaded products with the SAS, is better suited for bulk data downloads, supports wildcard characters, and includes examples documented in the SDSS website\footnote{\url{https://www.sdss.org}}.  For archival quality data transfers, access is also available via Globus Online\footnote{\url{https://www.globus.org}} through the SDSS globus endpoints such as sdss\#public.  Globus is significantly faster and more robust than using wget and provides automatic checksum verification which is not provided by rsync, but requires a local Globus endpoint to support the transfer\footnote{\url{https://www.globus.org/globus-connect/}}.  Access to data products via python is automated by the SDSS Access package\footnote{\url{https://sdss-access.readthedocs.io}} which allows easy scripting for bulk data downloads and is aware of the tree product and abstract paths for each data product on the SAS.  This allows a data product to be included in the download via its name and uses the same environmental variables and abstract path definitions defined in the SDSS tree product.

Access to catalogs of data products via a SQL based interface is available via the SciServer\footnote{\url{https://sciserver.org}} science platform.  SciServer hosts the following web-based apps that allow you to query and analyze your data in the cloud: the classic SkyServer\footnote{\url{https://skyserver.sdss.org}} and CasJobs\footnote{\url{https://skyserver.sdss.org/CasJobs/}} web portals that have been available since DR1, and the SciServer Compute application that provides Jupyter Notebook access to SDSS data in both interactive and batch modes in Python and R. Another new component, SciQuery, will also be available to provide scripted platform-independent access a la CasJobs. A good starting point for querying the SDSS data is the set of sample queries that are available on the SkyServer site. There are also several sample and tutorial Jupyter Notebooks for SDSS data access available in the ``Getting Started" and  ``SDSS SAS" folders in your SciServer Compute container once you create it. Help is, as always, an email away at helpdesk@sdss.org, please always Reply-All to helpdesk messages so that the helpdesk email account is cc-ed on all responses, as the Help Desk is staffed by multiple SDSS volunteers and cc-ing the help desk ensures that all of them see your request.

\begin{table}[h]
    \centering
    \begin{tabular}{|l|l|}

        \hline
        \textbf{File-based Data Products} & \textbf{Catalog Data Products} \\
        \textbf{(FITS format, served by SAS)} & \textbf{(SQL tables, served by CAS)} \\
        \hline
        Raw data transferred from the SDSS  &   \\
        observing sites at APO and LCO  &  \\
        \hline
        Targeting data and target selection & Targeting catalog SQL tables, with   \\
        pipeline outputs &  MOS Target tables new for SDSS-V\\
        \hline
        Imaging data and photometric pipeline & Photometric parameter SQL tables, \\
        reductions & JPEG images and cutouts \\
        \hline
        Spectroscopic pipeline reductions & Spectroscopic reduction summary SQL tables, \\
        & PNG images of spectra \\
        \hline
        Parameter pipeline analyses &  Spectroscopic parameter analysis SQL tables \\
        \hline
        Value-added catalogs (VAC) contributed & VAC parameter SQL tables \\
        by scientific working groups and  &  \\
        collaborative project teams & \\
        \hline
        & Cross-matches with non-SDSS data \\
        \hline

    \end{tabular}
    \caption{Data products available from SDSS, including those available from the SAS (column 1) and the CAS (column 2).  Note that the data products available in the CAS are derived from the flat-files available on the SAS, by a FITS to SQL CAS loading process, or by direct transfer in the case of the images, and the SQL tables in the CAS may be joined for deeper scientific inquiries than are directly available from single flat-files on the SAS. Note that, new for SDSS-V, there is a full copy of the SAS in SciServer Compute so that Jupyter notebooks can access FITS files to facilitate use cases that require both CAS and SAS data. }
    \label{tab:dataproducts}
\end{table}

\subsection{Data Visualization} \label{data-viz}

Well in advance of each SDSS public data release, there are science-driven projects within the SDSS collaboration that depend on high quality images and spectra that have been vetted and in many cases manually inspected.  This has driven the creation of a number of data visualization tools that have provided scientists within the SDSS collaboration the ability to manually inspect and in some cases comment on individual objects prior to their finalization and release to the public.  Many of these data visualization tools have themselves been released for the use of the public, but without requiring authentication or supporting manual inspection comments.

The first such tools were written in the {\tt IDL} programming language by the SDSS-II and SDSS-III (BOSS) pipeline developers, and are publically available in the {\tt idlspec2d} github repository, including {\tt plotspec}\footnote{\url{https://github.com/sdss/idlspec2d/blob/master/pro/spec1d/plotspec.pro}} and an extended version {\tt uuplotspec}\footnote{\url{https://github.com/sdss/idlspec2d/blob/master/pro/spec1d/uuplotspec.pro}} which included a backend database to collect comments on optical spectra, enabled by previous and next buttons for fast visualization of a list of optical spectra.

The first webapp released for public use, known as the {\tt dr10webapp}, is no longer supported and was deprecated in August 2024.   This imaging webapp provided single and bulk field searches, coverage checks, and mosaic image generation (downloadable script), and provided downloads of the image's frame files.  However, most of these functions were also provided by skyserver's imaging search\footnote{\url{https://skyserver.sdss.org/dr18/SearchTools/IQS}} including download options via {\tt wget} or {\tt rsync}, and there are tutorials coming soon to the public website~\footnote{\url{https://www.sdss.org/dr18/imaging/tools/}} to provide educators that previously depended on the dr10webapp with equivalent python notebooks.  Although the dr10webapp also displayed spectra, it was limited to spectra from the original SDSS spectrograph, and its development was frozen prior to SDSS-IV.

For SDSS-IV spectra, the dr10webapp was replaced by the Science Archive Webapp (SAW) and was designed with a modern architecture based on the python \texttt{flask} framework for better maintainability and longer sustainability; and to facilitate easy loading by using the same schema as used by the CAS to load the pipeline results.  The SAW\footnote{\url{http://dr18.sdss.org/home}} includes optical spectra from the SDSS and BOSS spectrographs, infrared spectra from the APOGEE and APOGEE2 spectrographs, and MaStar spectra which include a stellar library of observations from the MaNGA instrument.

For IFU data, such as from the SDSS-IV MaNGA survey, a suite of interconnected tools and services was developed, called \texttt{Marvin}\footnote{\url{https://magrathea.sdss.org/marvin}}~\citep{Cherinka2019}. \texttt{Marvin} consisted of a front-end web UI, a back-end API for programmatic data access, and a python package for access, visualization, and analysis of the MaNGA data products. \texttt{Marvin} was designed with a smart multi-modal data access system, which allowed for streamlined context switching between local and remote data sources, as well as between related data products for a given target. \texttt{Marvin} was developed completely open source on Github, and released on PyPI, available for public and collaboration users to access either public or proprietary SDSS data (with proper authentication credentials). The combination of the web, API, and python package allowed users to seamlessly move between different interaction nodes and optimize their scientific workflow. 

The inter-connectivity and power of \texttt{Marvin} was enabled because it was built on top of a variety of core tools that make up the SDSS Software Framework, i.e. tools that understand the SDSS environment, its products and data models, and allow for pythonic data access and downloads. For SDSS-V, these tools, along with the core components of \texttt{Marvin}, were further generalized for SDSS-V data, restructured to improve modularity, and re-built in modern languages to leverage the latest advancements in web and software technologies. 

To replace the SAW, a new web front-end, \texttt{zora}, is being developed in \texttt{Vue}, a modern, reactive, component-based UI framework. This interface is powered by a new back-end programmatic API, \texttt{valis}, built in the python \texttt{fastapi} framework, which provides distinctive endpoints for accessing SDSS information of interest.  The \texttt{sdss-brain} python package is being developed to provide \texttt{Marvin}-like python access to local or remote data products through context-aware classes.  By cleanly separating these components, users can access SDSS data in their preferred manner, or even build their own web interfaces on top of the API.     

In addition to improving the accessibility of SDSS-V data within the SDSS Software Framework, work has been done to make SDSS-V data accessible via community open-source tools like \texttt{specutils}~\citep{specutils2024} and \texttt{Jdaviz}~\citep{Jdaviz2022}, an open-source, data visualization and analysis package for astronomy data within both Jupyter and web-based environments.  In fact, \texttt{zora} leverages several of these tools to provide an enhanced user experience: interactive spectral visualization and analysis using \texttt{Jdaviz}, a dynamic data dashboard rendering millions of SDSS-V parameters using \texttt{solara} and \texttt{vaex}, and an interactive sky viewer overlaying SDSS-V data alongside other astronomical datasets and surveys using \texttt{aladin-lite}~\citep{aladinlite2024}.

\subsection{Data Releases}

An integral part of SDSS is to make its data publicly available, starting with the Early Data Release in 2001 \citep{2002AJ....123..485S}. As of date of writing, 19 public data releases have been published by SDSS, with DR18 being the most recent one \citep{sdss_dr18}. All data releases are documented on the SDSS wordpress website\footnote{\url{https://www.sdss.org/}}, which offers information for both experienced and novice astronomers to access its data products\footnote{SDSS also has a platform aimed at teachers and their students to learn about and work with SDSS data called Voyages: \url{https://voyages.sdss.org/}}. In particular, each data release contains a number of tutorials and examples for data access, with many of these available as Python notebooks. A full overview of each data release is given in its data release papers\footnote{\url{https://www.sdss.org/science/publications/data-release-publications/}}.

Along with the data products produced by the data reduction and analysis pipelines, the data releases also contain value added catalogs (VACs). These data products are generated by members of the SDSS collaboration based on SDSS data, and made available and accessible in the same format. An overview of all VACs released by SDSS so far can be found on the SDSS website\footnote{\url{https://www.sdss.org/dr18/data_access/value-added-catalogs/}}: each catalog has an entry providing links to its data files and data models, as well as further information on how the catalog was created, and instructions for usage. 

Creating a high-quality and well documented public data release is an extensive process, that is overseen by the data team with input from the science, software, and education and public outreach teams \citep{2019ASPC..521..177W}. Over the years, this process has been further refined, and effort has been made to lower and remove barriers for individual collaboration members to contribute to the data releases. In SDSS-IV we introduced a more streamlined and documented process to announce and prepare VACs for inclusion into data releases: this resulted in an increase of VACs and a more diverse group of VAC contributors, including students. We also offer more opportunities for collaboration members to contribute to documentation, which has led to an increase in the number of tutorials and Python notebooks available to access and work with the SDSS data \citep{2020ASPC..527..399W}.

\subsubsection{MOS Data Releases}

A small amount of SDSS-V single fiber optical spectra from eFEDS plates have already been released in DR18 \citep{sdss_dr18} from the BOSS spectrograph as part of the Black Hole Mapper's SPIDERS science program\footnote{\url{https://www.sdss.org/dr18/bhm/programs/spiders/}}.  With the replacement of manually plugged plates with robotic fiber positioners, future data release will include a fundamental change with respect to navigating individual MOS visits for both the BOSS and APOGEE spectrographs. DR19 (scheduled for 2025) will contain the first large set of BHM and MWM spectra and derived data products, as well as value added catalogs based on these data products. These data products will also be made available in the new for DR19 SDSS data visualisation webapp.

\subsubsection{LVM Data Releases}

Given that the commissioning and science operations for the LVM began in 2023, the first major public data release for the LVM is forseen for DR20, currently planned for 2026. Subsequent public releases will follow, with all data made public after the end of survey operations. 

Data releases will include the raw, reduced, and science-ready data and products, as well as all code used to reduce and analyse the data.  
For IFU data sets, the core science-ready data product is the row-stacked spectra (RSS files, see Section \ref{sec:lvm-data}). These consist of fluxes, uncertainties, LSF information, PSF information, fiber RA and DEC, and sky spectra associated with each individual survey tile observation.  We further plan to provide data products, produced by our DAP (see Section \ref{sec:lvm-dap}), including fluxes of more than 200 emission lines (or upper limits), and kinematic fits for subsets of those emission lines. 

In the long term, we aim to provide map and cube making routines, and directly provide maps of some of emission lines and kinematic data.  
Model-dependent quantities  (e.g. densities, temperatures, abundances, fits to nebular emission models) will be made available as Value Added Catalogs (VACs). All LVM data will also be made accessible through the SDSS Data Visualisation Webapp, and will be accompanied by tutorials and examples for common science cases.

\section{Project Organization} 
\label{sec:Org}

\begin{figure}[ht!]
\label{fig:orgchart}
\centering
  \includegraphics[width=\linewidth]{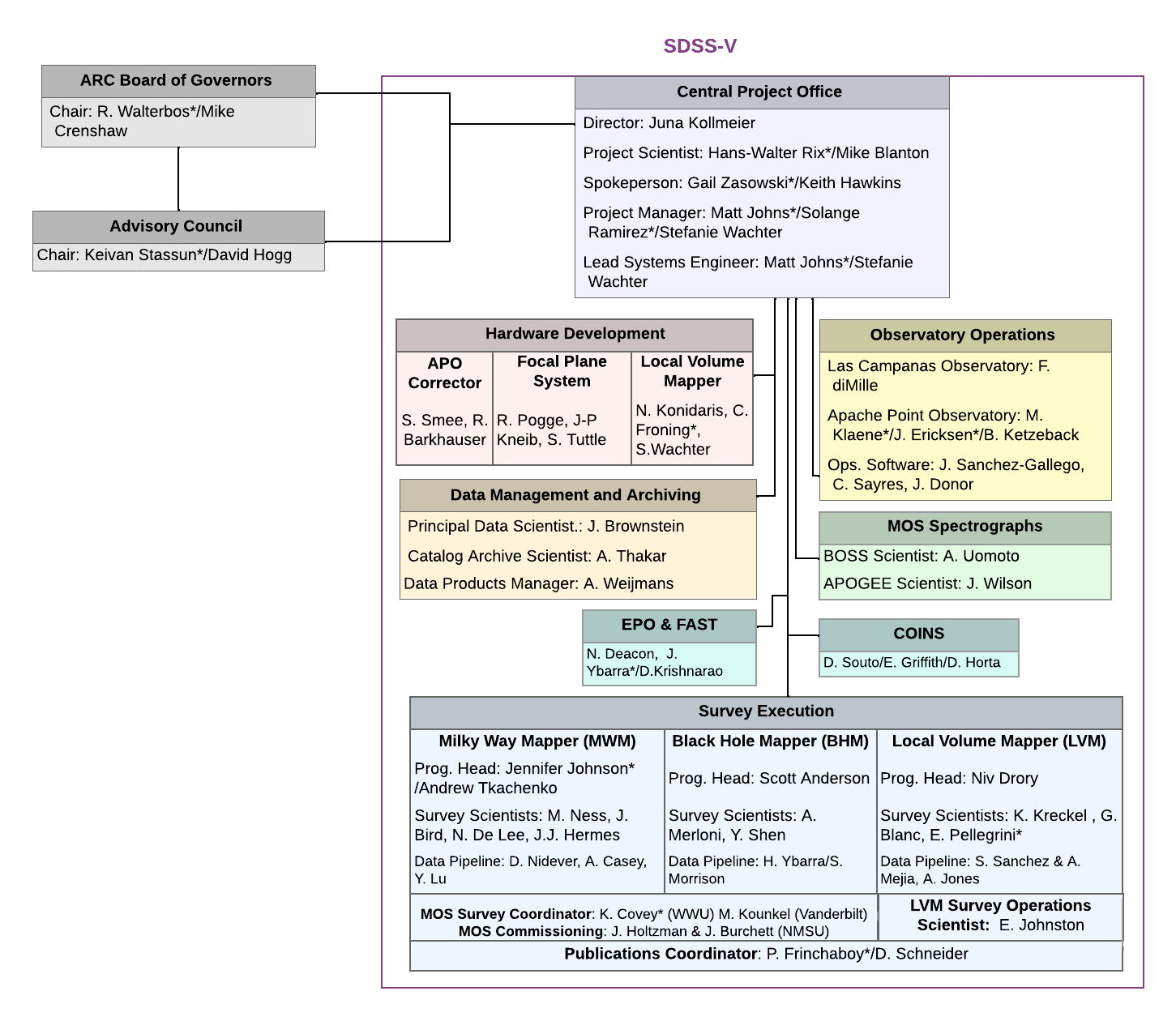}\caption{\footnotesize\linespread{1.2}\selectfont{} {\bf SDSS-V's Organizational Structure}. The shaded colored boxes show the SDSS-V Management Committee.  In cases where individuals have served out their terms, both the original and current individual are shown with individuals whose terms concluded are denoted by the asterisk. The roles of the central project office are described in detail in the SDSS-V Principles of Operation and in Section~\ref{sec:Org}.  The Scientific Working Groups within SDSS-V are numerous and their leaders coordinate multiple aspects of the program science with the collaboration.} 
  \label{fig:orgchart}
\end{figure}

The SDSS-V project benefits from the organizational and collaborative infrastructure and culture developed over three decades to remain scientifically vibrant throughout and draw upon a vast reserve of ``volunteer effort'' across the globe. At the same time, the SDSS endeavor and its policies had to evolve along with the landscape of its institutional partners and of astronomy as a whole.

\subsection{Oversight and Funding Model}
SDSS-V is overseen by the Astrophysical Research Consortium (ARC) and its Board of Governors.  The SDSS-V project operations, governance, and management structure are described in the SDSS-V Principles of Operation which have been developed over the lifetime of the SDSS surveys and updated at each transition.  SDSS is not a line-item in any institutional budget including ARC.  Rather, it is funded through a combination of foundation (public and private) funding as well as institutional contributions.  The contributions from member institutions are codified in Memoranda of Understanding (MOUs).  These include cash contributions as well as in-kind contributions. Survey membership grants data access rights according to membership level as codified in the SDSS-V Joining Document and the Principles of Operations.  

The ARC Board of Governors has established an Advisory Council (AC) consisting of representatives from the member institutions.   The AC provides oversight of the project and Director who is appointed by the ARC Board. 
ARC delegates to the SDSS Director the authority to organize and direct all aspects of the SDSS project for fixed renewable terms.  The SDSS Director delegates authority to a large number of individuals to lead different aspects of the SDSS program as it evolves.  The SDSS-V Central Project includes the Director, Project Scientist, Spokesperson, Project Manager and Lead Systems Engineer.  These positions are filled by PhD-level astronomers.   The organizational chart is shown in Figure~\ref{fig:orgchart}.

\subsection{Project and Infrastructure Management}
SDSS employs project management at multiple levels to ensure efficiency and progress-tracking across the multiple institutions where infrastructure activity is occurring.  During its major construction phase, project management and systems engineering tasks were handled by two full-time positions: the Project Manager and the Lead Systems Engineer, who worked closely with the CPO and the project managers of individual hardware development, software development, and survey design work packages. 

The project management practices followed included definition of a work breakdown structure (WBS), which establishes the project schedule and budget, and a risk register to manage identified risks.  The top-level WBS comprised fifteen groups, including groups developing hardware elements, groups associated with each of the survey mapper teams, groups representing each of the observing sites, and groups that perform development and coordination of the survey as a whole. 
The WBS is defined in several hierarchical levels, depending on the complexity of the tasks to be performed by each group. The information on the WBS serves as the basis for the schedule and the budget. The schedule includes milestones for each group, some of which are tagged to project reviews, others to activities occurring across groups. Reviews are scheduled throughout the life of the survey in order to evaluate maturity of the different sub-systems and keep stakeholders informed about progress.  Each group reports progress to the Project Manager/Lead Systems Engineer about work performed, typically on a weekly basis; performed work, schedule and risk update on a monthly basis; and performed work, schedule, risk and budget are updated on a quarterly basis.  All WBS group leads report to the Director and one another at a weekly Management Committee meeting. This serves to provide maximal transparency and communication across a globally distributed and large project team.

Systems engineering methodology includes definition and management of science requirements, system requirements flowed down from the science requirements, traceability of design elements to system requirements, definition and management of interfaces through interface control documents (ICDs), early development of testing and validation procedures and quality assurance plans. A key element of management and systems engineering is configuration control. A document repository system is available to host management (WBS, schedule, budget, risk documents), systems engineering (e.g. requirements, ICDs, plans) and project wide documents (e.g. review reports, design documents, policies, drawings).

\subsection{Scientific Management}
The SDSS-V Project Scientist provides the overall quality assurance for the project and ensures its scientific integrity.  The SDSS-V Spokesperson is responsible for fostering the scientific productivity of the collaboration, representing the project to the outside world, raising the visibility of SDSS-V within the astronomy and physics communities, maintaining good morale, and ensuring inclusiveness in the Collaboration.  The Spokesperson is the chair of the Collaboration Council (CoCo).

The three mapper programs are lead by a Program Head and Survey Scientists as needed by the Program.  These individuals ensure that the scientific goals of the mapper are achieved and are responsible for the mapper-level requirements.

The instrument and data scientists and site operations teams including the observers and engineers are responsible for developing, maintaining, and repairing the core scientific infrastructure that enables the survey. During Commissioning and Scientific Operations phases, we rely on coordinators to ensure that these activities are on-track and performing  smoothly.

Because SDSS-V is a geographically distributed team, communication plays a key role in keeping stakeholders informed and engaged, for which we use knowledge base wikis, video and teleconference tools, rapid messaging systems, and email lists. Design choices and trade studies are incorporated into the project documents that live in the document registry.
{\bf SDSS has a well-established culture of successfully managing complex, multi-faceted instrument development and survey science programs.}

Living policies and practices must be approved by the Advisory Council and ARC Board of Governors, and are publicly available across the consortium.  This set of policies principally includes our broad membership documents (Principles of Operation and Joining Document), as well as focused policies such as the Intellectual Property, Code of Conduct, External Participant, and Publication policies.  Our collaboration makes an effort to maintain best-practices for large multi-institution, multi-cultural collaborations and does its best to foster a welcoming environment for all.

\section{A Commitment to Opportunity}

From its inception, SDSS has maintained a commitment to society above and beyond the scientific papers produced by the collaboration.  This is most clearly manifested in an unwavering commitment to high-quality release of all datasets and analysis tools for discovery well-outside the boundaries of the relatively small circle of fortunate collaboration institutions.  In addition to the data itself, SDSS works to reach scientists wherever they are, professional and amateur alike.

\subsection{Pathways to Leadership}

A crucial aspect of the SDSS collaboration philosophy is to provide a wide range of gratifying scientific, technical, pedagogic, and management opportunities for students,  postdocs, and junior and senior faculty, that would not be available to participants at their institutions alone. 
Drawing from the institutional breadth of the survey and with mutual respect as part of the core policies, a combination of senior scientists providing continuity (many participating from the inception of SDSS) and junior scientists taking on roles of collaboration responsibility, often for the first time, creates not only a robust talent reservoir for projects across the spectrum of astronomy, but fertile ground where new talent can be cultivated.  We advance this program both within standing advisory committees and programs within SDSS-V as well as our policy repository, which contains the living documents that form the collaboration rules.  

\subsection{Education}
Education and outreach have been essential elements throughout SDSS I-V. This spans across multiple audiences, from teachers of school classes looking to access real astronomical data, to undergraduate students seeking an introduction to core astronomy topics, to researchers looking to take their first steps into using our data at institutions that do not have a research mission. We provide resources tailored to each of these audiences, allowing them to use our data for their own purposes. By doing so we broaden both our audience, and the scope of our project's impact. 

The Faculty and Student Teams (FAST) program enables access to SDSS data from organizations that do not otherwise have a research mission and, therefore, would not normally have access to the SDSS collaboration.  The program pairs SDSS mentors with faculty and students at these institutions and broadens research experience and training to these organizations.  This fills an important gap in enabling access to the survey data as broadly as possible -- one of the core missions of SDSS.

Its current flagship, Voyages\footnote{\url{https://voyages.sdss.org}}, offers a set of educational activities of various STEM topics related to astronomy and SDSS operations, such as astrometry, magnitudes, and spectra. These activities are aimed at K-12 educators and their students, and include different routes or voyages into exploring the solar system, stars, and galaxies. Many of these activities were previously offered through SkyServer. For undergraduate students and their educators, SDSS offers online notebooks integrated into the SciServer environment, to avoid downloading large data sets and to allow students to work on their labs independently. For both these notebooks and Voyages, the philosophy of SDSS has always been that we want pupils and students to work with real astronomical data, which allows them to explore and cultivate their interest in astronomy \citep{2019BAAA...61..261L}.

While no longer used for scientific observations, SDSS plates are still used in our ``Plates for Education" program that offers an alternative access path to interact with SDSS data.  These programs equip students and teachers with former plug plates along with information about the specific observations conducted with that plate.  

SDSS was also the data source for the first iterations of the pioneering Galaxy Zoo citizen science project (\citealt{Lintott2011}, \citealt{Willett2013}), providing imaging data that was then classified by an online community of volunteers.   

SDSS has also collaborated with several artists. The resulting artwork has offered a new and original view of SDSS, and extended the project's reach beyond an audience that already has a scientific interest in astronomy. These collaborations highlight that science can be approached (and enjoyed) from multiple angles, of which art is one that is accessible to a large number of people. SDSS data features in \emph{Pacific Standard Time: Art and Science Collide}, a collaboration with the Getty Museum and the Los Angeles County Museum of Art which has permanently acquired an SDSS-inspired artwork by the US artist Josiah McElheny. 

The SDSS plug plates have featured multiple times in artistic interpretations of SDSS. Examples include 'The Beginning of Infinity'\footnote{\url{https://whitespace.cn/exhibitions/the-beginning-of-infinity/}} by Chinese artist Jian Yang. In this work, an SDSS plate is used to represent the foot of an elephant, in reference to Chinese folklore stories. US artist Kathryn Cellerini Moore created an immersive space with her 'Galactic Ping-Pong'\footnote{\url{https://www.kmoostudios.com/copy-of-matter-splatter-spectrum-scatter}} installation, which featured SDSS plates alongside salt sculptures and Chandra X-ray data. SDSS-IV Scottish artist in residence\footnote{\url{https://press.sdss.org/artist-in-residence}} Tim Fitzpatrick explored the properties of spectra and SDSS plates in several of his art installations and murals, including at Las Cruces Space Festival.  A further selection of artwork is featured on the SDSS webpage\footnote{\url{https://www.sdss.org/education/sdss-in-art}}. 

SDSS has not only led to visual artworks, but also inspired musical composition: 'Symphonies of Galaxies' by Scottish composer Eddie McGuire is a response to the MaNGA survey in SDSS-IV. Its four movements, 'Dust-veiled Starlight', 'Embrace, Waltz, Merge', 'Dark Matter Echoes', and 'Galaxies, dancing', explore the structures of galaxies, and their evolutionary paths. This piece premiered in St Andrews, Scotland in 2015, and the sheet music is available from the Scottish Music Centre\footnote{\url{https://www.scottishmusiccentre.com}}. A performance report with notes from the composer is found in 
\citet{symphony_of_galaxies}.

\subsection{Inclusion}

The \emph{Committee On INclusiveness} in SDSS (COINS), established during the fourth phase of the Sloan Digital Sky Survey (SDSS-IV),emerged as the collaboration itself grew and in order to address issues of inclusion in the survey.  The main goals of COINS are to assess and understand the SDSS collaboration's climate and demographics, to recommend to the collaboration management new policies or practices with regard to removing barriers to participation, and to assist in the implementation of these recommendations where necessary. Periodically, the COINS team collects, analyzes and publicizes results to the SDSS collaboration regarding the demographic of the survey (e.g., Lundgren et al 2015, Jones et al 2023). COINS is responsible for ensuring that resources that promote inclusion are widely accessible to members of the SDSS collaboration. Working together with the Management Committee, the ultimate goal of COINS is to provide tools to the collaboration and its members that ensure a collaborative environment where members are able to achieve to their potential.

\section{Conclusion}
This paper has provided a general overview of the SDSS-V program, its science goals, the hard- and soft-ware infrastructure that is required to realize those goals, and the consortium organization.  Inevitably, such an overview is only ``top level" and incomplete in many pertinent details. We therefore encourage the reader to seek additional details about each of these elements in the referenced companion papers.  

While survey progress depends sensitively on each sub-program, as of January 2025, we are 60\% through with MOS observations at APO and 40\% through MOS observations at LCO

For black hole science, SDSS-V will provide the largest sample of robust and precise (supermassive) black hole masses until the LISA mission obtains science data.  For stellar science, SDSS-V will obtain the most comprehensive and uniform set of stellar spectra in the infrared across the HR diagram.  For Galactic science, SDSS-V will provide the best contiguous age map of the Galactic Disk for the foreseeable future.  For understanding the ISM physics of galaxy formation, SDSS-V will provide essential diagnostic benchmarks that mass-function-matching simulations must pass in order to be considered "realistic".

In addition to what SDSS-V will do on its own, the survey will provide the bootstrapping engine for other surveys to thrive outside of their originally conceived domains. This data set will prove to be important for extending the space and ground data sets of other projects further than could have otherwise gone.  Cosmological MOS programs such as e.g. DESI, PFS, 4MOST, and SPHEREx will benefit from our stellar library that now extends across the HR diagram, thereby making their "contaminants" and "ancilary targets" many times richer in their scientific treasure. We recognize that if SDSS-V is to be of maximal lasting scientific impact this sketch of its scientific scope will have to pale in comparison to the true impact through science beyond this initial scope.  

Finally, we hope that the researchers involved in SDSS-V  -- early career and senior alike --  have an enduring appreciation for how a heterogeneous admixture of individual scientists can work together in the absence of a governmental mandate or financial incentive, across cultural, linguistic, political and temporal differences, and make something extraordinary happen that is much larger than anything they could do under the force of their individual genius.   
%We hope these young people gift this knowledge to future generations of humans and AGIs so that we may have a future together.

\bigskip
%\begin{acknowledgements}
ACKNOWLEDGEMENTS:

Funding for the Sloan Digital Sky Survey V has been provided by the Alfred P. Sloan Foundation, the Heising-Simons Foundation, the National Science Foundation, and the Participating Institutions. SDSS acknowledges support and resources from the Center for High-Performance Computing at the University of Utah. SDSS telescopes are located at Apache Point Observatory, funded by the Astrophysical Research Consortium and operated by New Mexico State University, and at Las Campanas Observatory, operated by the Carnegie Institution for Science. The SDSS web site is www.sdss.org.

SDSS is managed by the Astrophysical Research Consortium for the Participating Institutions of the SDSS Collaboration, including Caltech, The Carnegie Institution for Science, Chilean National Time Allocation Committee (CNTAC) ratified researchers, The Flatiron Institute, the Gotham Participation Group, Harvard University, Heidelberg University, The Johns Hopkins University, L’Ecole polytechnique fédérale de Lausanne (EPFL), Leibniz-Institut für Astrophysik Potsdam (AIP), Max-Planck-Institut für Astronomie (MPIA Heidelberg), The Flatiron Institute, Max-Planck-Institut für Extraterrestrische Physik (MPE), Nanjing University, National Astronomical Observatories of China (NAOC), New Mexico State University, The Ohio State University, Pennsylvania State University, Smithsonian Astrophysical Observatory, Space Telescope Science Institute (STScI), the Stellar Astrophysics Participation Group, Universidad Nacional Autónoma de México, University of Arizona, University of Colorado Boulder, University of Illinois at Urbana-Champaign, University of Toronto, University of Utah, University of Virginia, Yale University, and Yunnan University.

%\end{acknowledgements}

\facilities{Du Pont, Sloan, Spitzer, WISE, 2MASS, Gaia, eROSITA, TESS, ZTF}

\bibliography{ms}{}
\bibliographystyle{aasjournal}

%% This command is needed to show the entire author+affiliation list when
%% the collaboration and author truncation commands are used.  It has to
%% go at the end of the manuscript.
%\allauthors

\end{document}